\documentstyle[12pt,epsf]{article}
\def\lb       {\left( }
\def\rb       {\right) }
\def\lmb      {\left\{ }
\def\rmb      {\right\} }
\def\lbb     {\left[ }
\def\rbb      {\right] }
\def\comma      { \, , }
\def\period     { \, . }
\def\bra#1      { \langle \, #1 \, \vert \, }
\def\ket#1      { \, \vert \, #1 \, \rangle \, }
\def\semiket#1  { \, #1 \, \rangle \, }
\def\del        {  \partial  }
\def\half       {  {1\over 2}  }

\def\defint#1#2 {  \int_{#1}^{#2}  }
\def\rootof#1   {  \left( #1 \right)^{1/2}  } 
\def\deldel#1   {  {\partial\over \partial #1}  }
\def\Tr      { \ \mbox{Tr} \  }
\def\abs#1      {  \, \vert #1 \vert \,   }

\def\evalat#1   {  \left\vert_{#1} \right. }

\def\when       { \biggm{\vert} }
\def\lsim    {\lower .65ex \hbox{\ $\stackrel{<}{\sim}$\ } }
\def\gsim    {\lower .65ex \hbox{\ $\stackrel{>}{\sim}$\ } }
\def\bfR     { {\bf R}}
\def\bfZ     { {\bf Z}}
\def\bfC     { {\bf C}}

\def\calB       { {\cal B} }

\def\calD       { {\cal D} }

\def\calL       { {\cal L} }

\def\calO       { {\cal O} }

\def\vecii#1#2      {  \left(\begin{array}{c}#1\\#2\end{array}\right)  }
\def\veciii#1#2#3   {  \left(\begin{array}{c}#1\\#2\\#3\end{array}\right)  }
\def\veciv#1#2#3#4  {  \left(\begin{array}{c}#1\\#2\\#3\\#4
                                 \end{array}\right)  }
\def\vecv#1#2#3#4#5 {  \left(\begin{array}{c}#1\\#2\\#3\\#4\\#5
                                 \end{array}\right)  }

\def\matrixii#1#2#3#4            {  \left(\begin{array}{cc}#1&#2\\#3&#4
                                       \end{array}\right) }
\def\matrixiii#1#2#3#4#5#6#7#8#9 {  \left(\begin{array}{ccc}#1&#2&#3\\
                                     #4&#5&#6\\#7&#8&#9\end{array}\right)  }
\def\mativ#1#2#3#4               {  \left(\begin{array}{cccc}
                                       #1\\#2\\#3\\#4\end{array}\right) }
\def\matv#1#2#3#4#5              {  \left(\begin{array}{ccccc}
                                     #1\\#2\\#3\\#4\\#5\end{array}\right)  }
\def\eqabegin         {  \begin{eqnarray}  }
\def\eqaend           {  \end{eqnarray}  }
\def\nn               {  \nonumber  }
\def\parn              {  \par\noindent }
\def\parbigskip        {  \par\bigskip  }
\def\parmedskip        {  \par\medskip  }
\def\parsmallskip      {  \par\smallskip  }
\def\parbigskipn        {  \par\bigskip\noindent  }
\def\parmedskipn        {  \par\medskip\noindent  }
\def\parsmallskipn      {  \par\smallskip\noindent  }
\def\boxit#1#2      {  \vbox{\hrule\hbox{ \hskip -4.1pt \vrule\kern3pt 
                       \vbox
                    {  \hsize #1 \strut\kern3pt #2 \kern3pt\strut  }
                       \kern3pt  \vrule} \hrule  } }
\def\centerbox#1#2  {  \mbox{  }\par\bigskip  \hfil \boxit{#1}{#2} \hfil
                       \par\bigskip\noindent }
\def\rightbox#1#2   {  \hfill\boxit{#1}{#2}  }
\def\leftbox#1#2    {  \boxit{#1}{#2}  }
\def\fullbox#1      {  \boxit{\textwidth}{#1}  }
\newenvironment{namelist}[1]{%
   \begin{list}{}
      {
       \settowidth{\labelwidth}{#1}
       \setlength{\leftmargin}{1.1\labelwidth}}
}{%
 \end{list}}
\def\titleandfile#1#2   {  \begin{center}{\Large\bf #1}\end{center}
                            \par\begin{flushright} #2 \end{flushright}  }
\def\sectionnumbering { \setcounter{equation}{0}
         \renewcommand{\theequation}{\arabic{section}.\arabic{equation}}}
\def\csection#1    { \par\medskip\noindent \begin{center} \let\boldface\bf \def\bf{\large\sc} 
               \section{#1} \let\bf\boldface \end{center} \sectionnumbering 
          \par\medskip\noindent}
\def\csectionast#1    { \begin{center} \let\boldface\bf \def\bf{\large\sc} 
               \section*{#1} \let\bf\boldface \end{center} \sectionnumbering }
\def\csubsection#1 {   \subsection{#1}   }
\def\isubsubsection#1 {  \let\boldface\bf \def\bf{\normalsize\it} 
               \subsubsection{#1}    \let\bf\boldface   }
\def\mcapsection#1   {  \begin{center} \csection{#1} \end{center} }
\def\SLtwo {  SL(2,R)  }
\def\cSLtwo{ \widetilde{SL}(2,R)  }
\def\sltwo {  sl(2,{R})  }
\def\AdS  {  AdS_3  }
\def\CAdS {  \widetilde{AdS}_3  }
\def\lp   { l_p }

\def\ginv {g^{-1}}
\def\delbar {\bar{\partial}}
\def\zbar   {\bar{z}}
\def\wbar   {\bar{w}}
\def\Jbar   { \tilde{J} }

\def\Lbar   {\tilde{L}}
\def\Nbar   {\tilde{N}}
\def\hatt   {\hat{t}} 
\def\hatphi {\hat{\varphi}}
\def\hatr   {\hat{r}}

\def\hatE  { \hat{E} }
\def\hatN  { \hat{N} }
\def\thetaL {\theta_L}
\def\thetaR {\theta_R}
\def\D#1     {{}^{#1} \! D}

\def\rp     {r_+}
\def\rmi    {r_-}
\def\deltap {\Delta_+}
\def\deltam {\Delta_-}
\def\alp {\alpha'}
\def\nw   { n_{ {\scriptscriptstyle W}} }
\def\nwn#1  { n_{ {\scriptscriptstyle W},#1} }
\def\gBHu  { g^{ {\scriptscriptstyle BH} } }
\def\gBHd   { g_{ {\scriptscriptstyle BH} } }
\def\mj   { m_J }

\def\Ichi { {\cal I}_\chi }

\def\ketpm#1  { \vert \ #1 \  \rangle {}_{\pm} }
\def\ketp#1 { \vert \ #1 \bigm)  }

\def\tilg  {\tilde{g}}

\def\Zphi  { Z_\varphi }

\def\rp          {r_+}
\def\rmi         {r_-}

\def\omh         {\Omega_H}
\def\th0         {\theta_0}
\def\Gf          {G_F}
\def\Gfe         {G_F^E}
\def\Gbh         {G_{\scriptscriptstyle BH  } }
\def\Gbhe        {G_{\scriptscriptstyle BH  }^E}
\def\tGf         { \tilde{G}_{F}^{E} }
\def\dH          {d_H}
\def\aH          {a_{\scriptscriptstyle H}}
\def\bH          {\beta_H}
\def\eH          {\epsilon_H}
\def\phip        {\varphi^+}

\def\deltau      { \Delta \tau }
\def\delvarphi      { \Delta \varphi }
\def\delphiE     { \Delta \varphi_{\scriptscriptstyle E } }
\def\delphiEp    { \Delta \varphi^{\scriptscriptstyle E+} }
\def\varphiE     { \varphi_E }
\def\JBH         { J_{ {\scriptscriptstyle BH  } } }
\def\MBH         { M_{ {\scriptscriptstyle BH  } } }
\def\sBH         { s_{ {\scriptscriptstyle BH  } } }
\def\SBH         { S_{ {\scriptscriptstyle BH  } } }
\def\nuBH        { \nu_{ {\scriptscriptstyle BH  } } }
\def\delom       { s }

\def\cnp         {  c_n^+  }
\def\cnm         {  c_n^-  }
\def\cn          {  c_n  }
\def\sn          {  s_n  }
\def\An          { A_n }
\def\Bn           { B  }
\def\bbh          {2 \pi \beta/\bH}
\begin{document}
%
%
\def\papertitlepage{\baselineskip 3.5ex \thispagestyle{empty}}
\def\Title#1{\baselineskip 1cm \vspace{1.5cm}\begin{center}
 {\Large\bf #1} \end{center}
\vspace{0.5cm}}
\def\Abstract {\vspace{1.0cm}\begin{center} {\large\bf Abstract}
           \end{center} \par\medskip}
\def\preprinumber#1#2#3{\hfill \begin{minipage}{4.2cm}  #1
              \par\noindent #2
              \par\noindent #3
              \end{minipage}}
\renewcommand{\thefootnote}{\fnsymbol{footnote}}
%
%
\papertitlepage
\preprinumber{March 1997}{UT-Komaba 97-9}{hep-th/9705209}
\vspace*{0.8cm}
\Title{ Study of Three Dimensional \\ \vskip 0.5ex 
 Quantum Black Holes\footnote{Ph.D. thesis at University of Tokyo} }
\begin{center}
 \vskip -1ex
   {\sc Yuji ~Satoh } \\
    {\sl Institute of Physics, University of Tokyo, Komaba \\
 \vskip -2ex
   Meguro, Tokyo 153, Japan\footnote[2]{address after April 25: 
  Department of Physics, Princeton
  University, Princeton, NJ 08544, USA \par \quad 
   e-mail: ysatoh@viper.princeton.edu } }  
\end{center}
\vskip -1ex
\baselineskip=3.1ex
\Abstract
We investigate 
quantum aspects of the three dimensional black holes
discovered by Ba\~{n}ados, Teitelboim and Zanelli.
The discussions are devoted to two subjects:
the thermodynamics of quantum scalar fields 
and the string theory in the three dimensional black hole backgrounds.
We take two approaches to the thermodynamics. 
In one approach we use mode expansion, and in the other we use Hartle-Hawking
Green functions. We obtain exact expressions of mode functions, 
Hartle-Hawking Green functions, Green functions on a cone geometry, 
and thermodynamic quantities. The entropies 
depend largely upon methods of calculation including
regularization schemes and 
boundary conditions. This indicates the importance of precise discussions 
on the definition of the thermodynamic quantities for understanding 
black hole entropy.
Then we investigate the string theory 
in the framework of conformal field theory. 
The model is described by an orbifold of the $ \cSLtwo $ WZW model. 
We discuss the spectrum by solving the level 
matching condition and obtain winding modes. We analyze the physical 
states and investigate the ghost problem. Explicit examples of 
negative-norm physical states (ghosts) are found.
Thus we discuss possibilities for obtaining a sensible theory. 
The tachyon propagation and the target-space geometry  
are also discussed.
This is the first attempt to quantize a string in 
a black hole background with an infinite number of propagating modes.
Although we cannot overcome all the problems, our results may 
provide a useful basis for both subjects.
%
\newpage
\pagestyle{empty}
\baselineskip=0.6cm
%
%
\csectionast{CONTENTS} \setcounter{section}{0}
\parn \parbigskipn
{\bf 1 \quad Introduction } \parn
\ \ \quad { 1.1 \quad Black holes and quantum gravity } \parn
\ \ \quad { 1.2 \quad Organization of the thesis } 
\parbigskipn
{\bf 2 \quad The Three Dimensional Black Hole } 
\parbigskipn
{\bf 3 \quad Thermodynamics of Scalar Fields in the Three Dimensional 
  Black Hole \\ \hspace*{1.5em} Background } \parn
\ \ \quad { 3.1 \quad Black hole entropy } \parn
\ \ \quad { 3.2 \quad Thermodynamics of scalar fields by summation over 
  states } \parn
\ \ \quad { 3.3 \quad Green functions in the three dimensional 
  black hole background } \parn
\ \ \quad { 3.4 \quad Thermodynamics of scalar fields by Hartle-Hawking 
  Green functions } \parn
\ \ \quad { 3.5 \quad Discussion } 
\parbigskipn
{\bf 4 \quad String Theory in the Three Dimensional Black Hole Geometry }
  \parn
\ \ \quad { 4.1 \quad The three dimensional black hole as a string 
  background } \parn
\ \ \quad { 4.2 \quad The spectrum of a string on 
  $ \mathaccent "0365 {SL}(2,R) / \Zphi $ orbifold } \parn
\ \ \quad { 4.3 \quad Investigation of unitarity } \parn
\ \ \quad { 4.4 \quad Tachyon and target-space geometry } \parn 
\ \ \quad { 4.5 \quad Discussion} 
\parbigskipn
{\bf 5 \quad Conclusion }
\parbigskipn
{\bf Acknowledgements }
\parbigskipn
{\bf Appendix } \parn
\ \ \quad { A \quad \ \ The Feynman Green function in 
  $ \mathaccent "0365 {AdS}_3 $} \parn
\ \ \quad { B \quad \ \ The Sommerfeld representation of 
  Green functions } \parn
\ \ \quad { C \quad \ \  Representations of $ SL(2,R) $ } \parn
\ \ \quad { D \quad \ \  Decomposition of the Kac-Moody module } 
\parbigskipn
{\bf References }
%
%
\newpage
\renewcommand{\thefootnote}{\arabic{footnote}}
\setcounter{footnote}{0}
\baselineskip = 0.6 cm
\pagestyle{plain}
\setcounter{page}{1}
%
%
%
\csection{INTRODUCTION}
%
\vskip 0.5ex
\csubsection{Black holes and quantum gravity}
In the early 1960s,  a rapid development began in general 
theory of relativity \cite{MTWW}. New mathematical techniques  simplified 
calculation of physical quantities. The progress of technology
enabled us to carry out various experiments.  
They changed general relativity into a tractable physics 
 from a profound but abstract theory.
By the end of the 1970s, we had obtained a deeper understanding of general 
relativity \cite{MTWW}-\cite{Wa}. 
This theory passed every experimental test. New astronomical 
objects (quasars, pulsars, black holes) 
and the cosmic background radiation were discovered. We found 
evidence of the existence of gravitational waves from a binary pulsar. 
Furthermore, global structure of space-time and properties 
of singularities and black holes were clarified. 
A number of exact solutions were also found.
\parsmallskip
In parallel with this progress, another excellent 
development took place in elementary particle physics. After a skeptic 
period over quantum field theory, the standard model was established. 
 Precise experiments became possible, and they 
revealed the nature of physics at very small scale. The experiments stimulated 
the progress of the theory, and vice versa. 
At last, based on simple principles, i.e., Lorentz invariance, 
the gauge principle and renormalizability, 
we found a beautiful theory of elementary particles. 
We obtained the basic law which describes fundamental 
interactions except for gravity in a quantum mechanical manner.
\parsmallskip
Through these developments, we acquired a profound insight into nature.
Natural questions then arise: Can we go beyond the standard model ?
Can we construct the theory of quantum gravity ?
These are important and challenging subjects in physics today.
\parsmallskip
In the investigation of quantum gravity, 
black holes are regarded as an excellent arena. 
They include a strong coupling region of gravity where quantum 
effects may become important. We can easily draw 
a physical picture of them. Moreover, close relations between 
black hole physics and quantum theory have been found through the study 
of black hole thermodynamics.
\parsmallskip
Black holes  obey analogs
of the laws of thermodynamics \cite{Wa}-\cite{Be},  
and this is called black hole thermodynamics. At first sight, the analogy
seems superficial since we have no reason that black holes are thermal.
However, we find arguments for this analogy by semiclassical analyses.
Hawking found that black holes emit thermal radiation 
by effects of quantum fields (Hawking radiation) \cite{Ha}. 
Bekenstein argued that the second law of the black hole 
thermodynamics is valid for a system of gravity and quantum matter
(generalized second law) \cite{Be,Ha2}.  
Thermal properties of black holes were further discussed  using 
the path integral and Green functions \cite{HH}-\cite{GibH}. 
These imply connections among black hole physics, quantum mechanics 
and statistical mechanics. In addition, the Hawking radiation 
indicates a contradiction to the unitary evolution of quantum mechanics
(the information paradox). 
Therefore, many works have been devoted to black hole thermodynamics 
for understanding it
from the microscopic, quantum mechanical and statistical mechanical 
point of view. We expect to get clues to quantum gravity
by the investigation of black holes. 
We also expect to understand fundamental 
problems concerning gravity, e.g., the problem of singularities and of the 
early universe.
%
\parsmallskip
Since string theory is the candidate of the fundamental theory including 
gravity, it should give the answer to the problems about 
quantum black holes. To this end, we have to develop analysis 
beyond low energy effective theory and the $ \alp $ (Regge slope)
expansion. The $ \SLtwo /U(1) $ black hole \cite{Witten,MSW} 
serves as a useful model in this respect. It gives an exact background 
of a string, and it is described by a simple Wess-Zumino-Witten
(WZW) model \cite{WZW}. Thus the properties of the $ \SLtwo /U(1) $ 
black hole have been  studied extensively 
(see, e.g., \cite{Witten}-\cite{BB}). 
  Nevertheless we need further 
investigations to clarify important issues of black hole physics.
The difficulties are rooted
in the fact that the target space is non-compact and curved in time
direction. Such difficulties are not characteristic of string theories 
in black hole backgrounds. They are typical of a string theory in a 
non-trivial background with curved time.
Although we have many consistent string theories on curved spaces, i.e., on
group manifolds, they are compact and must be tensored with Minkowski
spacetime. There have been a few previous attempts besides the
$ \SLtwo /U(1) $ case. For example, there are attempts
using the $ \SLtwo $ WZW model \cite{BOFW}-\cite{BN},
but it is known to contain
ghosts.\footnote{A resolution to the ghost problem has recently been  proposed 
in \cite{Bars}.}
So far, we have few consistent string theories with
curved time.\footnote{See, however, \cite{RT} for instance.}
\parsmallskip
In this thesis, we investigate quantum aspects of the three dimensional 
black holes discovered by Ba\~{n}ados, Teitelboim and Zanelli (BTZ) \cite{BTZ}.
The BTZ black hole is 
 a solution to the vacuum Einstein field equations with a negative
cosmological constant.   
It shares many characteristics with the $ (3+1) $-dimensional Kerr black hole 
(for a review, see, e.g., \cite{Carlip}).
Moreover, it provides a very simple system: it has a constant curvature and 
no curvature singularities.  
Therefore we can discuss many characteristics of 
the black hole physics in an explicit manner without mathematical complications.
Another importance of the BTZ black hole is the fact that
this is one of the few known exact solutions in string theory 
and one of the simplest solutions \cite{HW,Kaloper}. 
In addition, a string in three dimensions has an infinite number of
propagating modes, so it resembles a higher dimensional one. 
This theory has significance also as a string theory in non-trivial
curved spacetime. 
Thus we may obtain useful insights into important issues of quantum gravity
through the study of the string theory in the BTZ black hole background.
\parsmallskip
The aim of this thesis is to investigate quantum gravity by the analysis of 
quantum aspects of the three dimensional black hole.
In particular, we have two main purposes in this thesis.
One is to investigate the thermodynamics of 
quantum fields in the three dimensional black hole background \cite{IS}. 
Recently, black hole thermodynamics has attracted attention again 
\cite{Sr}-\cite{SF}. 
These works include suggestive arguments for the microscopic origin 
of black hole entropy and for the relation among the information 
paradox, the black hole entropy and the renormalization of the gravitational 
coupling constant. However, the arguments seem formal and deal mainly 
with the flat-space limit. In our case, we can study this issue
in a truly curved background and in an explicit manner. We get 
exact expressions and discuss the problems without ambiguity.
We find that thermodynamic quantities depend largely upon methods of 
calculation and that the results concerning divergences and the role of horizons 
do not necessarily agree with \cite{Sr}-\cite{SF}. 
These indicate the importance of curvature effects and 
precise discussions on the definition of 
the thermodynamic quantities including regularization schemes and boundary 
conditions. We need further investigations in these respects.
Our results, however, serve as a reliable basis for the quantum field theory 
and the thermodynamics of quantum scalar fields 
in the BTZ black hole background.
\parsmallskip
The other purpose is to investigate the string theory 
in the three dimensional black hole geometry \cite{NS}.
One of the motivations is to settle the open problems 
of the thermodynamics of the quantum fields and to clarify the 
microscopic origin of black hole thermodynamics.
However, the purpose here is more general as explained above.
In spite of its importance, detailed analyses have not been made so far.
This is partly because we do not know much about string theory in curved 
spacetime and it is not clear whether the string theory in the BTZ 
background satisfies consistency conditions as a sensible theory.
We investigate this string theory in detail 
in the framework of conformal field theory. 
We analyze the spectrum and obtain winding modes. We study the physical 
states. We examine the ghost problem and 
find explicit examples of ghosts. 
This means that our model is not unitary as it stands. 
Thus we discuss possibilities for obtaining a sensible theory.
The tachyon propagation and   
the target-space geometry are also discussed.
Although we cannot overcome all the problems, 
our work may provide a starting point
for further investigations of this issue.
\csubsection{Organization of the thesis}
This thesis is organized as follows.
\parsmallskip
In chapter 2, we review the three dimensional (BTZ) black holes.
\parsmallskip
In chapter 3, we discuss 
the thermodynamics of scalar fields in the BTZ black hole background. 
In section 3.1, we briefly review black hole thermodynamics and recent 
arguments about black hole entropy which are relevant to our discussion.
We take two approaches to the thermodynamics of the scalar field. 
In section 3.2, we utilize 
mode expansion and summation over states.
We find exact mode functions and thermodynamic 
quantities. In section 3.3, we construct Hartle-Hawking Green functions.
We investigate the thermodynamics based on the Euclidean 
Hartle-Hawking Green functions
in section 3.4. We obtain free energies and
Green functions on a cone geometry. By using them, we calculate 
entropies. These are also exact in the framework of quantum field theory in 
curve spacetime. Discussion on our results is given 
in section 3.5.
\parsmallskip
In chapter 4, we investigate the string theory in the three dimensional 
black hole geometry. 
In section 4.1, we review the BTZ black hole from the string-theory
point of view. In section 4.2, we construct the orbifold   
of the $ \cSLtwo $ WZW model which describes the string in the black hole 
background. We analyze the spectrum by solving the 
level matching condition. We obtain winding modes. In section 4.3,  
we investigate the ghost problem. We find explicit examples of 
negative-norm physical states. 
In section 4.4, we discuss the tachyon propagation and the target-space
geometry. We find a T-duality transformation reversing the black hole mass.
In section 4.5, we discuss possibilities for obtaining a sensible theory.
\parsmallskip
We conclude this thesis in chapter 5.
\parsmallskip
Technical details and mathematical backgrounds are collected in appendix A-D.
In appendix A, we summarize the derivation of the Feynman Green function
in $ \CAdS $. In appendix B, we derive Green functions on a cone geometry
by using the Sommerfeld integral representation. In appendix C, 
we collect basic properties of the representation theory of $\cSLtwo $.
Representations in the hyperbolic basis are explained in some detail.
Finally in appendix D, we show the Clebsch-Gordan decomposition of 
the $ \sltwo $ Kac-Moody module in the hyperbolic basis.
%
%
%
\csection{THE THREE DIMENSIONAL BLACK HOLE}
In this section, we review the three dimensional
black hole discovered by 
Ba${\rm \tilde{n}}$ados et al. (the BTZ black hole) \cite{BTZ}. 
\parsmallskip
We begin with the 
three dimensional anti-de Sitter space ($ \AdS $).
$ \AdS $ is a three dimensional hyperboloid 
embedded in a flat space with the metric
\eqabegin
      ds^2 &=& -dx_0^2 -dx_1^2 + dx_2^2 + dx_3^2 \comma 
    \label{flatG}
\eqaend
through the equation
\eqabegin
           - x_0^2 - x_1^2  + x_2^2 + x_3^2 &=& - l^2 
         \period \label{embedding}
\eqaend
It is a maximally symmetric space 
and forms a solution to Einstein gravity with a negative cosmological
constant $ - l^{-2} $. The curvature tensors are
\eqabegin
  &&  R_{\mu \nu } = -2 l^{-2} g_{\mu \nu } \comma \quad 
    R = -6 l^{-2} \period \nn
\eqaend 
Notice that the scalar curvature is  constant.
In order to decompactify the time direction of $ \AdS $, 
we go to the universal covering space $ \CAdS $.
We then consider three regions parametrized by 
\eqabegin
 \begin{array}{lllll}
\mbox{ Region I} &  ( \hatr ^2 > l^2) &:& 
   x_1 = \hatr \cosh \hatphi \comma & 
     x_0 = \sqrt{\hatr^2 -l^2} \sinh \hatt \comma 
     \\ 
  && & x_2 = \hatr \sinh \hatphi \comma & 
      x_3 = \sqrt{\hatr^2 -l^2} \cosh \hatt \comma 
   \\
\mbox{ Region II} &  ( l^2 > \hatr ^2 > 0) &:& 
   x_1 = \hatr \cosh \hatphi \comma &  
      x_0 = \sqrt{l^2 -\hatr^2} \cosh \hatt \comma 
    \\ 
 & && x_2 = \hatr \sinh \hatphi \comma &
      x_3 = \sqrt{l^2 -\hatr^2} \sinh \hatt \comma 
   \\
\mbox{ Region III} & ( 0 > \hatr ^ 2 ) &:& 
   x_1 = \sqrt{-\hatr^2} \sinh \hatphi \comma &
         x_0 = \sqrt{l^2 -\hatr^2} \cosh \hatt \comma 
    \\ 
 &&& x_2 = \sqrt{-\hatr^2} \cosh \hatphi \comma & 
         x_3 = \sqrt{l^2 -\hatr^2} \sinh \hatt \comma 
 \end{array}
  \label{xtohat}
\eqaend
where $ - \infty < \hatt \comma \hatphi < \infty $.
In every parametrization, substituting it into (\ref{flatG}) yields
the metric of $ \CAdS $ of the form
\eqabegin
 ds^2  
  &=&
     - \lb \frac{\hatr^2}{l^2} -1 \rb d \hatt^2 + \hatr^2 d \hatphi^2  
            + \lb \frac{\hatr^2}{l^2} -1 \rb^{-1} d \hatr^2   \period \nn
\eqaend
$ \del_{\hatt} $ and $ \del_{\hatphi} $ generate isometries. These
correspond to boost symmetries in the flat space. 
We make a further change of variables,  
\eqabegin
    \frac{\hatr^2}{l^2}  & = &  \frac{r^2 - \rmi^2}{\dH ^2} \comma 
                      \quad 
    \vecii{\hatt}{\hatphi} = \frac{1}{l}
     \matrixii{\rp}{-\rmi}{-\rmi}{\rp}
   \vecii{t/l}{\varphi}  \comma  \label{tphi} 
\eqaend
where $ r_\pm $ $ ( \rp > \rmi \geq 0) $ are positive constants, and
\eqabegin
   \dH ^2 &=& \rp ^2 - \rmi ^2 \period \nn
\eqaend
Then, by identifying  the points under a discrete subgroup of an isometry 
\eqabegin
   \varphi & \sim &  \varphi + 2 \pi n \quad (n \in \bfZ) \comma \nn
\eqaend
 one obtains the geometry of the three dimensional black hole:
\eqabegin
  d \sBH ^2 &=& \gBHu _{\mu \nu} d x^\mu d x^\nu \nn \\
   &=&  
        - N_\bot^2 dt^2 + N_\bot^{-2} dr^2 
           + r^2 (N^{\varphi} dt + d\varphi)^2 \comma 
          \label{BTZbh} \\
  &=&  
 - \left( \frac{r^2}{l^2} - \MBH \right) d t^2 - \JBH d t d \varphi 
   + r^2  d \varphi^2 
     + \left( \frac{r^2}{l^2} - \MBH 
    + \frac{\JBH ^2}{4r^2}\right)^{-1} d r^2 
     \comma \nn 
\eqaend
where 
\eqabegin
   N_\bot^2 & = &  \frac{(r^2 - \rp^2)(r^2 - \rmi^2)}{l^2 r^2 } \comma \quad 
   N^{\varphi} \ = \  - \frac{\rp\rmi}{l r^2} \comma \nn \\
  l^2 \MBH & =  & (\rp ^2 + \rmi ^2) \comma \qquad l \JBH \ = \ 2 \rp \rmi
    \period \nn 
\eqaend
The coordinates in (\ref{BTZbh}) now take 
$- \infty < t < + \infty $, 
$0 \leq \varphi < 2 \pi $ and $0 \leq r < + \infty$.  
Under the above identification, $ r^2 = 0 $ is the fixed point 
for $ \JBH = 0 $, but for $ \JBH \neq 0 $ one has no fixed points. 
%
%
\newpage
\hspace*{-1.0cm}
\begin{minipage}[c]{3.0in}
\epsfxsize = 2.5in
\hspace*{0.3in}
\epsfbox{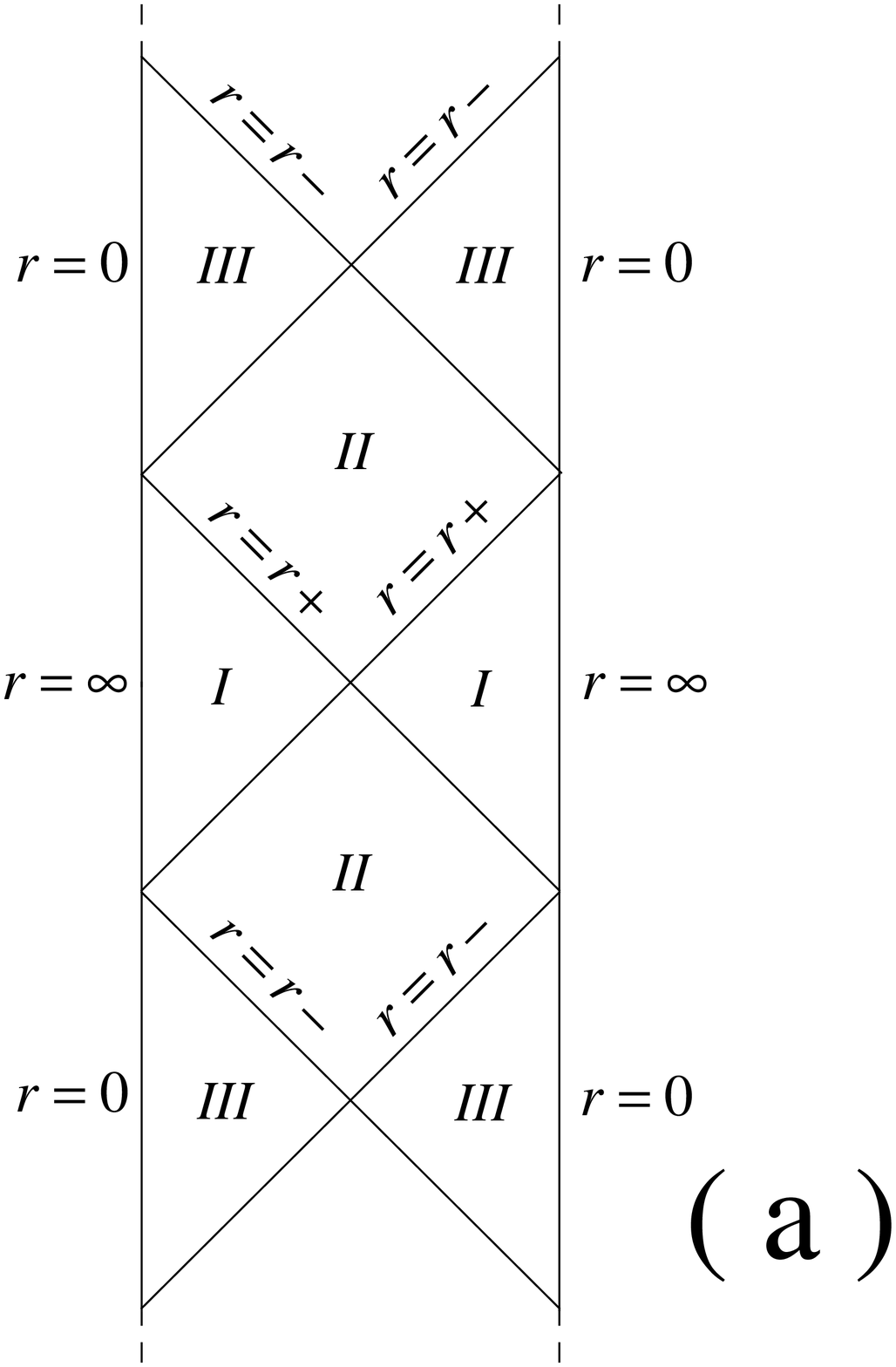}
\end{minipage}
\hfill
\begin{minipage}[c]{3.0in}
\epsfxsize = 2.5in 
\hspace*{1.0cm}
\epsfbox{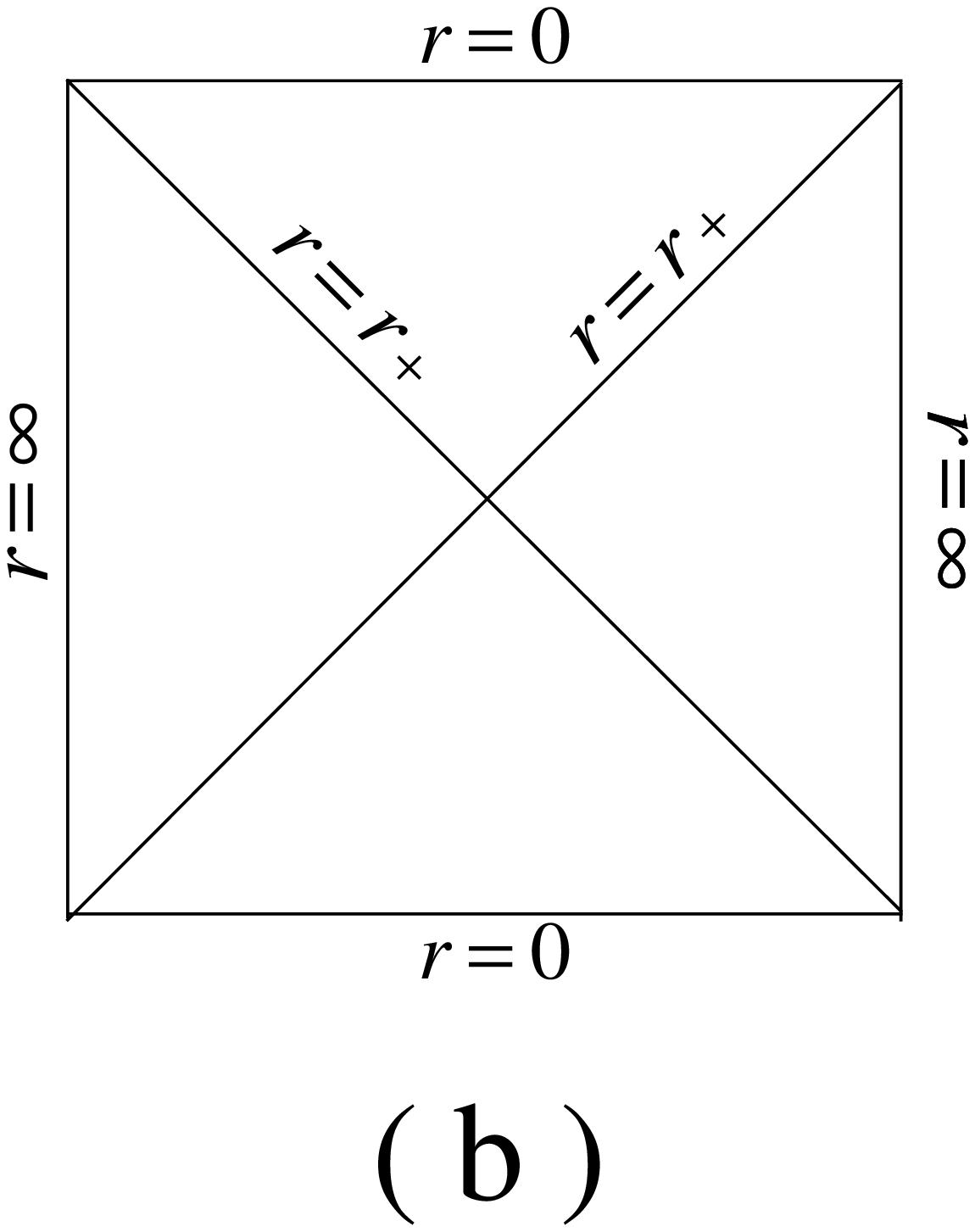}
\end{minipage}
\parn
\vfill
\begin{minipage}[c]{3.0in}
\epsfxsize 2.8in 
\epsfbox{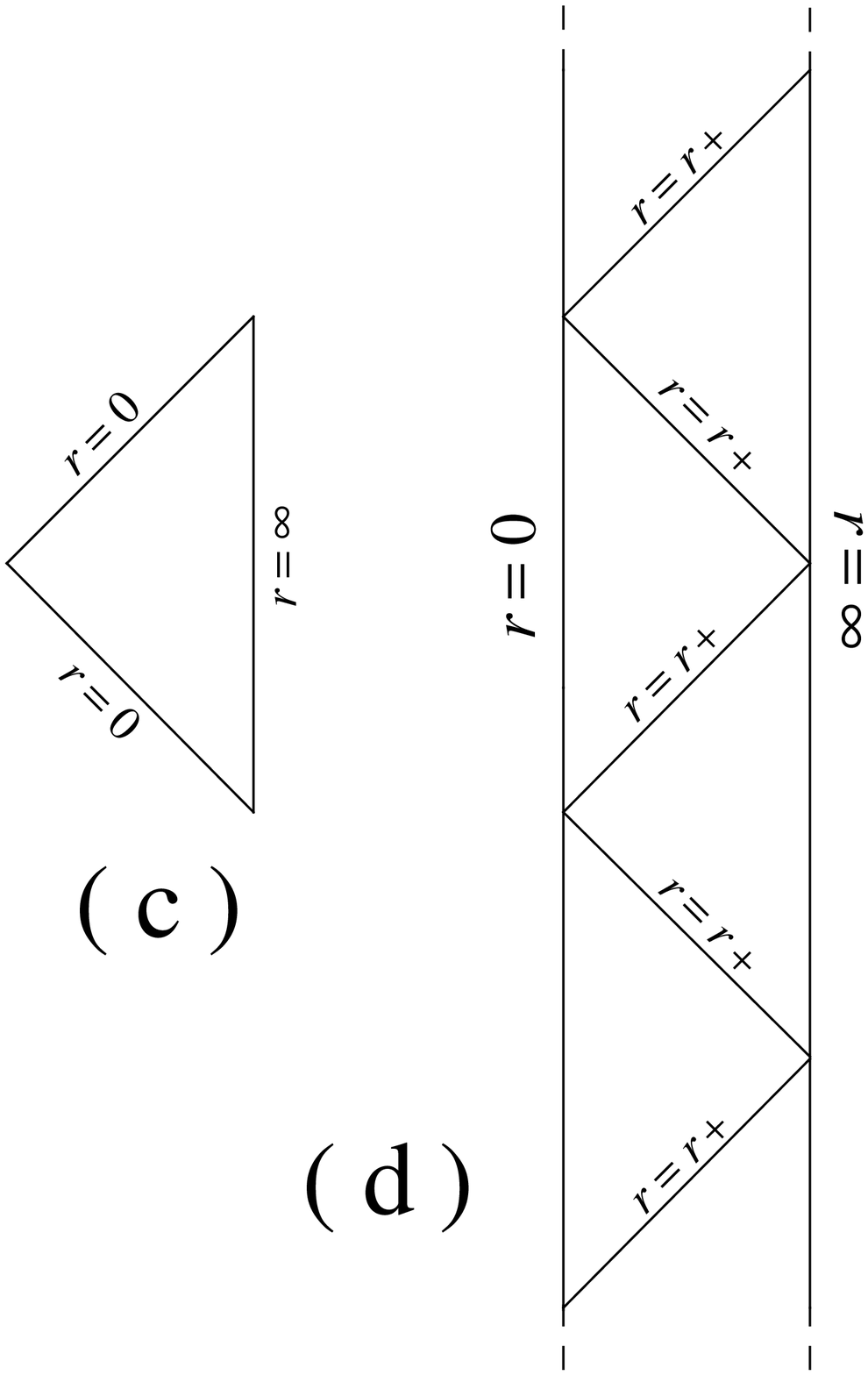}
\end{minipage}
\hfill
\begin{minipage}[c]{2.5in}
    \vspace*{-2.0in}
          {\bf  Figure 1a - 1d \ : }   \par \medskip \noindent
               Conformal diagrams for \par
          \quad  (a)  $ \JBH \neq 0  $ and $ \JBH/l \neq \MBH $, \par
          \quad  (b) $ \JBH = 0 $ and $ \MBH \neq 0 $, \par
          \quad  (c)   $ \MBH = \JBH/l = 0 $,  \par
          \quad  (d)   $ \MBH = \JBH/l \neq 0 $.
    \end{minipage}
\newpage
%
%
%
The above metric describes a non-extremal
 rotating black hole for $ \rmi \neq 0, \rp $.
$ \MBH $ and $ \JBH $ represent the mass and the angular momentum 
of the black hole, respectively. The black hole has two horizons
given by the surfaces $ r = r_\pm $. A conformal diagram for 
$ \rmi \neq 0 \comma \rp $ is shown in Figure 1a.
For $ \rmi = 0 $, the angular momentum $ \JBH $ vanishes and the black 
hole becomes non-rotating one. A conformal diagram for this case is given 
in Figure 1b. In the metric (\ref{BTZbh}),  we can take the limit
$ \rmi \to \rp \ (\JBH/l \to \MBH)$ although singular quantities
appear in the intermediate expressions. The resulting
geometry describes an extremal black hole (Figure 1c and 1d). The way  
to obtain the extremal black hole by an identification of $ \CAdS $
is quite different from the other cases.
In this thesis, we will focus on the non-extremal cases.
\parsmallskip
In the conformal diagrams, we have cut out the region $ r^2 < 0 $ where
closed timelike curves exist. There are arguments (i) that the inclusion  
or non-inclusion of this region is irrelevant to an observer outside 
the black hole because the surface $ r = \rp $ is the event horizon, 
and (ii) that the inclusion of matter produces a curvature singularity
at $ r = 0 $ (or $ r = \rmi$) and one {\it has to } drop that region
\cite{BTZ,GFs,LO}. We will briefly discuss this issue
later in the context of string theory.
\parsmallskip
Since the BTZ black hole is locally $ \CAdS $, it is
also a solution to Einstein gravity. 
The asymptotic region tends to be $ \CAdS $ 
instead of Minkowski spacetime.
The curvature is constant and there are 
no curvature singularities; namely, the BTZ black hole 
provides a very simple system. 
Thus, through its analysis, we can investigate many characteristics of 
black hole physics in an explicit manner without mathematical 
complications (for a review, see \cite{Carlip}).
Indeed, its properties have been
studied extensively in the classical theory. 
For example, 
the BTZ black hole shares many characteristics  
with the $ (3+1) $-dimensional Kerr black hole: it has an event horizon, 
an inner horizon and an ergosphere; it occurs as an 
end point of ``gravitational collapse''; it exhibits instability 
of the inner horizon; and it has a non-vanishing Hawking temperature
and various thermodynamic properties. 
The thermodynamic quantities are given by \cite{BTZ}
\eqabegin
   &&  \bH = \frac{ 2 \pi \rp l^2 }{\rp^2 - \rmi ^2 } \comma \quad 
       \SBH = \frac{4 \pi \rp}{l} \comma \quad \nuBH = \frac{\rmi}{\rp l}
   \comma  \label{clTS}
\eqaend
where
$ \bH $ is the inverse temperature; $ \SBH $ is the entropy; and 
 $ \nuBH $ is the chemical potential conjugate to $ \JBH $.   
Moreover, the utility of the BTZ black hole becomes evident 
in the quantum analysis. 
Quantum field theory in the BTZ black hole  background
has been explored and 
exact results have been obtained about Green functions for a  conformally 
coupled massless scalar.  
The thermodynamic and statistical mechanical properties of the black hole
have been investigated by the  
Chern-Simons formulation of the $ (2+1) $-dimensional general relativity.
\parsmallskip
The BTZ black hole is also
one of the few known exact solutions in string theory 
and one of the simplest ones. 
Since a string in a BTZ black hole background has an infinite number of
propagating modes, the string theory resembles a higher dimensional one. 
It has significance also 
as a string theory on a non-trivial curved spacetime. 
\parsmallskip
In the rest of this thesis, we will investigate the thermodynamics of quantum 
scalar fields \cite{IS} and the string theory \cite{NS} in 
the BTZ black hole background. 
%
%
%
%
\csection{THERMODYNAMICS OF SCALAR FIELDS 
IN THE THREE DIMENSIONAL BLACK HOLE BACKGROUND}
In this section, we discuss the thermodynamics of quantum scalar fields 
in the BTZ black hole background \cite{IS}. After a brief review of black hole 
thermodynamics,\footnote{ For a  review of black hole thermodynamics and 
recent arguments about black hole entropy 
in canonical quantum gravity, see, e.g.,\cite{Wa,BHTD,YS1}.
} 
we discuss the thermodynamics of the scalar field in two approaches.
One is to use mode expansion, and the other is 
to use Hartle-Hawking Green functions. In both approaches we obtain 
exact expressions. These enable us to discuss the black hole thermodynamics
without ambiguity.  
%
%
\parsmallskipn
\csubsection{Black hole entropy}
Through the study of black hole thermodynamics, close relations 
between black holes and quantum mechanics have been found 
\cite{Wa}-\cite{GibH}. In addition,
black hole entropy has lately attracted attention again \cite{Sr}-\cite{SF}.
In order to make the points of later discussions clear, 
we first review black hole thermodynamics and recent arguments
about the entropy of a quantum field in a black hole background.
\isubsubsection{Black hole thermodynamics}
In general relativity, 
black holes satisfy the following properties
under certain conditions \cite{Wa}-\cite{Be}:
\begin{namelist}{ 3. }
\item[ 0. ]
 $\kappa$ (surface gravity) is constant over the 
    horizon of a black hole. 
\item[ 1. ]
   In physical processes, the changes of physical quantities obey 
\parmedskipn
$ \qquad \qquad \delta M = (\kappa/8\pi) \delta A + \Omega_H \delta J 
             + \Phi_H \delta Q \comma 
$ 
\parmedskipn
where $ A,J $ and $ Q $ are the area, the 
angular momentum and the charge of the black hole, respectively. 
$\Omega_H$ is the angular velocity and  $\Phi_H$ is the potential 
at the horizon.
\item[ 2. ] The area of the horizon never decreases, 
$ \quad
 \delta A \geq 0 \period 
$
\item[ 3. ]
 It is impossible to achieve $\kappa = 0$ .
\end{namelist}
These remind us of the ordinary laws of thermodynamics. 
The correspondences are
$ M \leftrightarrow E$(energy), $ \alpha \kappa \leftrightarrow T$
(temperature) 
and $  A/8\pi \leftrightarrow S$(entropy), where $\alpha$ is 
some constant.
A gravitationally collapsing body
 rapidly settles to a black hole configuration specified only by 
three macroscopic quantities $M,J$ and $Q$. This is also suggestive 
of this analogy.   
But, the temperature 
of a black hole is classically zero, and the relationship between 
the area and the entropy is obscure. Thus the analogy seems
superficial at first.
Nevertheless, if one makes semiclassical analyses including quantum 
fields in black hole backgrounds, one finds evidence that black
holes are really thermal. Let us see the representative arguments.  
\parbigskipn
{\sl Hawking radiation} \cite{Ha} 
\parmedskip
By evaluating the vacuum expectation value of the number operator 
of a quantum field, Hawking 
showed that a black hole emits thermal radiation at temperature
$ T = \kappa /2\pi $.
This argument fixes the constant $ \alpha $ to be $1/2\pi$. The entropy 
of the black hole is then given by
\eqabegin
   \SBH  &= &  \frac{1}{4 \l_p^2} A 
    \ = \ \frac{1}{4 G} A \ = \ \frac{1}{4} A 
  \comma \nn
\eqaend
where $ \lp $ is the Planck length and $ G $ is the gravitational coupling 
constant. We have displayed the formula in various units.
(This is called Bekenstein-Hawking entropy.)
\parbigskipn
{\sl  The Hartle-Hawking Green function }\cite{HH} 
\parmedskip
One can define a Green function of a quantum field in a black hole geometry
by a generalization of the Feynman Green function in Minkowski spacetime.
This is specified by the analyticity or the boundary condition at the horizon. 
This is called the Hartle-Hawking Green function. By making use of this 
Green function, one can derive the Hawking radiation again.
\parbigskipn
{\sl  Euclidean black holes} \cite{GP} 
\parmedskip
By Wick-rotating the Schwarzschild metric by $ \tau \equiv i t $, one 
obtains a 
``Euclidean'' black hole geometry
\eqabegin
  d s_E^2 &=& R^2 d (\tau/4M)^2 + (r/2M)^4 d R^2 + r^2 d\Omega^2 
   \comma \nn 
\eqaend
where $ R \equiv 4M( 1-2M/r)^{1/2} $.
$ \tau $ then has the period 
$ \beta_H \equiv 8 \pi M $. Suppose that we define a Green function in the 
original Schwarzschild geometry  by an analytic continuation
(as in an ordinary field theory);
$
  G(x,y) \equiv G^E(x^E,y^E) \when_{\tau = it}
   \period
$
Then $G(x,y)$ has an imaginary period $ t \to t + i \beta_H $.
(This is nothing but the Hartle-Hawking Green function.)
An imaginary
period of the time direction is a characteristic feature of thermal
Green functions. Thus the above Green function suggests that the quantum
field is in thermal equilibrium at temperature
 $ T = \beta_H^{-1} = \kappa/2\pi$ ($\kappa = 1/4M $).
\parbigskipn
{\sl  Gibbons-Hawking entropy} \cite{GibH} 
\parmedskip
In Minkowski spacetime, the partition function of a quantum field $\Phi$    
at inverse temperature 
$ \beta  $ is given by 
\eqabegin
   Z(\beta) &=& {\rm Tr} \ e^{-\beta H} \ = \ 
   \int \calD \Phi \exp \lmb - \int_0^\beta d \tau 
   \int d \vec{x} \sqrt{g_E} \ \calL^E(\Phi)
    \rmb
     \period 
   \label{GHentropy}
\eqaend
Here $H$ is the  Hamiltonian and $\calL^E$ is the Euclidean Lagrangian density. 
The path integral is performed under the periodic
boundary condition $ \Phi(\tau) = \Phi(\tau + \beta) $.
Suppose that this expression is valid for the gravitational field in 
the Schwarzschild 
geometry, and evaluate the path integral at $ \beta = \beta_H $
in the saddle point approximation (the saddle point corresponds to the Euclidean 
Schwarzschild solution).
Then, one obtains  $ \SBH = A/(4G) $ again.
Since the action of the gravitational part is unbounded below, it is necessary
to regularize it.
\parsmallskip
 For a matter field, such a 
calculation corresponds to defining the 
partition function $ Z_{m} $ and the entropy $ S_{m}$ by 
\eqabegin
   \ln Z_{m} (\beta) & = & - \beta F_{m}(\beta) 
   \ = \  \half  \Tr  \ln G^E_{m} (\beta) \comma \nn \\
    S_{m} (\beta) &=& \beta^2 \frac{\del F_{m}}{\del \beta}
  \comma \label{ZG}
\eqaend 
where $ F_{m} $ is the free energy and $ G^E_{m} $ is the 
Euclidean Green function of the matter in the Schwarzschild background. 
Note that the Euclidean black hole geometry has a deficit angle with respect to 
$ \tau $ if  $ \beta \neq \bH $. In this case, $ (\tau, r) $-plane represents
a ``cone geometry''.
\parbigskipn
{\sl Generalized second law} \cite{Be,Ha2} 
\parmedskip
If the area of the horizon $ (1/4)A $  can be interpreted as 
the thermodynamic entropy, the total entropy of gravity plus matter 
should not decrease: 
\eqabegin
 \delta S^{tot}  &\equiv& \delta S^{m} + (1/4)\delta A \ \geq  \ 0
 \period \nn  
\eqaend
This is called the generalized second law.
Although the proof does not exist, there are arguments for 
the validity of this law.
\parbigskip

These semiclassical arguments materialize  black hole thermodynamics.
Again by an analogy to the ordinary thermodynamics, 
the microscopic meaning of which is given by statistical mechanics, 
we expect the existence
of some fundamental and microscopic mechanism of black hole thermodynamics.
In addition, the Hawking radiation implies that quantum coherence may not hold
in quantum gravity.
This is a serious problem for fundamental theory of physics.
In this way,  black hole thermodynamics indicates deep connection among 
gravity, quantum theory and statistical physics.  
Thus, this subject has been studied extensively to get clues 
to quantum gravity.
In particular, main problems have been
(i) to derive the black hole entropy by counting 
quantum states of black holes\footnote{In this respect, 
interesting results have been obtained recently 
for black holes in super string theory \cite{DEnt}.}
(ii) to explain 
the Hawking radiation in a quantum mechanical manner
and (iii) to prove the generalized second law.
%
%
%
\isubsubsection{Recent arguments about black hole entropy}
%
Let us turn to the recent arguments about black hole entropy.
Since we have no consistent theory of 
quantum gravity, it is difficult to make definite arguments about
black hole thermodynamics, in particular about the gravitational 
part. However, there are recent proposals to explain 
the black hole entropy in quantum mechanical manners for matter fields.
 In the following, we will list examples  
of these proposals \cite{Sr}-\cite{SF}
 relevant to the later discussions.\footnote{There are many other recent
arguments. We do not refer to them here.} 
They deal with the limit where black hole mass goes to infinity.
Note that the Schwarzschild geometry 
approaches Rindler space in that limit (Rindler limit).
\parbigskipn
{\sl  The brick wall model} \cite{tH},\cite{KS,SU} 
\parmedskip
Let us consider a quantum scalar field $ f(x) $ with mass  $ m \ll 1$ 
in a Schwarzschild black hole geometry
with mass $ M \gg 1  $. The horizon is at $ r = 2M $.
We impose the boundary condition $ f(x) = 0 $ at 
$ r = 2M + \epsilon \comma  L$, where $ \epsilon \ll 1 $ and $ L \gg M $.
It turns out that $ \epsilon $ and $ L $ play the role of regulators 
of the thermodynamic
quantities. This boundary condition is similar to 
putting a ``brick wall'' near the horizon. Then we calculate the 
density of states $ d g (E) $ by the WKB approximation, and the free energy 
by summation over states as  
$
 \beta F(\beta) =  \int_0^{\infty} d E \ (d g/d E) \ 
   \ln \lb  1 - \ e^{- \beta E}\rb  \period
$
Consequently, one finds that (i) the entropy of the matter is proportional 
to the area of the horizon $ A $, and 
(ii) the leading contribution to the entropy 
comes from the horizon and diverges as $ A/\epsilon^{2} $. This entropy 
is regarded as a quantum correction to the Bekenstein-Hawking 
entropy $(A/(4\lp^2))$.
These results are suggestive of the origin 
of black hole entropy, and they imply the importance of  
the horizon to quantum properties of black holes.
However, it is not clear 
why we should adopt such a boundary condition.
\parbigskipn
{\sl Geometric entropy} \cite{BKLS},\cite{Sr,KS,CW} 
\parmedskip
A black hole has an event horizon, and this separates the outside and 
the inside region; an observer outside the horizon cannot get any 
information from the inside. One can infer that this can be a source of 
the thermodynamic properties because the inside information may be 
averaged for the outside observer. This is the essential idea 
of geometric entropy. 
\parsmallskip
Let us consider a free scalar field in a {\it flat} spacetime 
separated into two parts by a boundary with thickness $ \epsilon $. 
We calculate the density matrix (by introducing appropriate cutoffs), 
and average the states inside the boundary by taking the trace of 
the density matrix. Then one finds that (i) the entropy calculated from 
the averaged density matrix is proportional to the area of the boundary
$ A $, and (ii) the leading contribution to the entropy 
comes from the boundary and diverges as $ A/\epsilon^{2} $. 
These results are suggestive again, and they imply the importance of
the correlation between the outside and the inside.
Notice that gravitation is absent in this calculation. Thus the relation 
to the real black hole physics is not clear although this argument is relevant
in the Rindler (flat) limit. 
\parbigskipn
{\sl Gibbons-Hawking entropy of matter} \cite{SU}-\cite{SF} 
\parmedskip
For a free scalar field in a Euclidean black hole geometry,
one can evaluate the path integral
in (\ref{GHentropy}) systematically by the heat kernel expansion.
We remark that one has to consider the heat kernel on a cone geometry
for $ \beta \neq \bH $, and hence for getting the entropy (see (\ref{ZG})).
In the Rindler limit, one then finds that (i) the entropy  
is proportional to the area of the horizon $ A $, and (ii) this 
is divergent like $ A/\epsilon^2 $ where $ \epsilon $ is an ultra-violet 
cutoff. This entropy 
is regarded as a quantum correction to the Bekenstein-Hawking entropy again.
From this point of view, 
one can regard the entropies as responses to the 
deficit angle of a geometry.
\parbigskip
 At first sight, the relation among these calculations is not obvious. But
one can show the equivalence of these three methods in the Rindler limit 
\cite{KS,CW}. This is reasonable because the three 
methods give the same result in Minkowski spacetime 
and Rindler space is a kind of a flat spacetime.
One can also generalize these calculation to a $D$-dimensional black hole 
geometry ($D$-dimensional Rindler space).
In summary, the claims are:
(i) the entropy of the matter is a quantum correction to
the Bekenstein-Hawking entropy;  (ii) the free energy takes the form
\eqabegin
    \beta F_{m}(\beta) &=& \frac{A_D}{\epsilon^{D-2}} C(\beta) \comma
   \label{Fbeta}
\eqaend
where $ A_D $ is the area of the horizon of a D-dimensional black hole, 
$ \epsilon $ is a short-distance cutoff and $ C(\beta) $ is a function 
of $ \beta $; (iii) the entropy takes a  form similar to (\ref{Fbeta}), 
in particular at $ \beta = \bH $ it takes the form 
\eqabegin
    S_{m}(\bH) & \propto &  \frac{A_D}{\epsilon^{D-2}} \ ; \nn
\eqaend
and thus (iv)  the entropy  is divergent
owing to the physics at short distance and/or the horizon.
Recall the form of the Bekenstein-Hawking entropy 
$ \SBH = A_D/(4G_D) $ where $ G_D $ is the gravitational coupling 
constant in $ D$-dimensions. Then we notice that if we define 
the ``renormalized gravitational coupling constant'' by
\eqabegin
   \frac{1}{G_R} &=& \frac{1}{G_D} + \frac{c}{\epsilon^2} \comma  \nn
\eqaend
where $ c $ is some constant, one can write the total entropy in the same 
form as $ \SBH $:
\eqabegin
   \SBH + S_{m}(\bH) &=& \frac{A_D}{4 G_R } \period \nn
\eqaend
From this observation,  
Susskind and Uglum  argued \cite{SU} that (i) 
the divergence of the entropy is closely related 
to the renormalization of the gravitational coupling constant,  (ii) one can 
deal with this divergence properly in string theory and (iii)
 the divergence 
might store an infinite amount of information.
\parsmallskip
However, these arguments are based on the results in the Rindler 
limit. 
We would like to investigate whether these are still valid for 
the finite mass cases, i.e., for real black holes.
In addition, the discussions are somewhat formal on  
boundary conditions and regularization schemes.
Since a system of a four dimensional black hole plus matter 
is complicated, it is difficult to discuss these issues in an explicit 
manner. 
One often encounters such problems in the study of quantum 
black holes. Without settling these problems, one cannot reach any definite
conclusions. Thus it is very important to make precise analyses as a starting 
point for getting reliable results.
In what follows, we will discuss the thermodynamics of scalar fields
in the BTZ black hole geometry to examine these issues \cite{IS}. 
It turns out that we can make definite 
arguments without ambiguity.
%
%
\csubsection{Thermodynamics of scalar fields by summation over states}
%
In this section, we consider the thermodynamics of scalar fields 
in the BTZ black hole background by 
mode expansion and direct computation of summation over states \cite{IS}. 
In order to study the dependence of the thermodynamic quantities upon 
boundary conditions, we consider two cases. In both cases, we require 
that the scalar field vanishes rapidly enough at spatial infinity.  
In one case, we further impose regularity at the origin. 
In the other case, we impose the condition that the scalar field 
vanishes near the outer horizon. This is an analog of the one in the  previous
brick wall model \cite{tH,KS,SU}. 
Although it is possible to consider other various boundary conditions, we 
do not consider them because their physical meaning is not clear in most cases.
%
\isubsubsection{Mode functions}
Let us consider a scalar field with mass squared $ m^2 $ 
in a BTZ black hole background.
The field equation is given by
\eqabegin
 &&  \lb \Box - \ \mu l^{-2} \rb  f(x) \ = \ 0 \period
   \label{TFE} 
\eqaend
Since $ R = -6 l^{-2} $,
$ \mu l^{-2} = m^2 $ for a scalar field minimally coupled to 
the background metric and
$ \mu l^{-2} = m^2 + (1/8) R = m^2 - 3/4 l^{-2} $ 
for a conformally coupled scalar
field. 
To solve the equation, it is useful to use the coordinate system 
$ (\hatr, \hatt, \hatphi) $. 
Then we expand the field as 
\eqabegin
    f(r, t, \varphi) &=& \sum_{N \in \bfZ} \int d E \ f_{E N } (r) 
    \ e^{-i E t} e^{i N \varphi } 
    \ = \  \sum_{\hatE , \hatN} f_{\hatE \hatN } (\hatr) 
    \ e^{-i \hatE \hatt} e^{i\hatN \hatphi } \comma 
    \label{Tmodes2}
\eqaend
where
\eqabegin
   && l^2 E =   \rp \hatE + \rmi \hatN \comma \quad 
      l N =  \rmi \hatE + \rp \hatN \nn \period \label{ENhat}
\eqaend
In the latter expansion, the field equation reads as
\eqabegin
   && (\hatr ^2 - l^2)  \del_{\hatr}^2 f_{\hatE \hatN }
    + (3\hatr - \frac{l^2}{\hatr}) \del_{\hatr} f_{\hatE \hatN} 
  + ( \frac{l^2 \hatE ^2 }{\hatr ^2  - l^2}
    - \frac{l^2 \hatN ^2}{\hatr ^2 } - \mu l^{-2} ) 
  f_{\hatE \hatN} = 0 \period \nn
\eqaend
This equation has three regular singular points at 
$ \hatr = 0, 1, \infty \ (r = \rmi, \rp, \infty)$ 
corresponding to the inner horizon, the outer horizon 
and the spatial infinity, respectively. Thus the solution is given 
by the hypergeometric function. 
To see this, we make further changes of variables, 
$ u = 1 - \hatr ^2/l^2 $ and 
\eqabegin
   f_{\hatE \hatN} (u) &=& 
  (-u)^{i \hatE/2} ( 1- u)^{- i \hatN/2 } g_{\hatE \hatN} (u)
  \period \nn
\eqaend
Consequently,
the field equation is reduced to the hypergeometric equation \cite{IS}
\eqabegin
  && u(1-u) \del_u^2 g_{ \hatE \hatN } + \lmb  c- (a + b + 1) u 
   \rmb \del_u g_{ \hatE \hatN } - ab \ g_{ \hatE \hatN } \ = \ 0
 \comma \nn
\eqaend
where
\eqabegin
   a &=& \half ( 1 + \sqrt{1+\mu}) + i \lb  \hatE - \hatN \rb /2 \comma \nn \\
   b &=& \half ( 1 - \sqrt{1+\mu}) + i \lb  \hatE - \hatN \rb /2 \comma 
    \label{abc}  
  \\
   c &=& 1 + i \hatE  \period \nn 
\eqaend

The hypergeometric equation has two independent solutions around each regular
singular point. The independent solutions around $ u = \infty $ are
\eqabegin
  U_{ \hatE \hatN }&=& (-u)^{i\hatE /2} (1-u)^{-i\hatN/2} (-u)^{-a}
       F(a,a-c+1,a-b+1;1/u) \comma \nn \\
  V_{ \hatE \hatN } &=& (-u)^{i\hatE /2} (1-u)^{-i\hatN/2} (-u)^{-b}
       F(b,b-c+1,b-a+1;1/u) \comma 
    \label{modeUV}
\eqaend
where $ F $ is the hypergeometric function.
In order to specify the solution, we have to impose boundary conditions.
First, notice that  
$ U_{ \hatE \hatN } $ vanishes as $ r \to \infty $ for an arbitrary 
$ \mu $, but $ V_{ \hatE \hatN } $ becomes divergent there 
for $ \mu > 0 $.
Second, 
$ \AdS $ has timelike spatial infinity
and requires special boundary conditions there. 
The authors of \cite{AIS}-\cite{BL}  
have discussed  the quantization of scalar fields in anti-de Sitter 
spaces or their covering spaces in various dimensions. 
If we follow them and require the Cauchy problem to be well defined, 
the surface integral of the energy momentum tensor at spatial infinity 
must vanish:
\eqabegin
&&  \lim_{r \to \infty} \int d S_i \sqrt{-\gBHd} \ T^i_{\ t}  = 0 
      \period \label{BCAdS}
\eqaend
This means
$ \sqrt{r} f_{\hatE \hatN}  \to 0 \quad ( r \to \infty ) $.
Third, in chapter 3, we deal with the case $ \mu \geq - 3/4 $, 
i.e., the case of non-negative mass squared for both 
minimally and conformally  coupled scalar.\footnote{However,  
since we can deal with the case $ -3/4 > \mu \geq -1 $ on the same footing, 
we will include this case in the following discussion.}
Then, 
only $  U_{ \hatE \hatN } $ 
satisfies the condition (\ref{BCAdS}). Therefore we choose the solution 
$ f_{ \hatE \hatN } = U_{ \hatE \hatN } $, which 
vanishes rapidly enough at spatial infinity for arbitrary $ \mu $.
%
\parsmallskip
The above mode functions have been obtained independently of \cite{IS} 
by Ghoroku and Larsen \cite{GL}. Using these modes, 
they have discussed the tachyon scattering in the string theory 
in the three dimensional black hole geometry. In their case, $ \mu = -24/23 $.
As discussed later, these mode functions are closely related to 
the representation theory of $ \SLtwo $.
%
\isubsubsection{Case I}
%
To examine the dependence of the thermodynamic quantities 
on boundary conditions, we will consider two further boundary conditions.
For usual radial functions in quantum mechanics, one requires 
square integrability, and this condition leads to regularity
at the origin. In a BTZ background, the meaning of the square integrability 
is not clear until we specify the inner product.
However, we do not 
know what inner product we should adopt.
Thus as an exercise for the above purpose, we first impose 
on $ f_{ \hatE \hatN } $ regularity 
at the origin $ (r = 0) $.
\parsmallskip
It is easy to solve this boundary condition.
Indeed, we readily find that 
$ U_{ \hatE \hatN } $ 
is regular at the origin because $ r = 0 $ corresponds to none of 
$ \hatr/l = 0, 1, \infty $. 
Thus we need no restriction on the value of $ E $. 
\parsmallskip
Here let us recall that 
a system of a rotating black hole plus a scalar field has a chemical 
potential $ \omh $. This is the angular velocity of the outer horizon:
\eqabegin
  \omh &=& - N^\varphi (\rp)
        \ = \ \frac{\rmi}{l\rp }
        \period \nn
\eqaend
In addition, the system has superradiant scattering modes given by the 
condition
\eqabegin
  E - \omh \ N \leq 0 \comma \nn
\eqaend
where $ E  $ and $ N $ are the energy and angular momentum of the 
scalar field, respectively. 
Thus we have to regularize the (grand) partition function
by introducing a cutoff $ \Lambda_1 $ for the occupation number of 
particles for 
each superradiant scattering mode.
\parsmallskip
With this remark in mind, the remaining calculation is straightforward
and, by introducing appropriate cutoffs, 
we will obtain explicit results \cite{IS}.
First, the partition function for a single mode 
labeled by $ E,  N $ and the inverse temperature $ \beta $ 
is give by\footnote{
The value $ E = \omh N $ is similar to the value where 
Bose condensation occurs. Then we might have to omit the summation 
over the superradiant scattering modes because they may 
be a signal for instability. However, 
we can get the quantities omitting this summation 
simply by setting $ \Lambda_1 = 0 $.
}
\eqabegin
 Z_o( \beta; E, N ) &=& \sum_{n = 0}^{\infty}  
   \ e^{- n \beta ( E - \omh N )}
    \nn \\
  &=& \left\{ \begin{array}{cl}
       {\displaystyle \left( 1- \ e^{-\beta(E - \omh N )} \right)^{-1} }
                         & {\rm for} \quad E - \omh N > 0 \\
       {\displaystyle  \Lambda_1 }
                         & {\rm for} \quad E - \omh N = 0 \\ 
        {\displaystyle  \frac{1 - \ e^{- \Lambda_1 \beta(E - \omh N )}}
                  {1 - \ e^{- \beta( E  - \omh N )}}
         }
                         & {\rm for} \quad  E  - \omh N < 0 
       \end{array} \right.
         \period \nn
\eqaend
Then we obtain the total partition function, 
\eqabegin
 Z_o(\beta) &=& \prod_{ E, N} \ Z_o(\beta;  E, N) \comma \nn
\eqaend
and the free energy, 
\eqabegin
 - \beta F_o(\beta) &=& \sum_{ E, N} \ \ln \ Z_o(\beta;  E, N ) \nn \\
                  &=& - \sum_{ \abs{N} = 0 }^{\Lambda_2} \frac{1}{\delom}
               \int_0^{\infty} d  E  \ 
                   \ln  \left( 1- \ e^{-\beta( E  - \omh N )} \right)
                    + \sum_{N=0}^{\Lambda_2} \Lambda_1 \nn \\
       && \qquad 
         + \sum_{N = 0}^{\Lambda_2} \frac{1}{\delom}
            \int_0^{N \omh} d  E  \ 
            \ln \left( 1- \ e^{-\beta \Lambda_1 ( E  - \omh N )} \right)
                   \comma \nn
 \eqaend
where $ \Lambda_2 $ is  the cutoff for 
the absolute value of quantum number $ N $, and $ \delom $ is 
the minimum spacing of $  E  $. Note that  
 $ \delom^{-1} $ is the density of states and  
the above result is divergent as $ s \to 0 $ 
regardless of the existence of the horizon. 
Furthermore, by making the change of variables $ y = \beta ( E  - \omh N ) $ 
for the 
 first term and $ y = \Lambda_1\beta ( N \omh -  E  ) $ 
for the third term, we obtain
 \eqabegin
  - \beta F_o(\beta) 
       & =  & \frac{1}{s} \lbb 
        \frac{\pi^2}{6 \beta} ( 2 \Lambda_2 + 1)
  + \frac{\beta}{12} \omh^2 (\Lambda_1 - 1) 
      \Lambda_2 (\Lambda_2  + 1)(2\Lambda_2  + 1)   
             \right. \nn \\ 
       &&   \qquad  \left. 
                  + \frac{1}{\Lambda_1 \beta} \sum_{N=1}^{\Lambda_2}
          \int_0^{\Lambda_1 \beta \omh N} d y \ \ln \left( 1 - \ e^{-y} \right)
          \rbb
        + \Lambda_1 ( \Lambda_2 +1 )    
      \period \nn
\eqaend
In the limit 
$ \Lambda_1 \to \infty $, the last term in the bracket is simplified to  
$ - \Lambda_2 \zeta(2)/(\Lambda_1 \beta) $.
Finally, by the formula (\ref{ZG}),  
we get the entropy 
\eqabegin
  S_o(\beta) &=& \frac{1}{\delom} 
    \lbb   
       \frac{\pi^2}{3 \beta} ( 2 \Lambda_2 + 1)
     - \omh \sum_{N = 1}^{\Lambda_2} N \ 
      \ln \left( 1 - \ e^{-\Lambda_1 \beta \omh N } 
       \right) 
             \right. \nn 
 \\  
       &&   \qquad  \left. 
                  + \frac{2}{\Lambda_1 \beta} \sum_{N=1}^{\Lambda_2}
          \int_0^{\Lambda_1 \beta \omh N} d y \ \ln \left( 1 - \ e^{-y} \right)
          \rbb
        + \Lambda_1 ( \Lambda_2 +1 )    
     \period \nn 
\eqaend 
Thus, the entropy is not proportional to 
the area (perimeter) of the outer horizon $ (2 \pi \rp) $ for a generic 
$ \beta $ including $ \beta = \bH $ in (\ref{clTS}). In addition, its divergence
is not due to the existence of the outer horizon. It comes from 
the cutoff $ s $. 
%
%
%
\isubsubsection{Case II}
%
Let us consider another boundary condition.
One may expect that something singular occurs at the horizon 
since, e.g., the redshift factor of the black hole becomes divergent 
there.
Thus, as the second case, we require regularity at the outer 
horizon. This boundary condition is an analog of the one in the brick 
wall model \cite{tH,KS,SU}. In our case, we can solve this condition
and obtain thermodynamic quantities
without the WKB approximation \cite{IS}. 
%
\parsmallskip
To solve the boundary condition,
we first study the behavior of $  U_{ \hatE \hatN } $ 
near the outer horizon 
($ r = \rp $, i.e., $ \hatr/l = 1 $ ). 
By making  use of a linear transformation formula with respect to the 
hypergeometric function, we get 
\eqabegin
   U_{ \hatE \hatN } &\propto&  (-u)^{i\hatE/2}( 1 - u )^{-i \hatN/2}
    F(a,b; c; u ) \nn \\
      && \qquad     - \ \Theta \ (-u)^{-i\hatE/2}( 1 - u )^{-i \hatN/2}  
          F(a-c+1,b-c+1; 2-c ; u) 
          \comma \nn
\eqaend
where 
\eqabegin
 \Theta &=& 
   \frac{\Gamma(1 - b)\Gamma(c) \Gamma(a-c+1) }
  {\Gamma(a)\Gamma(2-c)\Gamma(c-b)} \period \nn
\eqaend
From $ (\Gamma(z) )^\ast = \Gamma(z^\ast) $ and (\ref{abc}), we find that 
$ \abs{\Theta} = 1 $ for $ \mu \geq -1 $. Thus we may set
\eqabegin
 \Theta &=&  \ e^{-2\pi i \theta_0} \qquad ( 0 \leq \theta_0 < 1 )
 \comma \nn
\eqaend
where $ \theta_0 $ is determined by $  E  $ and $ N $ through $ a $, $ b $
and $ c $.   
Choosing an appropriate normalization constant, we can write 
\eqabegin
  U_{ \hatE \hatN } &=&
 (-u)^{i\hatE/2}( 1 - u )^{-i \hatN/2} 
   \ e^{ \pi i \theta_0} F(a,b; c; u ) \nn \\
   && \qquad 
          -  \ (-u)^{-i\hatE/2}( 1 - u )^{-i \hatN/2}  
         \ e^{ - \pi i \theta_0}  F(a-c+1,b-c+1; 2-c ; u) \period \nn
\eqaend
Then by introducing an infinitesimal constant $ \eH $ and 
substituting $ -u =  \eH^2/ l^2  $ (namely, 
$  \eH^2/l^2 =  (r^2 - \rp^2)/\dH ^2 $ ),
we find the behavior 
of $  U_{ \hatE \hatN } $ near the outer horizon;
\eqabegin
 U_{ \hatE \hatN } &\stackrel{\eH/l \to 0 }{\longrightarrow}& 
      \ e^{ i \hatE \ln (\eH/l)  + \pi i \th0 } - 
      \ e^{- ( i \hatE \ln (\eH/l)  + \pi i \th0 ) } \period \nn
\eqaend
Here, we impose 
the boundary condition $  U_{ \hatE \hatN } = 0 $
 at $  - u = \eH^2/l^2  $. This condition yields
\eqabegin
 E  &=& \omh \ N + C 
               ( K + \th0 )  \quad  ( K \in {\bf Z} ) \comma \quad 
 C \ = \   \frac{ \pi \dH^2 }{\rp l^2 \ln (l/\eH) } 
               \period \nn
\eqaend
This shows that $  E  $ and $ \theta_0  $ are labeled by two integers 
$ K $ and $ N $, i.e., $ E = E (K,N) $ and  
$ \theta_0 = \theta_0(K,N)$. $ C^{-1} $ becomes singular as 
$ \eH \to 0 $. 
In this way,  we could solve the ``brick wall'' boundary condition 
without the WKB approximation. 
\parsmallskip
Then, we can obtain the partition function and the free energy 
similarly to the previous case: 
\eqabegin
 - \beta F_h(\beta) &=& \sum_{K, N} \ \ln \ Z_h(\beta; K, N ) \nn \\
                  &=& - \sum_{ \abs{N} = 0 }^{\Lambda_2} 
                  \sum_{ { K+ \theta_0 \neq 0 }
                  \atop{ C(K + \theta_0) \geq - \omh  N }}
                   \ln \left( 1- \ e^{-C \beta(K + \theta_0)} \right)
         + \Lambda_1 \sum_{N=0}^{\Lambda_2} \delta_{\theta_0(0, N), 0} \nn \\
                  && \qquad 
                  + \sum_{N = 1}^{\Lambda_2} \quad  
                  \sum_{ 0>  C(K + \theta_0) \geq - \omh N}
              \ln \left( 1- \ e^{- \Lambda_1 C \beta (K + \theta_0)} \right)
                   \period \nn
 \eqaend
Since $ C<<1 $ in the limit $ \eH/l \to 0 $, 
the summation with respect to $ K $ can be approximated 
by integrals. 
Notice that 
\eqabegin
 \frac{d K}{d E} &=& \frac{1}{C}
             - \frac{d \theta_0}{d K }\frac{d K }{d E}
          \ \sim \ \frac{1}{C} 
       \period \nn
\eqaend
This shows that the density of states diverges because of the existence of 
the outer horizon, i.e., as $ \eH/l \to 0 $.
In this limit, we get 
 \eqabegin
  - \beta F_h(\beta) & \sim &   
             \frac{\rp \delom l^2 \ln (l/\eH) }{\pi \dH^2}
             \lmb -\beta F_o(\beta) -  \Lambda_1(\Lambda_2  +   1) \rmb 
            +
               \Lambda_1 \sum_{N=0}^{\Lambda_2} \delta_{\theta_0(0,N), 0} 
                \period \nn
\eqaend
Then the entropy is 
\eqabegin
 S_h(\beta) &\sim&  
    \frac{\rp \delom l^2 \ln (l/\eH)}{\pi \dH^2}
                  \lmb  S_o(\beta) - \Lambda_1(\Lambda_2  +   1) \rmb 
                 +
                 \Lambda_1 \sum_{N=0}^{\Lambda_2} \delta_{\theta_0(0,N), 0} 
                \period  \nn
\eqaend
Therefore, 
(i) the leading term of the entropy as $ \eH/l \to 0 $ is proportional 
to $ \rp l^2 /(\dH ^2 \beta) $, and   
(ii) the entropy diverges owing to the outer horizon 
like $ \ln (l/\eH) $ as $ \eH/l \to 0 $. Thus again the entropy is not 
proportional to the area for a generic $ \beta $ including 
$ \beta = \bH $.
\parsmallskip
In section 3.2, we calculated the thermodynamic quantities in a truly 
curved spacetime without 
the heat kernel expansion nor the WKB approximation. This enables us to make 
definite discussions on the thermodynamics of a scalar field. 
From the results, we find that 
the physical quantities depend largely upon boundary conditions as expected.
In case II where the boundary condition is closely related to the horizon, 
the thermodynamic quantities are expressed by the parameters related to the 
horizon, e.g., $ \rp $ and $ \eH $. But they are not so in case I.
The divergence of the entropy is due to the horizon in case II, but 
this is not so in case I. These indicate that one must justify 
the brick wall boundary condition if one adopts it. 
Otherwise the quantities obtained in that manner
 may be artifacts of the boundary condition.
The entropies are not proportional to the area of the horizon. This may 
be related to the special properties of the BTZ black hole. 
The BTZ black hole has a fundamental scale $ l$ and asymptotically 
$ \AdS $ instead of Minkowski spacetime. However, the naive expansion 
in the literature 
with respect to the inverse black hole mass may not converge, and the 
entropy may receive a large correction. Thus further investigations 
in a truly curved case may be necessary.
%
%
\csubsection{Green functions in the three dimensional black hole background}
Now, let us turn to a new approach to the thermodynamics using Green functions. 
Comparing the results with the previous ones, 
we can examine the equivalence among various methods for 
the thermodynamics of a scalar field in a black hole geometry. 
In the literature, Green functions are evaluated 
by the heat kernel expansion. However in our case, we can construct the exact 
Green functions for scalar fields with a generic mass squared \cite{IS}.
Making use of them, we can obtain exact expressions of the thermodynamic 
quantities \cite{IS}.
As a first step, we will discuss the Green functions in this section.
%
\isubsubsection{Construction of Green functions}
%
For a conformally coupled massless scalar in a BTZ 
background, the Green functions have been obtained in \cite{GFs,LO} by
making use of the Green functions in $ \AdS $ and of the method of images.
Here we will generalize this construction to a generic mass squared.
\parsmallskip
Quantization  of a scalar field in the universal covering space of 
$ D$-dimensional anti-de Sitter space ($ \widetilde{AdS}_D  $) 
has been discussed in 
\cite{AIS}-\cite{BL}. For a generic value of $ D $, 
the Feynman Green function is given 
in terms of the hypergeometric functions \cite{BL}. 
But for $ ( D=3 ) $, it is expressed by a simple form \cite{IS}
\eqabegin
 - i \Gf (x,x') &=& - i \Gf(z) \ \equiv \ 
           \frac{1}{4 \pi l} ( z^2 - 1)^{-1/2} 
           \lbb z + ( z^2 - 1 )^{1/2} \rbb ^{1-\lambda} \comma \label{Gf} 
\eqaend
where 
\eqabegin
 z &=& 1 + l^{-2} \sigma (x, x') + i \varepsilon \comma \nn \\
 \lambda &=&  \lmb
           \begin{array}{ll}
              \lambda_\pm \, \equiv \, 1 \pm \sqrt{1 + \mu} 
              & {\rm for } \ \ 0 > \mu > -1 \comma \\
              \lambda_+  & {\rm for} \ \     \mu \geq 0 \comma \mu = -1 \comma  
           \end{array}
              \right.
              \comma \nn
\eqaend
(two $ \lambda $'s are possible for $ 0 > \mu > -1  $) and $\varepsilon$
is a positive and infinitesimal constant. 
$ \sigma(x,x') $ is half of the distance between $ x $ and $ x'$ in the four 
dimensional embedding space,
\eqabegin
  \sigma (x, x') &=& \half \eta_{\alpha\beta} (x-x')^\alpha (x-x')^\beta 
              \comma \nn
\eqaend
where $ \eta_{\alpha\beta} $ and $ x^\alpha $ 
$ \ (\alpha, \beta = 0$-$3)$ are given by 
(\ref{flatG}) and (\ref{embedding}).
Since the derivation of this result is technical, we relegate it to appendix A.
\parsmallskip
By making use of the above result, 
Green functions for a generic mass squared in a BTZ  background are obtained 
by the method of images \cite{IS};
\eqabegin
 - i \Gbh (x,x') &=& - i \sum_{n = - \infty}^{\infty} \Gf (x, x'_n)
                         \nn \\
    &=& \frac{1}{4 \pi l} \sum_{n = - \infty}^{\infty} ( z_n^2 - 1)^{-1/2} 
           \lbb z_n + ( z_n^2 - 1 )^{1/2} \rbb ^{1-\lambda} 
           \comma \nn 
\eqaend
where 
\eqabegin
  x_n &\equiv& x \ \Bigm{\vert}_{\varphi \to \varphi - 2n \pi} \comma \qquad 
  z_n(x, x') \, = \, z(x, x'_n) 
  \period \nn
\eqaend
From (\ref{GfEq}), we can check
\eqabegin
 ( \Box - \mu l^{-2} ) \Gbh =  \frac{1}{\sqrt{-\gBHd}} \delta(x-x') 
  \comma 
 \label{GbhEq}
\eqaend
where $ \delta(x-x') $ is the delta function 
in the black hole geometry. Note that the Green functions are functions of 
\eqabegin
z_n(x,x') - i \varepsilon &=& \frac{1}{\dH^2} 
         \lbb \sqrt{r^2 -\rmi^2} \sqrt{r'^2 - \rmi^2} 
         \cosh \lb \frac{\rmi}{l^2} \Delta t - \frac{\rp}{l} \Delta \varphi_n
               \rb \right.
               \nn \\
      & & \left. \qquad \qquad - 
          \sqrt{r^2 -\rp^2} \sqrt{r'^2 - \rp^2} 
         \cosh \lb \frac{\rp}{l^2} \Delta t - \frac{\rmi}{l} \Delta \varphi_n
               \rb \rbb
              \comma \label{zn} 
\eqaend
where 
\eqabegin
 \Delta t &=& t - t' \comma \qquad 
    \Delta \varphi_n \, = \, \varphi - \varphi' + 2 n \pi
 \period \nn
\eqaend

For the conformally coupled massless scalar field, namely, 
for $ \mu = -3/4 $, we have $ \lambda_\pm = 3/2, 1/2 $, and the 
 Green functions are reduced to 
\eqabegin
 -i \Gbh(x,x') &=& \frac{1}{2^{\lambda + 1} \pi l} \sum_{n = - \infty}^{\infty}
             \lbb \frac{1}{\sqrt{z_n -1}} \pm \frac{1}{\sqrt{z_n + 1}}
             \rbb
             \period \nn
\eqaend
These coincide with the Green functions discussed in \cite{GFs,LO} 
which have the ``Neumann" or the ``Dirichlet" boundary condition.
%
\isubsubsection{Boundary conditions and the vacuum}
%
We have constructed the Green functions in a BTZ 
 black hole background. However, the physical meaning of the Green functions 
is not clear unless we specify its boundary conditions 
and identify the vacuum with respect to which they are defined.
It turns out \cite{IS} that $ \Gbh $   
satisfies the boundary conditions:
(i) to be regular at infinity, (ii) to be analytic in the upper 
half plane on the past complexified outer horizon, and 
(iii) to be analytic in the lower half plane 
on the future complexified outer horizon. These conditions 
fix $ \Gbh $ as a solution of the inhomogeneous wave equation  
\cite{HH}. This means that the vacuum is defined
 by the Kruskal modes, i.e., it is the Hartle-Hawking vacuum \cite{HH,GP}.
Thus the Green function is regarded as the Hartle-Hawking Green function
which is important to discussions on 
thermodynamics of black holes and Hawking radiation.
\parsmallskip
For the conformally coupled 
massless scalar field ($\mu = -3/4 $) in a non-rotating 
black hole geometry ($ \JBH = 0 $), the statements (i)-(iii) have been verified 
\cite{LO}.
Thus we follow the  strategy in \cite{LO}. For brevity, we concentrate 
on the case $ r, r' \geq \rp $ in the following. 
\parsmallskip
First, from the definition of $ \Gbh $, we easily find that 
the boundary condition (i) is satisfied .
\parsmallskip
Let us turn to the condition (ii).
We first introduce Kruskal coordinates \cite{BTZ} by 
\eqabegin
 V &=& R(r) \ e^{\aH t} \comma \qquad U \, = \, - R(r) \ e^{-\aH t} 
   \comma \nn \\
 R(r) &=& \sqrt{\lb \frac{r-\rp}{r + \rp} \rb 
     \lb \frac{r + \rmi}{r -\rmi}\rb ^{\rmi/\rp}} 
     \comma  \label{defKruskal} 
\eqaend
where 
$$  
    \aH = \frac{\rp^2 - \rmi ^2 }{ \rp l^2 } = \frac{2\pi}{\bH}
   \period
$$
In this coordinate system, the metric becomes
\eqabegin
 d \sBH ^2 &=& \Omega^2(r) d U d V + r^2 \lb N^\varphi d t + d \varphi \rb ^2 
  \comma \nn \\
 \Omega^2(r) &=& \frac{\rp ^2(r^2 - \rmi^2 )(r + \rp)^2}{\dH^4 r^2 } 
                 \lb \frac{r-\rmi}{r + \rp}\rb ^{\rmi/\rp}
 \period \nn
\eqaend
From (\ref{defKruskal}), the past complexified outer horizon 
is given by $ V = 0 $ and Re$(-U) > 0 $.
In terms of $ (t, r) $, this reads as
\eqabegin
 \lmb 
   \begin{array}{l}
    r \longrightarrow \rp \\
    t \longrightarrow - \infty \comma
   \end{array}
 \right.
   &{\rm with} & \sqrt{r- \rp} \ e^{-\aH t} \longrightarrow  \gamma_H 
  \period \nn
\eqaend
Here $ \gamma_H $ is a constant determined by the value of $ U $ and has  
the property Re $ \gamma_H > 0 $. Im $ \gamma_H >0 $ and Im $ \gamma_H < 0 $ 
correspond to the lower
and the upper half plane of $ U $, respectively. 
\parsmallskip
Then let us examine the regularity of $ \Gbh $. For this purpose, we further 
introduce a new 
angle coordinate rotating together with the outer horizon;
\eqabegin
 \phip &=& \varphi - \omh t   \period \nn 
\eqaend
Note that $ N^{\varphi} d t + d \varphi = d \phip $ on the outer horizon.
This is an analog of the angle coordinate in the Kerr geometry which 
is used for obtaining the regular expression 
of the metric on the outer horizon 
and for maximally extending the spacetime \cite{Cart}. 
Using $ \phip $,  $ z_n $ are written as 
\eqabegin
 && z_n(x,x') - i \varepsilon \ \equiv \ 
      z_n(w_n, \Delta \phip_n;r,r') - i \varepsilon \label{znp}  \\
  && \qquad   =
   \frac{1}{\dH^2} 
         \lbb \sqrt{r^2 -\rmi^2} \sqrt{r'^2 - \rmi^2} 
         \cosh \lb \frac{\rp}{l} \Delta \phip_n
               \rb  - 
          \sqrt{r^2 -\rp^2} \sqrt{r'^2 - \rp^2} 
         \cosh \lb i w_n
               \rb \rbb 
               \comma \nn
\eqaend
where
\eqabegin
  \Delta \varphi^+_n &=&= \varphi^+ - \varphi'^+ + 2 n \pi  \comma \nn \\ 
  i w_n &=& \aH \Delta t - \frac{\rmi}{l} \Delta \phip_n \period \nn
\eqaend
Thus, on the past complexified outer horizon
we have
\eqabegin
z_n(x,x') - i \varepsilon &\longrightarrow& \frac{1}{\dH^2} 
         \lbb \sqrt{\rp ^2 -\rmi^2} \sqrt{r'^2 - \rmi^2} 
         \cosh \lb \frac{\rp}{l} \Delta \phip_n
               \rb \right.
               \nn \\
      & & \left. \qquad \qquad - 
          \sqrt{ \rp l/2 } \sqrt{r'^2 - \rp^2} 
         \ e^{ \aH  t'  +  (\rmi/l) \Delta \phip_n } \ \gamma_H \rbb
                \period \nn
\eqaend
Let us recall that each component
in the summation in $ \Gbh $ has singularities at $ z_n = \pm 1 $.
From the above expression of $ z_n $, we find that
the points $ z_n = \pm 1 $ on the past complexified outer horizon 
correspond to  
\eqabegin
 \gamma_H  &=& \alpha^\pm_0 + i \varepsilon \comma \nn
\eqaend
where $ \alpha^\pm_0 $ are some positive numbers.
Consequently, each component in $ \Gbh $ is regular 
in the upper half plane of $ U $. 
\parsmallskip
Then we will use Weierstrass' theorem \cite{Ah}: if a series 
with analytic terms converges uniformly on every compact subset of a region, 
then the sum is analytic in that region, and the series can 
be differentiated term by term. Since  
the series in $ \Gbh $ converges uniformly,    
$ \Gbh $ is analytic in the upper half plan of the past complexified outer
horizon. This completes the proof of (ii).
\parsmallskip
The proof of (iii) is similar, and we omit it.
\parsmallskip
The Hartle-Hawking Green function in a 
black hole geometry was originally defined in the path-integral formalism 
as a generalization of the Feynman Green function in Minkowski 
spacetime \cite{HH}. In our case, $ \Gf $ is also defined so as to 
conform to the Feynman Green function in the flat limit 
(see (\ref{epGf}) and the comment below it). Thus it is natural for  
$ \Gf $ to satisfy the Hartle-Hawking boundary condition. 
%
%
\csubsection{Thermodynamics of scalar fields by Hartle-Hawking Green functions}
We have constructed the Hartle-Hawking Green functions in the BTZ 
background. In this section, 
we discuss thermodynamics of a scalar field by using them \cite{IS}. 
%
\isubsubsection{Euclidean Green functions}
%
For calculating the thermodynamic quantities, we will introduce 
Euclidean Green functions and study their properties \cite{IS}.
First, we define the Euclidean time  by $ \tau = i t $ and the 
`` Euclidean angle "  by $ \varphiE = - i \varphi $ for $ \JBH \neq 0 $ and 
$ \varphiE = \varphi $ for $ \JBH = 0 $. Furthermore, we 
introduce a new coordinate
$$
    i \delphiEp_n  \equiv \delvarphi ^+_n = \lmb
      \begin{array}{ll}
                  i ( \delphiE + \omh \deltau ) + 2 \pi n
                & {\rm for } \ \JBH = 0 \\
                   \delphiE + i \omh \deltau ) + 2 \pi n 
                & {\rm for } \ \JBH \neq 0 

      \end{array}
   \right. \period
$$
Then we have 
\eqabegin
 &&  - w_n = \aH \deltau + \frac{\rmi}{l} \delphiEp _n \period \label{wn2}
\eqaend
Using the Euclidean coordinates, we obtain the Euclidean black hole 
geometry 
\eqabegin
 d s_E^2 &=& g^E_{\mu \nu} d x^\mu d x^\nu \ =  \
          \lmb 
          \begin{array}{ll}
               N_\bot^2 d \tau^2  + r^2 d \varphiE ^2 + N_\bot^{-2} d r^2
                & {\rm for } \ \JBH = 0 \\
               N_\bot^2 d \tau^2  - r^2 ( N^\varphi d \tau - d \varphiE )^2 
               + N_\bot^{-2} d r^2 & {\rm for } \ \JBH \neq 0 
          \end{array}
          \right. \period \nn
\eqaend
Notice that the Euclidean metric becomes
complex unless we use the Euclidean angle.\footnote{
We can make the Euclidean metric real by a continuation
$ \JBH ^E  = -i \JBH $ instead of $ \varphiE = -i \varphi $ \cite{Carlip}.
The following discussions are valid also in this case 
with slight modifications. } 
Then the Green function in the Euclidean geometry (the Euclidean Green function)
is given by
\eqabegin
  \Gbhe (x,x') &\equiv&
       \sum_{n = - \infty}^{\infty}  \Gfe (x,x'_n)
       \comma \nn \\
   \Gfe (x,x'_n) &\equiv& 
     i \Gf (z(w_n(\deltau,\delphiEp _n), i\delphiEp _n; r,r'))
       \comma \label{GBHE}
\eqaend
(recall (\ref{znp})).
The factor in front of $ \Gf $ was
chosen so that the physical quantities calculated later will 
have real values and  appropriate signs.
From (\ref{GbhEq}), we can check
\eqabegin
 ( \Box^E - \mu l^{-2}) \Gbhe &=& 
            \frac{a}{\sqrt{ \abs{g^E} }} \ \delta^E (x-x') 
                        \comma \nn
\eqaend
where $ a = -1 $ for $ \JBH = 0 $ and $ a = i $ for $ \JBH \neq 0 $.
\parsmallskip
Next, we consider thermal properties of $ \Gbhe $. 
The Green function $ \Gbh $ was a function of $ z_n $ given by (\ref{zn}).
Thus $ \Gbhe $ for $ \JBH \neq 0 $ is periodic under 
\eqabegin
   \delta \lb \rmi  \frac{\tau}{l^2} + \rp  \frac{\varphi}{l} \rb &=&
      2 \pi m  \  \nn \\
   \delta \lb  \rp \frac{\tau}{l^2} +  \rmi \frac{\varphi}{l} \rb &=&
      2 \pi n  \  \qquad 
     ( m, n \in {\bf Z} ) \comma \nn
\eqaend
where $ \delta ( ... ) $ means the variation of the arguments.
Namely, $ \Gbhe $ is of double period
\eqabegin
  \vecii{\delta (\tau/l) }{ \delta \varphi} &=& 
  \frac{2 \pi  l}{\dH^2} \matrixii{-\rmi}{\rp}{\rp}{-\rmi}
   \vecii{m}{n}   \period \nn
\eqaend
If we require that 
the chemical potential vanishes as $ \JBH \to 0 $ $ (\rmi \to 0 ) $, 
the fundamental period is determined uniquely as 
\eqabegin
  \tau  &\to & \tau + \bH  \comma \qquad
  \varphi  \, \to \, \varphi -  \nuBH \bH  \comma \nn
\eqaend
where 
\eqabegin
 \nuBH  & =  & \frac{\rmi}{l\rp } \, = \, \omh 
 \period \nn
\eqaend
This periodicity is also valid for $ \JBH = 0 $.
Recall that a thermal Green function at temperature 
$ \beta^{-1} $ and with a chemical potential $ \nu $ conjugate to 
angular momentum is defined by 
\eqabegin
 G_{\beta}^E (x, x'; \nu) &=& \Tr   
          \lbb  \ e^{ -\beta( \hat{H}- \nu \hat{L})} \ T 
          \lb \psi(x) \psi(x') \rb \rbb 
          \biggm{/} \Tr
          \lbb  \ e^{ -\beta( \hat{H}- \nu \hat{L})} \rbb 
      \comma \nn 
\eqaend
where $ T $ denotes the (Euclidean) time ordered product and $ \hat{H} $ and 
$ \hat{L} $ are 
the generators of time translation and rotation, respectively.
From the definition, we have
\eqabegin
  G_{\beta}^E (\tau, \varphi, r; \tau', \varphi', r'; \nu) &=& 
  G_{\beta}^E 
    (\tau + \beta , \varphi -  \nu \beta , r; \tau', \varphi', r'; \nu)
  \period \nn
\eqaend
Comparing this with $ \Gbhe $,  
$ \Gbhe $ can be regarded as 
a thermal Green function with the inverse temperature  $ \beta = \bH $
and  the chemical potential 
$ \nuBH $.  This is consistent with the classical result in (\ref{clTS}).
In the following, we will explicitly denote the period of Green 
functions, for example,  
as $ \Gbhe (x,x'; \bH ) $. 
\parsmallskip
It is instructive to consider  the behavior of the metric near the 
outer horizon. We introduce a coordinate $ \eta $ by 
\eqabegin
   r &=& \rp  + \frac{2 }{\aH}  \eta^2 
  \period \nn
\eqaend  
Then, for small $ \eta $ , the metric becomes
\eqabegin
 d s^2 & \sim & - ( 2\pi/\bH )^2
    \ \eta^2 \  d t^2 + d \eta^2 + \rp^2 (d \phip)^2 
   \period \label{metho}  
\eqaend 
In terms of the Euclidean time $ \tau =  i t $ , 
$ ( \tau, r ) $ (or $( \tau, \eta)$) space represents a plane 
with the origin $ r = \rp $ ($ \eta = 0 $). 
Therefore $ \beta_H $ is  nothing 
but the period around the outer 
horizon of the Euclidean black hole. 
The periodicity of 
 $ \Gbhe $  gives an explicit example to 
the arguments in the literature of thermodynamics of black holes. 
%
\isubsubsection{Free energy}
%
By making use of the Euclidean Green function, 
we calculate the free energy in this subsection. In terms of 
the Euclidean Green function, the free energy is given by 
the formula (\ref{ZG}).
In our case, the trace is defined by 
\eqabegin
  \Tr  (\  ... \ ) &=&  \int d^3 x  \sqrt{\abs{g^E} } \lim_{x \to x'} 
                   ( \ ... \ )
                     \nn \\
               &=& 
          \lmb 
          \begin{array}{ll}
                   \int_0^{\beta} d \tau \int_0^{2 \pi} d \varphiE
                     \int_{\rp}^{\infty} d r \cdot r \ \lim_{x \to x'}
                     ( \ ... \ ) 
                & {\rm for } \ \JBH = 0 \\
                   \int_0^{\beta} d \tau \int_0^{\omh \beta} 
                   d \varphiE
                     \int_{\rp}^{\infty} d r \cdot r \ \lim_{x \to x'}
                     ( \ ... \ ) 
               & {\rm for } \ \JBH \neq 0 
          \end{array}
          \right.  \period \nn
\eqaend
Here we have set  
the lower end of the integration with respect to $ r $ to be  
$ \rp $. The reasons are (i) in the Euclidean geometry, 
the topology of $ (\tau, r ) $
space is $ {\bf R}^2 $ and the origin corresponds to $ r = \rp $, 
and (ii) it turns out that the entropy becomes complex if we perform  
integration below $ \rp $.
\parsmallskip
For flat spacetime, an expression for free energy like (\ref{ZG}) is 
divergent, and we have to regularize it by differentiating it 
 with respect to mass squared. 
To get the right answer, we then integrate the differentiated expression.
Thus we will follow the prescription 
for flat spacetime. In our case, the parameter $ \mu $
corresponds to mass squared and we get 
\eqabegin
  \frac{\del}{\del \mu} \lb \beta  \ F(\beta) \rb 
  &=& - \frac{1}{2 l^2} \Tr  \Gbhe (\beta) 
  \period \nn
\eqaend
where we have used $ \Gbhe = ( \Box^E - \mu l^{-2})^{-1} $ (up to a factor).
\parsmallskip
Then the remaining calculation is straightforward. 
We first consider $ \JBH \neq 0 $ case. In this case, we have 
\eqabegin
  \frac{\del}{\del \mu} \lb  \beta \ F(\beta) \rb  
          \biggm{\vert}_{\beta = \bH} &=& 
          - i \ \frac{\omh}{4 l^2} \beta^2 \sum_{n = - \infty}^{\infty}
          \int_{\rp^2}^{\infty} d (r^2) 
           \lim_{r \to r'}  \Gfe ( z_n^0; \bH)
           \ \when_{\beta = \bH}
           \comma \label{delF}
\eqaend
where $ z_n^0 = z_n \vert _{\deltau = \delphiE = 0} $, and  we have used 
the fact that the integrand is independent of $ \tau $ and $ \varphiE $.
Recall the expression of $ \Gf $ and $ z_n $, i.e., (\ref{Gf}) and 
(\ref{zn}). Then
the integrand with $ n = 0 $ in the summation
diverges as $ r  \to r ' $.  So we remove this term 
for a  moment.
Notice that $ \Gf(z_n; \bH) $ and $ z_n^0 $ are written 
as 
\eqabegin
  && - i \Gf (z_n; \bH) = 
          \lmb
            \begin{array}{ll}
              {\displaystyle \frac{1}{4 \pi l (1 - \lambda)}  \frac{d}{d z_n}
              \ e^{(1-\lambda) \coth^{-1} z_n } }
              & {\rm for } \ \lambda \neq 1 \\
              {\displaystyle \frac{1}{4 \pi l} } \lb  z_n^2 -1 \rb ^{-1/2} 
              & {\rm for } \ \lambda = 1 \\
            \end{array}
          \right. \comma \nn \\
   &&      z_n^0 \when_{r = r'} = \frac{1}{\dH^2} 
            \lmb (r^2 - \rmi^2) \cnp - (r^2 - \rp^2 ) \cnm \rmb
            \comma \nn
\eqaend
where 
\eqabegin
 \cn^\pm &=& \cosh \lb 2 \pi n \frac{r_\pm}{l} \rb 
  \period \nn
\eqaend
Here and in the following, we omit the infinitesimal imaginary part 
of $ z_n $ except when it is relevant for the  
discussion. 
Then by making the change of variables from $ r^2 $ to $ z_n^0 $, we get 
\eqabegin
 && -i \int_{\rp^2}^{\infty} d (r^2)  \lim_{r \to r'} \Gfe (z_n^0; \bH) \nn \\
 && \qquad = \ \lmb
      \begin{array}{ll}
      {\displaystyle  \frac{1}{4\pi l} \frac{\dH^2}{(\cnp - \cnm)(1 -\lambda)} 
         \lb  z + \sqrt{z^2-1} \rb ^{1 -\lambda} \when_{\cnp}^{\infty} }
       & {\rm for } \ \lambda \neq 1 \\
       {\displaystyle \frac{1}{4\pi l } \frac{2 \dH^2}{(\cnp - \cnm)} 
         \ln  \lb  z + \sqrt{z^2-1} \rb \when_{\cnp}^{\infty} }
       & {\rm for } \ \lambda = 1 
      \end{array}
     \right. \period \nn
\eqaend
The integral diverges at the upper end for $ \lambda \leq 1 $, 
but for $ \lambda > 1 $ 
 we get 
\eqabegin
 &&  \frac{\del}{\del \mu} \lb \beta  F(\beta) \rb 
          \biggm{\vert}_{\beta = \bH}  
     =
  \frac{\omh \bH^2 \dH^2 }{8 \pi l^3 (\lambda-1)}  
   \sum_{n \geq 1}^{\infty} \frac{1}{\cnp - \cnm} \ 
   e^{ - 2 \pi (\lambda-1) n \rp/l} \ + c_0  \comma \nn
\eqaend
where $ c_0 $ is the divergent term coming from $ n = 0 $.
Consequently, by integrating the above expression, we obtain \cite{IS}
\eqabegin
   \beta  F(\beta)  
          \biggm{\vert}_{\beta = \bH} 
 &=& 
  - \frac{\omh \bH^2 \dH^2 }{8 \pi^2 \rp l^2 } 
     \sum_{n \geq 1}^{\infty} 
    \frac{1}{n (\cnp -\cnm)} 
    \ 
   e^{ - 2 \pi (\lambda-1) n \rp/l}  \nn \\
 & & \qquad \qquad \qquad \qquad + \ \ {\rm const.} 
     \qquad \qquad ({\rm for} \ \lambda > 1 ) \period \nn
\eqaend

Similarly, we get the result 
for $ \JBH  = 0 $ ;
\eqabegin
   \beta  F(\beta)  
          \biggm{\vert}_{\beta = \bH} 
 &=& 
  -   \frac{\rp \bH}{4 \pi l^2} 
    \sum_{n \geq 1}^{\infty} \frac{1}{n (\cnp - \cnm)}
    \ 
   e^{ - 2 \pi (\lambda-1) n \rp/l}  \nn \\
 & & \qquad \qquad \qquad + \ \ {\rm const.} 
     \qquad \qquad ({\rm for} \ \lambda > 1 ) \period \nn
\eqaend
This can be obtained also by the replacement $ \omh \bH \to 2 \pi  $ in the 
expression for $ \JBH \neq 0 $. 
%
\isubsubsection{Green functions on a cone geometry}
%
In order to calculate the entropy, we have to differentiate the
Euclidean Green function with respect to $ \beta $ with the chemical 
potential fixed. 
Namely, we need the Green functions with a
period different from $ \beta_H $ with $ \omh $ fixed.
These Green functions are regarded as those on $ (\tau, r) $ plane
with a deficit angle around the origin, i.e., on a cone geometry. 
In this subsection, we will construct the Green functions on a cone geometry
with an arbitrary period \cite{IS}.
\parsmallskip
To get them, we first take $ \Gbhe $ as a function of $ r^{(')}$,   
$ \deltau $ and $ \delphiEp $.
Then they are obtained by (i) fixing 
the values of $ r^{(')}$ and $ \delphiEp $ and (ii) changing
the period with respect to $ \deltau $.
The chemical potential is surely fixed by this procedure. 
We denote these Green functions by 
\eqabegin
    && \Gbhe(x,x'; \beta)  
   =  \sum_{n = - \infty}^{\infty} \Gfe( x,x'_n; \beta) 
  \comma \nn
\eqaend
where 
$ \beta $ is the period of $ \deltau $, and 
$ \Gfe(x,x'_n; \beta) $ are Green functions on a cone geometry 
corresponding to $ \Gfe(x,x'_n; \bH) $. $ \Gbhe(x,x'_n; \bH) $ and 
$ \Gfe(x,x'_n; \bH) $ are the Green functions in (\ref{GBHE}), 
which we have considered so far.
\parsmallskip
The authors of \cite{SoCa} discussed how to construct 
Green functions with an arbitrary period for certain differential equations.
Their method was also 
applied to field theory on curved spaces \cite{Do,SF}.
By following them,  we can obtain the explicit form of $ \Gfe $ \cite{IS}:
\eqabegin
  \Gfe ( x, x'_n; \beta ) &=& \frac{\bH}{2 \pi \beta} \int_\Gamma d \zeta \ 
  \tGf( \zeta ; 2 \pi ) 
  \frac{ e^{ i \bH \zeta/\beta}}
{ e^{ i \bH \zeta/\beta}- \ e^{ i \bH w_n /\beta}}
  \comma \nn \\
  &=& \frac{\bH}{ 4 \pi i \beta} \int_\Gamma  d \zeta \ 
       \tGf(\zeta; 2 \pi)
        \cot \lmb \frac{\bH}{2 \beta}( \zeta - w_n ) \rmb
        \period \nn 
\eqaend
where $ \tGf(\zeta ; 2 \pi )  $ is defined by 
\eqabegin
  \tGf ( \zeta ; 2 \pi ) &\equiv& \Gfe ( z(\zeta, \delphiEp _n; r,r'); \bH ) 
                    \when_{ \delphiEp_n, r, r' : \ {\rm fixed} } 
    \period \nn
\eqaend
%
%
From (\ref{znp}) (or (\ref{zzeta})),  
$ \tGf(\zeta ; 2 \pi )  $ is periodic under
$ \zeta \to \zeta + 2 \pi $ $ (\tau \to \tau + \bH) $.
The contour $ \Gamma $  is given by the solid lines in Figure 2.
Making use of this expression, we can get the derivative of the Green function
with respect to $ \beta $. At $ \beta = \bH $, it is given by \cite{IS}
\eqabegin
  && \frac{\del}{\del \beta} \Gfe(x,x'_n; \beta) \when_{\beta = \bH}  \nn \\
  &=& - \frac{B}{\bH} \int_{\An + B }^{\infty} d z 
   \ \Gfe (z; \bH) \frac{1}{\sqrt{( z - \An )^2 - B^2}}
   \frac{\cn (z  -  \An)  +   B }{( z  -  \An  +   \cn B )^2}
  \comma  \label{delGf}
\eqaend
where 
\eqabegin
  \An &=& \frac{1}{\dH^2} \sqrt{r^2 - \rmi^2}\sqrt{r'^2 - \rmi^2}
    \cosh \lb \frac{i\rp}{l} \delphiEp_n \rb
  \comma \quad  B \, = \, 
    \frac{1}{\dH^2} \sqrt{r^2 - \rp^2}\sqrt{r'^2 - \rp^2}
  \comma \nn \\
    c_n  &=& \cosh( iw_n ) \period \nn
\eqaend
The derivation of these results is given in appendix B.
\parsmallskip
Note that  $ B>0  $ for $ r, r' > \rp $ and $ B < 0 $ 
for $ r, r'< \rp $. Thus, for $ r, r' < \rp $, 
$ \An + B  $ can be less than $ 1 $, and 
the Green function $ -i\Gfe(z; \bH) $  becomes complex.
Because of 
the contribution from this region,  the entropy also becomes complex. 
This indicates that we should consider only the region $ r > \rp $ in the 
calculation of thermodynamic quantities as we have done.
%
\isubsubsection{Entropy}
%
Making use of the Green functions with a generic $ \beta $, we can
obtain the entropy \cite{IS}.
Since the definition of the trace is different between $ \JBH = 0 $ and 
$ \JBH \neq 0  $ case, we first consider 
%
%
\newpage
\begin{minipage}{3.5in}
\epsfxsize 2.5in
\epsfbox{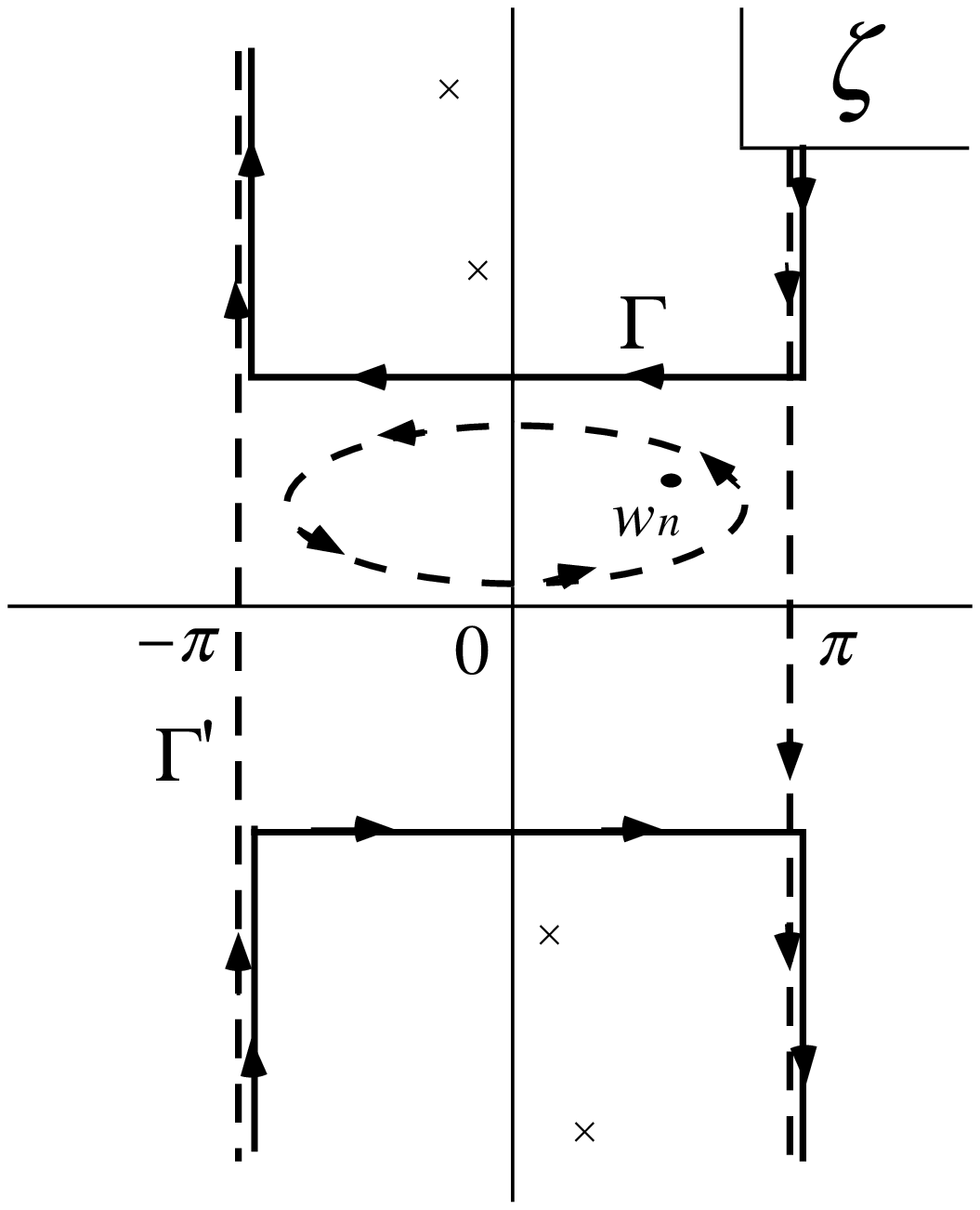}
\end{minipage}
\hspace*{0.2in}
\hfill
\begin{minipage}{3.0in}
\vspace{0.5in}
{\footnotesize {\bf Figure 2 \ :} 
       Contour $ \Gamma $ (solid lines) and Contour $ \Gamma' $ 
      (dashed lines) in $ \zeta  $ plane. The crosses $ (\times) $ 
      indicate the singularities of $ \tGf(\zeta;2\pi)$ in the region 
      $ - \pi < {\rm Re } $ 
       $ \zeta \leq \pi $ for $ r,\ r' \geq \rp $. The dot $ (\bullet) $
      indicates $ w_n $ up to $ 2 \pi m $.
      In this figure, we 
      show the contour $ \Gamma $ for small $ \abs{ \mbox{Im} \ w_n } $. 
      For 
      large  $ \abs{ \mbox{Im} \ w_n } $, 
      the line $ \mbox{Im} \ \zeta  = \mbox{Im} \ w_n  $ 
      is, for example, above the crosses, and 
      we can not take a contour like $ \Gamma $ in this figure.  
      However, in this case we have only to deform $ \Gamma $ and 
      take a contour topologically equivalent to $ \Gamma' $.   
 }
\end{minipage}
%
\parbigskipn
\parbigskipn
$ \JBH \neq 0 $ case.
First, from (\ref{delF}) and (\ref{delGf}), we get  
\eqabegin
  && \frac{\del}{\del \mu} S(\beta) \when_{\beta = \bH}
   = - \frac{i}{4 l^2} \omh \bH^2 
  \sum_{n = - \infty}^{\infty} \    
\int_{\rp^2}^{\infty} d(r^2) \  
     \lim_{r \to r'} 
        \Biggl[ \Gfe(z_n; \bH) 
     \nn \\
  && \quad     \left.
     -  B \int_{\An  +   B }^{\infty} 
     d z \ 
     \Gfe(z; \bH) \   
     \frac{1}{\sqrt{( z  -  \An )^2  -  B ^2}}
\frac{\cn (z  -  \An) +   B }{( z  -  \An  +   \cn B )^2} \rbb
       \when_{\deltau = \delphiE = 0}
 \period \nn
\eqaend
The first term is nothing but 
$ \del_{\mu}( \beta F(\beta) \vert _{\beta = \bH} ) $. 
By integrating the above expression, we get the exact expression of 
the entropy
\eqabegin
  && S(\bH) = \bH F(\bH) 
   +   \frac{\omh}{8 \pi l^3 } \ \bH^2 \ 
  \sum_{n = - \infty}^{\infty} \ 
 \int_{\rp^2}^{\infty} d (r^2) 
  \ \lim_{r \to r'} 
    \  \Biggl[ B \int_{\An  +   B }^{\infty} 
     d z  
 \label{SJ} \\
  && \qquad  \times \left.
     \frac{X^{1  -  \lambda} \lmb 1 + (\lambda  -  1) 
     \ln  X \rmb }
      {\ln ^2 X \sqrt{z^2   -  1}}
      \frac{1}{\sqrt{( z  -   \An )^2  -  B^2 }}
\frac{\cn (z  -  \An) +   B }{( z  -  \An  +   \cn B )^2} \rbb
       \when_{\deltau = \delphiE = 0}
   + \ c  \period \nn
\eqaend
Here  
\eqabegin
 X &=& z + \sqrt{z^2 -1}
 \comma \nn
\eqaend
and $ c $ is a constant independent of $ \mu $, which 
is dropped in the flat case. 
\parsmallskip
For $ \JBH = 0 $, we similarly get 
\eqabegin
  && S(\bH)  
   =   \frac{1}{4 l^3 } \ \bH \ 
\sum_{n = - \infty}^{\infty} \
\int_{\rp^2}^{\infty} d (r^2)  \
 \lim_{r \to r'} \
     \Biggl[ B \int_{\An  +   B }^{\infty} 
     d z  
   \label{SJ0} 
  \\
&& \quad     \left. \times
\frac{X^{1  -  \lambda} \lmb 1  +   (\lambda  -  1) \ln  X \rmb }
      {\ln ^2 X \sqrt{z^2   -  1}}
      \frac{1}{\sqrt{( z  -   \An )^2  -   B^2}}
\frac{\cn (z  -  \An )  -  B }
{( z  -   \An  +    \cn B )^2} \rbb
       \when_{\deltau = \delphiE = 0}
   + \ c  \period \nn
\eqaend
This can be obtained also from the second term in (\ref{SJ}) by 
the replacement $ \omh \bH \to 2 \pi $. The term 
corresponding to the first term in (\ref{SJ}) is absent because the 
power of $ \beta $ in Tr $ \Gbhe  $ is different.
\parsmallskip
As we have exact expression of the entropy, 
we can study its properties without any ambiguity \cite{IS}.
Here, we will concentrate on the structure of the divergences 
coming from short-distance behavior.
In the recent arguments \cite{KS}-\cite{SF}, 
these divergences are considered to be important for understanding
the black hole entropy and quantum aspects of gravity. 
We find that the short-distance divergences
come from (i) the contribution from taking 
the trace (ordinary ultraviolet divergences in statistical field theory),
or (ii) the integration near the outer horizon. 
\parsmallskip
To see this, we introduce a variable $ \rho $  defined by 
$
   \rho^2 = r^2 - \rp^2 \comma 
$
and cutoffs   
$$
 s^2  =  r'^2 - r^2 
$$ 
for the trace, and 
$$
\rp ^2 \to \rp ^2 + \eH ^2
$$ 
for the integration near the horizon.
In the limit $ \rho, \ s \to 0 $, we have 
\eqabegin
  \An &\sim& \cnp \lb 1 + \frac{1}{\dH^2} (\rho^2 + s^2/2 ) \rb
 \comma  \quad B \ = \  \frac{\rho}{\dH^2}  \sqrt{\rho^2 + s^2} 
 \comma \nn \\
  z^0_{n=0} & = &  z(x,x') \when_{\deltau = \delphiE = 0} 
    \sim  \ 1 + \frac{s^4}{8(r^2-\rp^2)(r^2 - \rmi^2)}
    \quad  \qquad \qquad  \qquad {\rm for} \ \rho^2 \gg s^2 \nn \\
   & & \qquad \qquad \qquad \qquad  \sim  \ 
   1 + \frac{1}{\dH^2} 
   \lb \rho^2 + \frac{s^2}{2}- \rho \sqrt{\rho^2 + s^2} 
   \rb  \ \qquad  \quad \mbox{ otherwise }
      \period \nn 
\eqaend
Then we find that the ultraviolet divergences  
come only 
from the term with $ n = 0 $  in (\ref{SJ}) and (\ref{SJ0}) 
(including the $ n=0 $ term of $ F(\bH) $ in (\ref{SJ})).  
\parsmallskip
The integrand in the $ n = 0 $ term of $ F(\bH) $ 
becomes divergent for small $ s $ like
\eqabegin
 && \frac{1}{\sqrt{\sigma(x,x')}} \sim \frac{1}{s^2} 
    \sqrt{(r^2 - \rp^2)(r^2 - \rmi^2)}
    \period \nn
\eqaend
This can be regarded as an ordinary ultraviolet divergence of 
field theory since $ \sqrt{\sigma(x,x')} $ is a distance.
\parsmallskip
The term relevant to the other divergences for $ \JBH \neq 0 $ is 
the $ n = 0 $ term in the summation in (\ref{SJ}).
We rewrite this  term by 
$
  z' \equiv  z - A_0  $ 
and 
$
  \delta  =  A_0 -1 
  \sim \dH^{-2}\lb \rho^2 + {s^2}/{2} \rb
$.
Noting that $ \ln  X \sim \sqrt{2(z'+ \delta)} $ up to 
$ \calO (z', \delta) $, we find the expression of the divergences; 
\eqabegin
 I &\sim& 
 \frac{\omh}{8 \pi l^3 } \ \bH^2  \int_{\eH^2} d(\rho^2) B \int_{B} d z'
 \frac{1  +   ( \lambda  -  1) \sqrt{2(z'  +   \delta )}}
 {( z'  +   \delta)(z'  +   B)
 \sqrt{z'  +   \delta}\sqrt{z'^2  -  B^2}} \nn \\
 &=& 
 \frac{\omh}{8 \pi l^3 } \ \bH^2  \int_{\eH^2} d(\rho^2)  \ B \ \int^{1} d v
  \nn \\
  && \qquad \times \
 \frac{1}{\lb 1  +   (\delta/B) v \rb (1  +   v) 
  \sqrt{1  -  v^2}}  \
   \lmb 
  \frac{1}{\sqrt{1  +   (\delta/B) \ v} } \ \lb 
  \frac{v}{B} \rb ^{3/2}
  + \sqrt{2} ( \lambda  -  1 ) \frac{v}{B}
 \rmb
 \nn \comma
\eqaend
where $ v = B/z' $. Here we have two cases.
First, for $ \eH, \rho \simeq s $ or $ \eH, \rho \gg s $, 
we obtain
\eqabegin
I & \sim & \omh \bH^2 l^{-3} \int_{\eH^2} d ( \rho^2 ) 
        \lb B^{-3/2} + \ c \ B^{-1} \rb \nn \\
   & \sim &
    \omh \bH^2 \dH^2 l^{-3} \lb \dH / \eH + \ c' \ \ln ( \dH/\eH )  \rb
    \comma \nn 
\eqaend
where $ c $ and $ c' $ are constants.
On the other hand, for $ \eH, \rho \ll s $, 
we obtain
\eqabegin
I & \sim & \omh \bH^2 l^{-3} \int_{s^2} d ( \rho^2 ) 
        \lb \delta^{-3/2} + \ c \ \delta^{-1} \rb \nn \\
   & \sim &
    \omh \bH^2 \dH^2 l^{-3} \lb \dH / s 
    + \ c' \ln ( \dH/s )  \rb
    \period \nn 
\eqaend
Therefore the divergences are given in terms of the larger  
cutoff, i.e.,  max$ \{ \eH, s  \} $. Namely, we  have only to introduce 
$ s $ {\it or } $ \eH $ in order to regulate the divergences.
\parsmallskip
Similarly, the divergences for $ \JBH = 0 $ is obtained from the above 
by the replacement
$ \omh \bH \to 2 \pi $. 
\parsmallskip
From the expression of the entropy, 
we find  various divergences 
coming from short 
distance such as $ \eH^{-1}, \ln  \eH^2, $ and $ s^{-2} $. However, 
all of them 
are not due to the existence of the outer horizon.
The expression of the divergences depends largely upon the regularization
schemes. If we adopt the regularization related to the horizon, they 
are expressed by the cutoff related to the horizon. But if not, this is 
not the case. In addition,  
the leading divergent term is proportional to $ \omh \dH^3 \bH ^2 l^{-3} $
$= \rp\rmi/\dH $ 
for $ \JBH \neq 0 $ and to $  \dH^3 \bH l^{-3} $ $=\rp^2/l $ for $ \JBH = 0 $.
Therefore, they are not proportional to the area of the horizon.
\parsmallskip
These results might be due to the special properties of the BTZ 
black hole. But they seem to indicate the importance 
of regularization schemes and of curvature 
effects in a truly curved spacetime.
We should carefully examine what regularization scheme is physical
so that we justify the claim in the literature that the horizon 
plays an important role in black hole thermodynamics. 
%
%
\csubsection{Discussion}
In this chapter, 
we explored the thermodynamics of scalar fields 
in the BTZ black hole background in the framework of quantum field 
theory in curved spacetime. We took two approaches. One  
was based on 
 mode expansion of the scalar field and summation 
over states.
We obtained mode functions and explicit forms of densities 
of states, free energies and entropies. We did not need the
 WKB approximation. We found that   
the thermodynamic quantities  depended largely upon boundary conditions. 
In particular, divergent terms of the entropy were not necessarily due to 
the existence of the outer horizon. 
In addition, the partition function did not take the form (\ref{Fbeta}), 
and the 
entropy was not proportional to the area of the outer horizon.
\parsmallskip
In the other approach, we used Hartle-Hawking Green functions. 
For  scalar fields with 
a generic mass squared, we obtained the 
Hartle-Hawking Green functions and Green functions 
on a cone geometry. Moreover, we obtained 
exact expressions of free energies and entropies. 
We did not need the heat kernel expansion.
We again found that the free energy did not take the from (\ref{Fbeta})
and the entropy was not proportional to the area of the horizon.
The divergences of the entropy were not necessarily due to the horizon.
They depended upon regularization schemes. The thermodynamic quantities 
were different from those in the previous approach.
\parsmallskip
We can imagine several reasons that our results do not agree with 
those in the literature:
(i) Three dimensional spacetime and four dimensional spacetime 
    (and $D$-dimensional spacetime considered in \cite{KS}-\cite{SF}) are 
    different;  in the former, gravitons do not exist and 
   the meaning of gravitational coupling is not clear. 
(ii) The BTZ black hole is different from a four dimensional black hole;
 it is not asymptotically flat, but it approaches $ \AdS $, and it 
 has a cosmological constant (fundamental scale). 
(iii) The results in \cite{KS}-\cite{SF} are obtained in the Rindler
 (flat) limit, and they may receive large corrections by curvature effects.
\parsmallskip
It is worthwhile investigating that reason further.
However, we should note that our results indicate the importance 
of curvature effects and 
of precise discussions on boundary conditions, regularization schemes, 
and relations among methods of calculation.
We saw that the claims in the literature were largely affected 
by them.\footnote{
Recently, entropies of a scalar field in a BTZ black hole background
 were discussed
in the WKB approximation \cite{LKK} and in the heat kernel expansion
\cite{MSo}. In the former approach, the entropy is proportional to the 
area. In the latter, they argued that the divergences can be 
absorbed into the renormalization of the gravitational coupling constant
and the cosmological constant. But the entropy is not proportional to the 
area. The relation among them and ours are not clear.
} 
Thus without settling these problems we cannot understand 
the meaning of the entropy and its relation to the information paradox 
and the renormalization of the gravitational coupling constant.
\parsmallskip
At present, we have no ``definition'' of thermodynamics of a quantum 
field in a black hole background. We have to look for it taking into account
 physical consistency as a clue 
(if black hole thermodynamics is truly sensible).
We have to determine the ``correct'' boundary condition and prescription 
to calculate the thermodynamic quantities.
Admittedly, this is not an easy task. 
In the case of the BTZ black hole, one possibility is to utilize the 
result of $ \CAdS $. Quantization of a scalar field 
in $ \CAdS $ has been well studied \cite{AIS}-\cite{BL}. One knows
the inner product and the complete basis. As discussed in the next chapter, 
$ \CAdS $ is the same as $ \cSLtwo $ and the mode functions are expressed by 
the matrix elements of $ \cSLtwo $ representations in the elliptic basis 
(see appendix C). On the other hand, our mode functions in (\ref{modeUV})
are the matrix elements in the hyperbolic basis. Thus we may obtain
a sensible quantum theory from the one in $ \CAdS $ by changing the basis 
of $ \cSLtwo $ representations. In turn this may settle the problems. 
Another possibility is to investigate the string theory in the BTZ black hole 
background 
because string theory is regarded as the fundamental theory including gravity.
This is the subject in the next chapter.
\parsmallskip
In the study of quantum black holes, it has been hardly possible to
make precise discussions and to examine various claims. This is 
because we have no quantum theory of gravity and a system of quantum 
fields plus a black hole is quite complicated. 
In our model, we could discuss the thermodynamics of a quantum scalar
field in the BTZ background without ambiguity. 
Therefore we believe that 
our results in this chapter  may provide reliable bases 
for further investigations of this subject.

\csection{STRING THEORY IN THE THREE DIMENSIONAL BLACK HOLE GEOMETRY}
Now we begin a new approach to 
the three dimensional quantum black holes. 
In this chapter, we will discuss the 
string theory in this black hole geometry in the framework of conformal field 
theory \cite{NS}.  One of the motivations 
is to consider the open problems in the previous chapter and understand 
the microscopic origin of black hole entropy. 
However, our main purpose here is more general and
 it is to pursue fundamental problems of quantum gravity
as discussed in the introduction. 
Since the string theory is regarded as the fundamental theory 
including gravity, we expect to get a deeper understanding of quantum aspects 
of the three dimensional black hole. 
From the string-theory point of view, 
this work also has a significance as the first attempt 
to quantize a string  in a black hole
background with an infinite number of propagating modes.
Moreover, the analysis here serves as
an investigation of a string theory in a non-trivial background.
%
\parsmallskip
After the BTZ black hole was found, it was soon realized that 
a slight modification of the solution yields a solution (an exact background)
of the bosonic string theory \cite{HW,Kaloper}. 
This is one of the 
few known exact solutions in string theory and one of the simplest
solutions. However, any detailed analyses of this  string theory
 had not been made.
In addition, a string in a curved spacetime 
has not been well understood and 
it is not clear whether a sting in a black hole background can 
be physically sensible. Therefore we will make detailed analyses 
and study consistency conditions of the string in the three dimensional 
black hole geometry \cite{NS}.
It turns out that we can investigate 
this model in an explicit manner owing to its simplicity. 
We analyze the spectrum by solving the level matching condition and obtain
winding modes. We then study the ghost problem and show explicit
examples of physical states with negative norms. We discuss general 
properties of the tachyon
propagation
and the target-space geometry which are irrelevant to the details of the
spectrum. We find a self-dual T-duality
transformation reversing the black hole mass.
The existence of the ghosts indicates that our model is not physical as it is.  
Thus we also discuss possibilities to obtain a sensible string theory.
%
%
\vskip 3ex
\csubsection{The three dimensional black hole as a string background}
First, we review the three dimensional black hole from the string-theory
point of view. In the context of string theory, 
the three dimensional black hole  is 
described by  an orbifold of the $ \widetilde{SL}(2, R) $
WZW model \cite{HW,Kaloper}.
%
\isubsubsection{Description using the $ \cSLtwo $ WZW model}
In string theory, 
we start from the $ \SLtwo $ WZW model
\footnote{There are difficulties to construct a CFT based on a
non-compact
group manifold. In this thesis, we will simply assume the existence of the
$ \cSLtwo $ WZW model.} with action
\eqabegin
&& \frac{k}{8\pi} \int_{\Sigma} d^2 \sigma \sqrt{h} h^{\alpha \beta}
      \Tr \left( \ginv \del_\alpha g \ginv \del_\beta g  \right)
    + i k \Gamma (g) \comma \nn 
\eqaend
where $ h_{\alpha \beta } $ is the metric of a Riemann surface $ \Sigma $, 
$ g $ is an element of $ \SLtwo $ and $ k $ is the level of the WZW model.
$ \Gamma $ is the Wess-Zumino term given by
\eqabegin
   && \frac{1}{12\pi} \int_{B_\Sigma}
      \Tr \left( \ginv d g \wedge \ginv d g \wedge\ginv d g \right)
   \comma \nn 
\eqaend
where $ B_\Sigma $ is a three manifold with boundary $ \Sigma $.
We parametrize $ g $ by
\eqabegin
  &&g = \matrixii{X_1 + X_2}{X_3 + X_0}{X_3 - X_0}{X_1 - X_2}
     \comma \nn \\
  && \nn \\
  && {\rm det} \ g  = X_0^2 + X_1^2 - X_2^2 - X_3^2 = 1
      \period \nn 
\eqaend
The latter equation is nothing but the embedding equation
of $ AdS_3 $ in a flat space.
\footnote{Note that $ X_i $ are dimensionless.}
Thus
$ \SLtwo $ and $ AdS_3 $ are the same manifold. This is the essence of 
the reason why the BTZ black hole is described by using the
$ \SLtwo $ WZW model. By setting $ X_i = x_i/l $, we obtain 
the direct correspondence to chapter 2.
\parsmallskip
As before, 
in order to decompactify the time direction of $ \SLtwo $,
we go to the universal covering group $ \cSLtwo $, 
and consider three regions parametrized by (\ref{xtohat}).
Furthermore, we make the change of variables (\ref{tphi}), identify 
$ \varphi $ with $ \varphi + 2 \pi $, and cut out the region $ r^2 < 0 $.
Consequently, the WZW action takes the form
\eqabegin
  S &=& \frac{1}{4\pi \alp} \int d^2 \sigma \sqrt{h}
     \left( h^{\alpha \beta} G_{\mu \nu}
       + i \epsilon^{\alpha \beta} B_{\mu \nu} \right)
          \del_\alpha X^\mu \del_\beta X^\nu \comma \nn
\eqaend
where $ ( 2 \pi \alp )^{-1} $ is the string tension. 
$ G_{\mu \nu} $ and $ B_{\mu \nu} $
are given by 
\eqabegin
 ds^2_{\rm string}  &=& G_{\mu \nu} d X^\mu d X^\nu
    \ = \ \frac{\alp k}{l^2} d \sBH ^2
    \nn \\
  B &=&   \frac{\alp k}{l^2}  r^2 d \varphi \wedge d (t/l)
   \period \label{GB}
\eqaend
$ B  $ is defined up to exact forms.
Comparing the above metric with $ \gBHu _{\mu \nu} $ 
in (\ref{BTZbh}), we find that
$ G_{\mu \nu} $ represents the three dimensional black hole 
with  $ l^2 = \alp k $.
\parsmallskip
Since the model is described by a WZW model, the background geometry 
maintains the conformal invariance of the world sheet to all orders in 
$ \alp $; the geometry 
gives an exact background (a solution to string theory).
The exact geometry, which is read off from the full quantum effective action, 
is given simply by the replacement $ k \to k - 2 $ \cite{BST}. 
Here $ -2 $ is the second Casimir of the
adjoint representation of $ \sltwo $. Then one has  
\eqabegin
   l^2 &=& (k-2) \alp  \period \label{alpkl}
\eqaend

One can confirm that the above geometry is a solution to the 
low energy effective theory \cite{HW}.
In three dimensions, the low energy string action is 
\eqabegin
  S_{LEET} &=& \int d^3 x \sqrt{-G} \ e^{-2\phi} 
     \left[  \frac{2(26 - D)}{3 \alp} + R + 4 (\nabla \phi)^2 
     - \frac{1}{12} H_{\mu \nu \rho} H^{\mu \nu \rho}
     \right]  \comma \nn 
\eqaend 
where $ \phi $ is the dilaton, $ H = d B $,  and $ D = 3 $. 
The equations of motion derived from this action are
\eqabegin
  &&  R_{\mu \nu} +  2 \nabla_\mu \nabla_\nu \phi 
   - \frac{1}{4} H_{\mu \rho \sigma} H_{\nu}^{ \  \rho \sigma} = 0 \nn \\
   && \nabla^\mu \lb  e^{-2 \phi} H_{\mu \nu \rho } \rb  = 0 \nn \\
  && 4 \nabla^2 \phi - 4 (\nabla \phi)^2 + \frac{2(26 - D)}{3 \alp} + R
   - \frac{1}{12} H^2 = 0 \period \nn 
\eqaend
A special property of three dimensions is that $ H_{\mu \nu \rho } $ must be
proportional to the volume form $ \epsilon_{\mu \nu \rho} $. 
Then, by setting $ \phi = 0 $, the second equation yields
$ H_{\mu \nu \rho } = (2/l) \epsilon_{\mu \nu \rho} $.
Substituting this into  
the first equation gives
\eqabegin
   R_{\mu \nu } &=& - \frac{2}{l^2} G_{\mu \nu} \period \nn
\eqaend
This is exactly the Einstein's equation with a negative cosmological constant
$  - l^{-2} $. The third equation is satisfied if 
\eqabegin
 &&  l^2 = \frac{6 \alp}{23} \period \label{alpl}
\eqaend
Therefore every solution to three dimensional general relativity with 
negative cosmological 
constant is a solution to low energy string theory with $ \phi = 0 $, 
$ H_{\mu \nu \rho } = (2/l) \epsilon_{\mu \nu \rho} $, and 
$ l^2 = 6 \alp/23 $. In particular, the geometry (\ref{GB}) is a 
solution with 
\eqabegin
  && B_{\varphi t} =  \frac{r^2}{l} \comma \quad \phi = 0 \period \nn 
\eqaend
The level $ k $ is determined by (\ref{alpkl}) and (\ref{alpl}):
\eqabegin
   && k = \frac{52}{23} \period \label{kcr}
\eqaend

In the following, we will set $ l = 1 $ for brevity.
It can be recovered simply by counting of dimension.
\isubsubsection{Chiral currents and the stress tensor}
Next, let us summarize the properties of currents for later use.
The $ \cSLtwo $ WZW model has a chiral $\cSLtwo_L \times \cSLtwo_R$
symmetry. 
The currents associated with this symmetry are given by
\eqabegin
   J(z) &=& \frac{i k}{2} \del g  g^{-1} \comma \quad
   \Jbar (\zbar) \ = \ \frac{i k}{2} g^{-1} \delbar g
  \comma  \label{currents}
\eqaend
where $ z = \ e^{\tau + i \sigma } $ and $ \zbar = \ e^{\tau - i \sigma} $.
The currents act on $ g $ as
\eqabegin
  J^a(z) g(w,\wbar) & \sim & \frac{- \tau^a g }{ z - w } \comma \quad
  \Jbar^a(\zbar) g(w,\wbar) \sim \frac{- g \ \tau^a  }{ \zbar - \wbar }
  \period \label{Jg} 
\eqaend
Here, we have defined $ J^a $ $ (a = 0,1,2) $ by
$ J(z) = \eta_{ab} \tau^a J^b(z)$ and similarly for $ \Jbar ^a$,
where $ \eta_{ab} = $ diag $ (-1, 1, 1)$.
$ \tau^a $ form a basis of $ \sltwo $ with
the properties
\eqabegin
  && \left[ \tau^a \comma \tau^b \right] = i \epsilon^{a b}_{\ \ c} \ \tau^c
   \comma    \quad
 \Tr \left( \tau^a  \tau^b \right) = - \half \ \eta^{ab}. \nn 
\eqaend
In terms of the Pauli matrices,
$\tau^0 = - \sigma^2/2 , \tau^1 = i \sigma^1/2$ and $\tau^2 = i \sigma^3/2$.
From the currents, one obtains the stress tensor 
\eqabegin
   T(z) &=& \frac{1}{k-2} \eta_{ab}J^a(z)J^b(z) \nn 
  \period
\eqaend
Then the conformal modes of the currents and the stress tensor satisfy
the commutation relations
\eqabegin
   \left[ J^a_n \comma J^b_m \right]
    &=&  i  \epsilon^{ab}_{\ \ c} J^{c}_{n+m} + \frac{k}{2} n
       \eta^{ab} \delta_{m + n} \comma \nn \\
  \left[ L_n \comma J_m^a \right] &=& - m J^a_{n+m} \comma \nn \\
  \left[ L_n \comma L_m \right] &=& (n-m) L_{n+m} + \frac{c}{12}n(n^2-1)
   \delta_{n+m} \comma \nn
\eqaend
where $ c = 3k/(k-2) $. For the critical value $ c = 26 $, we have
$ k = 52/23 $. This is consistent with (\ref{kcr}).
The above Kac-Moody algebra is expressed in the basis
$ I^\pm_n \equiv  J^1_n \pm i J^2_n $ and $ I^0_n \equiv J^0_n $ as
\eqabegin
  \left[ I^+_n , I^-_m \right] &=& -2 I^0_{n+m} +  k n \delta_{n+m}
  \comma \quad \left[ I^\pm_n , I^\pm_m \right] \ = \ 0 \comma \nn \\
  \left[ I^0_n , I^\pm_m \right] &=& \pm I^\pm_{n+m} \comma
  \qquad \qquad \qquad
   \left[ I^0_n , I^0_m \right] \ = \ - \frac{k}{2} n \delta_{n+m}
   \period \nn 
\eqaend
On the other hand,
in the basis $ J^\pm_n \equiv J^0_n \pm J^1_n  $ and
 $ J^2_n $, the algebra is written as
\eqabegin
  \left[ J^+_n , J^-_m \right] &=& -2 i J^2_{n+m} - k n \delta_{n+m}
  \comma \quad \left[ J^\pm_n , J^\pm_m \right] \ = \ 0 \comma \nn \\
  \left[ J^2_n , J^\pm_m \right] &=& \pm i  J^\pm_{n+m} \comma
  \qquad \qquad \qquad
   \left[ J^2_n , J^2_m \right] \ = \  \frac{k}{2} n \delta_{n+m}
   \period \label{JJcom}
\eqaend
Note that the Hermitian conjugates for the latter basis are given by
\eqabegin
  && \left( J^\pm_m \right)^{\dag}  = J^\pm_{-m} \comma \quad
     \left( J^2_m \right)^{\dag} = J^2_{-m} \period \label{dagJ}
\eqaend

Similar expressions hold for the anti-holomorphic part.
\isubsubsection{Twisting}
As explained before,
in order to get the
three dimensional black hole,
we have (i) to go to the universal covering space of
$ \SLtwo $,
(ii) to make the identification $ \varphi \sim \varphi + 2 \pi $ and
(iii) to drop the region $ r^2 < 0 $.
We can take (i) into account by considering the representation theory
of $ \cSLtwo $ instead of $ \SLtwo $.
The point (iii) was related to
the problem of closed timelike curves (see chapter 2 and \cite{BTZ,HW});
we will discuss this point in section 4.5.
In terms of string theory, (ii) represents a twist of $ \cSLtwo $, and 
we will concentrate on (ii) for now.
\parsmallskip
In order to express the identification in (ii) by the $ \sltwo $
currents, it is convenient to parametrize the group manifold by
analogs of Euler angles; we parametrize Region I-III by \cite{NS,DVV}
\eqabegin
\begin{array}{lllllll}
   \mbox{ Region I}
     & \! : \!&
   g
   &\! = \!&  e^{- i \thetaL \tau^2} e^{- i \rho \tau^1}
           e^{-i \thetaR \tau^2}
    &\! = \! &
  \matrixii{ e^{\hatphi} \cosh \rho/2}{ e^{\hatt} \sinh \rho/2}
         { e^{-\hatt} \sinh \rho/2}{ e^{-\hatphi} \cosh \rho/2}
     \comma  \\
  \mbox{ Region II}
   &\! : \! &
  g &\! =\! &  e^{-i \thetaL \tau^2} e^{-i \rho \tau^0}
       e^{-i \thetaR \tau^2}
    & \! = \!&
  \matrixii{ e^{\hatphi} \cos \rho/2}{ e^{\hatt} \sin \rho/2}
         { - e^{-\hatt} \sin \rho/2}{ e^{-\hatphi} \cos \rho/2}
    \comma   \label{gmat} \\
   \mbox{ Region III}
   & \! : \! & g
    & \! = \!
   &  e^{-i \thetaL \tau^2} \ s \ e^{-i\rho \tau^1} e^{-i \thetaR \tau^2}
    & \! = \! &
\matrixii{ e^{\hatphi} \sinh \rho/2}{ e^{\hatt} \cosh \rho/2}
  { - e^{-\hatt} \cosh \rho/2}{ - e^{-\hatphi} \sinh \rho/2}
     \comma
    \end{array}
\eqaend
where $ s = \matrixii{0}{1}{-1}{0} $,
\eqabegin
  \thetaL &=& \hatphi + \hatt \comma \quad \thetaR \ = \ \hatphi - \hatt
   \comma \label{thetaLR}
\eqaend
and
\eqabegin
  \begin{array}{lllll}
   \mbox{ Region I} &:&
     \hatr = \cosh \rho/2 \comma &
   \sqrt{ \hatr ^2 - 1} = \sinh \rho/2 \comma
    & ( \rho > 0 ) \comma \\
 \mbox{ Region II} &:&
   \hatr  = \cos \rho/2 \comma &
  \sqrt{1 - \hatr ^2 } = \sin \rho/2 \comma
    & ( \pi > \rho > 0 ) \comma \label{rhor} \\
    \mbox{ Region III} &:&
    \sqrt{- \hatr ^2 } = \sinh \rho/2 \comma &
   \sqrt{1 - \hatr ^2 } = \cosh \rho/2 \comma
    & ( \rho > 0 ) \period
  \end{array} \nn 
\eqaend
The currents (\ref{currents}) then take the form, e.g.,
\eqabegin
    J^2 &=&
     \frac{k}{2} \left( \del \thetaL + (2\hatr^2 -1) \del \thetaR \right)
   \comma \qquad
  \Jbar^2  \ = \
    \frac{k}{2}
   \left( \delbar \theta_R + (2\hatr^2 -1) \delbar \theta_L \right)
  \period \label{J2}
\eqaend

From (\ref{tphi}), we find that 
the translation of $ \varphi $ is given by a linear combination of
those of $ \hatt $ and $ \hatphi $.
From the $ \AdS $ point of view, the translations of
$ \hatt $ and $ \hatphi $ corresponded to boosts in the flat spacetime
in which $ \AdS $ was embedded. 
On the other hand, in the context of the
$ \SLtwo $ WZW model, the translations of $ \hatt $ and $ \hatphi $
correspond to a vector and an axial symmetry generated by 
$ J^2_0 \pm \Jbar^2_0 $ \cite{HW} (see (\ref{Jg})).
Therefore the
translation of $ \varphi $ is generated by
$ Q_{\varphi} \equiv \deltam J_0^2 + \deltap \Jbar^2_0 $,
where 
$$
\Delta_\pm = \rp \pm \rmi \period 
$$
Then $ \delta \varphi = 2 \pi $ with fixed $ t $ is expressed by 
\eqabegin
   && \deltap \delta \thetaL = \deltam \delta \thetaR = 2 \pi \deltap \deltam
   \period \label{deltheta} 
\eqaend
For describing the black hole,
we have to twist (orbifold) the WZW model
with respect to this discrete group.  In the following, 
we denote it by $ \Zphi $ and hence 
we will call our black hole the $ \cSLtwo / \Zphi $ black hole.
\parsmallskip
Note that, if one gauges the vector or the axial symmetry, 
the resulting coset theory describes
the $ \SLtwo / U(1) $ black hole \cite{Witten}.
%
%
\csubsection{The spectrum of a string on $ \cSLtwo/ \Zphi $ orbifold}
As a consequence of the identification
$ \varphi \sim \varphi + 2 \pi $,
twisted (winding) sectors arise in the theory.
In this section, we will discuss the spectrum including
the twisted (winding) sectors \cite{NS}.
One difficulty here is that the field
$ \varphi $ is not a free field.
We are working in a group manifold, so we cannot use
the argument for flat theories.
However, a similar orbifolding has been discussed
in \cite{GPS} to construct a $ SU(2)/ Z _N $ orbifold. Thus we will follow
that argument and  solve the level matching
condition; this is required from
various kinds of consistency of string theory, for example, 
modular invariance and the invariance under
the shift of the world-sheet spatial coordinate.
Other consistency conditions such as unitarity should also be checked.
We will discuss them in section 4.3 and 4.5.
These consistency conditions are closely related
to each other.  
\parsmallskip
%
%
\isubsubsection{Kac-Moody Primaries in the $ \cSLtwo $ WZW model}
Before discussing the orbifolding, let us consider Kac-Moody
primaries in the $ \cSLtwo $ WZW model. Operators are
Kac-Moody primary if they
form irreducible representations of global
$ \cSLtwo _L $ $ \times \cSLtwo _R $ and if they are annihilated by
the Kac-Moody generators $ J^a_n $ and $ \Jbar^a_n $ for $ n > 0 $.
For WZW models, they are also Virasoro primary.
For a unitary theory based on a compact group, local fields
(wave functions) on the group correspond to Kac-Moody primaries 
and they are given by the matrix elements of
the unitary representations of the group  \cite{GPS,GW}.
We are also interested in a unitary string theory.  
Thus, we start from the Kac-Moody primary fields which correspond to
the matrix elements of 
the unitary representations of $ \cSLtwo $. They have 
local expressions in $ \theta_L , \theta_R $
and $ \rho $ without derivatives of these fields,  
\eqabegin
  && V \lb \theta_L (z, \zbar) , \theta_R (z, \zbar) ,
       \rho (z , \zbar ) \rb
  \period \nn 
\eqaend

\parsmallskip
For $ \cSLtwo $,  we have five types of unitary representations 
\cite{VK}, namely, the identity representation,   
the principal continuous series, the complementary
series, the highest and the lowest discrete series. 
In order to express the matrix elements of these representations, 
we have to further specify the basis of the representation.
In representations of $ \cSLtwo $, one has three types of basis.
Let us denote the generators of $ \sltwo $ by $ J^0 $, $ J^1 $ and $ J^2 $.
Then, the bases diagonalizing $ J^0 $, $ J^2 $ and $ J^0 - J^1 $ are
called elliptic, hyperbolic and parabolic, respectively.
Since we are interested in the orbifolding related to the action
of $ J_0^2 $ and $ \Jbar_0^2 $, we consider representations
in the hyperbolic basis. 
This basis has been used in the
study of the Minkowskian $ \SLtwo /U(1) $ black hole \cite{DVV,DN}.
Consequently, the Kac-Moody primaries 
other than the identity are 
expressed by the matrix elements as
\eqabegin
 & \D{P(C)} _{J \pm ,J' \pm}^{\chi} \lb g \rb  &
     \mbox{ for the principal continuous $(P)$ and the complementary 
          $(C)$series} \comma   \nn  \\
  &\D{H(L)} _{J,J'}^{j} \lb g \rb &
     \mbox{ for the highest $(H)$ and the lowest $ (L)$ weight series} \comma 
   \label{matrix} 
\eqaend
where $ j $ labels the value of the Casimir; $ J $ and $ J' $ refer to
the eigenvalue of $ J^2 $. For the principal continuous and the complementary 
series,
one has additional parameters, $ 0 \leq  m_0  < 1 $
specifying the representation,
and $ \pm$ specifying the base state. $ \chi $ is the pair $ (j,m_0) $.
Under this construction, the primary fields have the common $j$-value in the
left and the right sector.
Note that the spectrum of $ J^2 $
ranges all over the real number, i.e., 
$ J , J' \in \bfR $. For the details, see appendix C.
\parsmallskip
Here some remarks may be in order. First, one can explicitly construct 
the primary fields belonging
to the unitary representations of $ \cSLtwo $ using free field realizations
of the $ \sltwo $ Kac-Moody algebra \cite{GH}. 
Second, in section 4.4, we will find the correspondence 
between the above primary fields and the 
Klein-Gordon fields discussed in chapter 3.
Third, suppose \cite{DVV} that 
 the Kac-Moody primary fields lead to normalizable operators, and that
the CFT inherits the natural inner product of the $ \cSLtwo $ representations.
Then the Kac-Moody primaries should be given by the matrix elements of 
the unitary representations (except for the complementary series)
because a complete basis
for the square integrable functions on $ \cSLtwo $ is given by them.
Fourth, most of our discussions below do not change
even if we start from other representations at the base. 
We easily find that the theory becomes
non-unitary if we start from non-unitary ones as in \cite{DN}. 
Finally, we cannot deny the possibility
of the Kac-Moody primaries which are non-local and/or contain 
derivatives of the coordinate fields. 
However, in our understanding, such a possibility has not been found so far.
\parsmallskip
\isubsubsection{Vertex operators in the $ \cSLtwo / \Zphi $ theory}
We now turn to the $ \cSLtwo / \Zphi $ theory and consider the vertex operators
\cite{NS}.
First, we construct the operator which expresses the twisting.
Let us recall that the chiral currents $ J^2(z) $ 
and $ \Jbar ^2 (\zbar) $ have the
operator product expansions (OPE)
\eqabegin
 && J^2(z) J^2(0) \sim \frac{k/2}{z^2} \comma \quad
  \Jbar^2(\zbar) \Jbar^2(0) \sim \frac{k/2}{\zbar ^2} \period \nn 
\eqaend
So, we represent them
by free fields $ \theta_L^F(z) $ and $ \theta_R^F(\zbar) $ as
\eqabegin
  && J^2(z) = \frac{ k}{2} \del \theta_L^F \comma \qquad
     \Jbar^2(\zbar) = \frac{k}{2} \delbar \theta_R^F \period \nn 
\eqaend
The normalization of the fields is fixed by
\eqabegin
   \theta_L^F (z) \theta_L^F (0) \sim + \frac{2}{k} \ln z \comma &&
   \theta_R^F (\zbar) \theta_R^F (0) \sim + \frac{2}{k} \ln \zbar
  \period \nn
\eqaend
The signs are opposite to the usual case because of the negative metric of
the $ J^2 $ direction.
The explicit forms of $ \theta_L^F $ and $ \theta_R^F $ are obtained by
integration of (\ref{J2}).
The local integrability is assured by the current conservation.
In addition, we introduce
$ \theta_L^{NF} (z,\zbar) $ and $ \theta_R^{NF} (z,\zbar) $ by
\eqabegin
  \theta_L (z, \zbar) = \theta_L^F (z) + \theta_L^{NF}(z,\zbar) \comma \qquad
  \theta_R (z, \zbar) = \theta_R^F (\zbar) + \theta_R^{NF}(z,\zbar)
 \period \nn
\eqaend
Note $\theta_L^{NF}$ and $\theta_R^{NF}$ are not free fields.
%
%
Then, the twisting operator with winding number 
$ \nw \in \bfZ $ is 
given by 
\eqabegin
  W (z,\zbar;\nw) & \equiv & \exp \lmb - i\frac{k}{2} \nw
       \lb \deltam \theta_L^F - \deltap \theta_R^F \rb \rmb
  \period \nn
\eqaend
Indeed, this has the OPE's
\eqabegin
  \theta_L^F(z) W (0,\zbar;\nw)
 & \sim & - i \nw \deltam \ln z \cdot W (0,\zbar;\nw)
  \comma \nn \\
 \theta_R^F(\zbar) W (z,0;\nw)
 & \sim &  + i \nw \deltap \ln \zbar \cdot W (z,0;\nw)
  \period \nn 
\eqaend
Thus, $ \theta_L^F $ and $ \theta_R^F $ shift
by $ 2 \pi \deltam n $ and
$2 \pi \deltap n $, respectively,
under the translation of the world-sheet coordinate
$\sigma \to \sigma + 2 \pi $, i.e.,
$z \to e^{2\pi i } z $ and $ \zbar \to e^{ - 2\pi i } \zbar $. Namely,
$ \delta \varphi = 2 \pi \nw  $ and $\delta t = 0 $
on $ W (z,\zbar;\nw) $ under
$ \delta \sigma = 2 \pi $. Hence, $ W (z,\zbar;\nw) $
expresses the correct twisting.
\parsmallskip
Then we readily obtain the primary fields in our model.
First, a general untwisted primary field takes the form (\ref{matrix}).
In our parametrization (\ref{gmat}), it is given by
\eqabegin
 V^{j}_{J_L,J_R}(z,\zbar;0)
 & = & D_{J_L,J_R}^j \lb g'(\rho) \rb \ e^{-i J_L \theta_L - i J_R \theta_R }
  \comma \label{V0}
\eqaend
where we have omitted irrelevant indices of the matrix elements.
The explicit form of $ g'(\rho) $ depends on which region we consider.
Second, combining the untwisted primary field and the twisting operator $W$,
we obtain a general primary field in the $ \cSLtwo / \Zphi $
black hole CFT \cite{NS}:
\eqabegin
  V^{j}_{J_L,J_R}(z,\zbar;\nw) &=& V^{j}_{J_L,J_R}(z,\zbar;0)
      W (z,\zbar;\nw)  \comma  \label{Vn} \\
 &=& D_{J_L,J_R}^j \lb g'(\rho) \rb
   \ \exp \lmb - i \lb J'_L \theta_L^F + J_L \theta_L^{NF} +
   J'_R \theta_R^F + J_R \theta_R^{NF} \rb \rmb  \comma \nn
\eqaend
where
\eqabegin
  J'_L &=& J_L + \frac{k}{2}\deltam \nw  \comma \quad
  J'_R = J_R - \frac{k}{2}\deltap \nw  \period \nn 
\eqaend
\parsmallskip
From this primary field, we find that a general vertex operator
takes the form
\eqabegin
  && J_N \cdot \Jbar_{\Nbar} \cdot  V^{j}_{J_L,J_R}(z,\zbar;\nw)
  \comma \nn 
\eqaend
where $ J_N $ and $ \Jbar_{\Nbar} $ stand for generic products
of the Kac-Moody generators $ J^a_{-n} $ and $ \Jbar^a_{-n} $, respectively.
Here we have a restriction on the above form because of the orbifolding.
Note that 
the untwisted part depends on $ \theta_L^F $ and $ \theta_R^F $ as
$ \exp ( -i \omega_L \theta_L^F -i \omega_R \theta_R^F ) $
and the full operator as
$ \exp ( -i \omega'_L \theta_L^F - i \omega'_R \theta_R^F ) $, where
\eqabegin
   \omega^{(')}_L &=& J^{(')}_L + i ( N_+ - N_- ) \comma \quad
   \omega^{(')}_R \ = \ J^{(')}_R + i ( \Nbar _+ - \Nbar _- ) \ ;
  \nn 
\eqaend 
$ N_\pm $ and $ \Nbar _\pm $ are the number of $ J^\pm_{-n} $ and
$ \Jbar ^\pm_{-n} $, respectively. This follows from the fact that 
 $ J^\pm_{-n} (\Jbar^\pm_{-n}) $ shifts $ \omega_L (\omega_R) $
by $ \pm i $ because of  the commutation relation (\ref{JJcom}).\footnote{
      This seems to contradict
     the Hermiticity of $ J^2_0 (\Jbar ^2_0)$.
      However, this is not the case because
     the spectrum of $ J^2_0 (\Jbar ^2_0)$
     is continuous. Representations of $ \SLtwo $
      in the hyperbolic basis are described in appendix C.}
The vertex operator cannot be
single-valued on $ \cSLtwo / \Zphi $ orbifold if 
$ \omega^{(')}_{L,R} $ are complex. Thus
$ N_+ = N_- $ and
$ \Nbar _+ = \Nbar _- $, namely, $ \omega^{(')}_{L,R} = J^{(')}_{L,R}$ should 
hold. Consequently, the vertex operators in our model are given by 
\eqabegin
   && K_{-n}^{a} \cdots \tilde{K}^b_{-m} \cdots  V^{j}_{J_L,J_R}(z,\zbar;\nw)
   \comma \label{vertex}
\eqaend
where $ K^a_{-n} $ and $ \tilde{K}^a_{-n} $ $\ (a = +,-,2) $ are defined by
\eqabegin
  && K^+_{-n} = J^+_{-n} J^-_{0} \comma \qquad K^-_{-n} = J^-_{-n} J^+_{0}
   \comma \qquad K^2_{-n} = J^2_{-n}  \comma \nn 
\eqaend
and similar expressions for $ \tilde{K}^a_{-n} $.
%
%
%
\isubsubsection{Level matching}
Now we consider the level matching condition to further
discuss the spectrum.
To obtain the expressions of $ L_0 $ and $ \Lbar _0 $ for the vertex operators,
we first decompose the stress tensor following 
GKO (Goddard-Kent-Olive) \cite{GKO}. For the holomorphic 
part, we then have
\eqabegin
   T (z) &=& T^{\sltwo/ so(1,1)} (\rho, \theta_L^{NF}, \theta_R^{NF})
    + T^{so(1,1)} (\theta_L^F) \comma \nn \\
  && T^{so(1,1)} (\theta_L^F) = + \frac{k}{4} \del \theta_L^F
\del \theta_L^F,
  \quad  T^{\sltwo/ so(1,1)} = T  - T^{so(1,1)} \period \nn 
\eqaend
Since $ T^{so(1,1)} $ acts only on $ \theta_L^F $, the weight
with respect to $ T^{so(1,1)} $ is given by
$ \Delta^{so(1,1)} (J'_L) \equiv - J_L^{'2}/k  \  +
$ (the grade of $ J^2_{-n}$'s ). Moreover, for the untwisted sector, 
$ L_0 $ is given by the Casimir
plus the total grade;
\eqabegin
  \Delta^{\sltwo/so(1,1)} (j,J_L) + \Delta^{so(1,1)} (J_L)
   &=& - \frac{j(j+1)}{k-2} + N \comma \nn 
\eqaend
where $ -j(j+1) $ is the Casimir and $ N $ is the total grade of $ J^a_{-n} $'s.
Therefore, we find a general expression of $ L_0 $;
\eqabegin
  L_0 &=&  \Delta^{\sltwo/so(1,1)} (j,J_L)
   + \Delta^{so(1,1)} (J'_L)  \nn \\
 &=&  \frac{-j(j+1)}{k-2} + \frac{J_L^2 - J_L^{'2}}{k} + N
 \comma \nn
\eqaend
where $ \Delta^{\sltwo/so(1,1)} $ is
the weight with respect to  $ T^{\sltwo/ so(1,1)} $.
Similarly, we obtain 
\eqabegin
  \Lbar _0 &=& \frac{-j(j+1)}{k-2} + \frac{J_R^2 - J_R^{'2}}{k}
   + \Nbar \period \nn 
\eqaend

Using these expressions, the level matching condition
is given by
\eqabegin
   L_0 - \Lbar _0 &=&
  -   \nw \left[ \left( \deltam J_L + \deltap J_R \right)
    - \frac{k}{2} \nw  \JBH   \right]  + N - \Nbar
   \ \in \ {\bf Z} \period \label{LM} 
\eqaend
To proceed, let us consider the OPE
of two vertex operators with quantum numbers 
$ ( \nwn{i} , J_{L,i} , J_{R,i} ) $ $ (i = 1,2) $.
Since $ J_{L,R} $ and
$ \nw $ are conserved, the level matching condition for the resulting
operator reads
\eqabegin
 &&
- ( \nwn{1} + \nwn{2} ) \ \sum_{i=1}^2 \
  \left[ \left(  \deltam J_{L,i} + \deltap J_{R,i} \right)  -
   \frac{k}{2}  \nwn{i} \JBH   \right]
 \ \in \ {\bf Z} \period \nn 
\eqaend
Thus, if $ J_{L(R),1(2)} $ and $ n_{W,1(2)} $ satisfy (\ref{LM}),
the closure of the OPE requires \cite{NS}
\eqabegin
&& \left(  \deltam J_L + \deltap J_R \right)
   - \frac{k}{2} \nw \JBH \ \equiv \mj 
 \ \in \ {\bf Z}  \period \label{solLM} 
\eqaend
This is the solution to the level matching condition.
The spectrum of the theory is specified by this condition.
\parsmallskip
We can check the single-valuedness of the vertex operator which satisfies
this condition. Let us denote by $ \exp \left( - i \Theta \right)$ the
$ \theta_{L,R}^{F,NF} $-dependence of (\ref{Vn})
 and recall
(\ref{deltheta}).
Then, under $\delta \varphi = 2 \pi $,
\eqabegin
 \delta \Theta
  & = & 2 \pi \mj + \frac{k}{2} \pi \nw \left[ \frac{1}{\pi}
    \left( \deltam \delta \theta_L^F - \deltap \delta \theta_R^F  \right)
   + \lb \deltap ^2 - \deltam ^2 \rb  \right]
  \period \nn 
\eqaend
Hence, the vertex operator is invariant
under
\eqabegin
 && \delta \theta_L^{NF} = \delta \theta_L^F = \pi \deltam \comma \quad
     \delta \theta_R^{NF} = \delta \theta_R^F = \pi \deltap \period \nn 
\eqaend
Single-valuedness is guaranteed in this sense.
\parsmallskip
In our twisting, only the free field part seems relevant.
In the untwisted sector, only the combinations
$ \theta_{L,R} = \theta_{L,R}^F + \theta_{L,R}^{NF} $ appear, so this
does not matter. On the other hand, for a twisted sector, this is
curious; we were originally considering the orbifolding
with respect to
$ \varphi \sim \varphi + 2 \pi  $ including the non-free part.
However, the non-free part {\it is } relevant in the
above sense. This is related to the Noether current ambiguity
in field theory \cite{GPS}.
In any case, one can take the point of view that we are just considering
possible degrees of freedom represented by the twisting
with respect to $ \theta_{L,R}^F $.
\parsmallskip
So far we have dealt with a generic value of $ \Delta_\pm $ corresponding
to a rotating black hole. For the non-rotating black hole,
we have only to set $ \deltap = \deltam  = \rp $ in the above discussion.
In addition, we can formally take the limit $ \deltam \to 0 $ at the
end. However, we have to
examine whether this limit in our result correctly represents
the extremal limit as discussed before.
\isubsubsection{Physical states}
Let us turn to the discussion on physical states \cite{NS}. 
We use the old covariant approach. The states
corresponding to the vertex operators in (\ref{vertex}) are
written as
\eqabegin
    &&  K^a _{-1} K^b_{-1} \cdots  \tilde{K}^c _{-1}  
    \tilde{K}^d_{-1} \cdots \ket{ j; J_L , \nw } \ket{ j; J_R , \nw } 
   \period \label{KMmodule}  
\eqaend
Here we have used the fact that the operators with higher grade are
generated by those with grade $1$.
%
%
Then
the physical states are given by  
the physical-state conditions
\eqabegin
      \lb L_n - \delta_n \rb \ket{ \Psi }
   &=& \lb \Lbar _n - \delta_n \rb \ket{ \Psi }
  \ = \ 0 \quad (n \geq 0) \period \nn 
\eqaend
In particular, the on-shell condition yields 
\eqabegin
  J_L &=& - \frac{k}{4} \deltam \nw + \frac{1}{\deltam \nw }
   \left( N - 1   - \frac{j(j+1)}{k-2} \right)  \comma \nn \\
 J_R &=& + \frac{k}{4} \deltap \nw - \frac{1}{\deltap \nw }
   \left( \Nbar - 1   - \frac{j(j+1)}{k-2} \right) \comma \label{L0n} \\
  N &=& \Nbar + \nw \mj \nn
\eqaend
for twisted sectors ($ \nw \neq 0 $),
and
\eqabegin
  1 &=& - \frac{j(j+1)}{k-2} + N \label{L04} \comma \qquad
  N \ = \  \Nbar \label{L00}
\eqaend
for the untwisted sector ($ \nw = 0 $). 
Therefore, for a given $ j $, an arbitrarily
excited state is allowed in the twisted sectors.
On the contrary, in the untwisted sector, $j$-value is
completely determined by grade $ N $:
\eqabegin
  j &=& j(N) \ \equiv
  \ \frac{1}{2} \left\{  - 1 - \sqrt{1 + 4(k-2) (N-1)} \right\}
  \comma \label{jN}
\eqaend
where we have chosen the branch Re $j \leq -1/2 $ (see appendix C).
This result is the same as in the string theory on $ \SLtwo $.
%
%
\csubsection{Investigation of unitarity}
We have discussed the spectrum
of the $ \cSLtwo / \Zphi $ model by solving the level matching condition.
But other consistency conditions remain to be discussed,
and as a result, the spectrum in the previous section
may be further restricted.
\parsmallskip
In this section, we will investigate the ghost problem.
The unitary (ghost)
problem for the string on $ \SLtwo $ has been discussed 
and it has been shown to contain ghosts \cite{BOFW},\cite{BN,Bars}.
However, there is a recent proposal for a unitary $ \SLtwo $ theory
using modified currents \cite{Bars}. Thus it may be worth 
studying our case.
Because of the orbifolding and the use of
representations in the hyperbolic basis, we cannot apply 
the argument in the $ \SLtwo $
theory to our case. 
Nevertheless, we can still utilize
a tool developed for the $ \SLtwo $ theory with a slight modification.
Here, we will first summarize the argument in the $ \SLtwo $ case.
This may also make the later discussion clear.
We then find explicit examples of negative-norm physical states;
the string theory on $ \cSLtwo / \Zphi $ orbifold is not unitary \cite{NS}.
\isubsubsection{The unitarity problem of a string on $ \SLtwo $}
Let us briefly review the unitarity problem of the $ \SLtwo $ case
\cite{BOFW},\cite{BN,Bars}.
The holomorphic and the anti-holomorphic
part are independent in the $ \SLtwo $ WZW model
until we consider the modular properties,
so we focus on the holomorphic part.
For the unitarity problem of the $ \SLtwo $ theory,
it is useful to notice the following facts:
\parbigskipn
1. \qquad The on-shell condition is
   the same as (\ref{L00}).
\parmedskipn
2. \qquad Let $ V^a $ be an operator satisfying
\eqabegin
  \lbb I_0^a , V^b \rbb &=& i \epsilon^{ab}_{\ \ c} V^c \comma \nn 
\eqaend
(an example is $ V^a = I^a_{-n}$) and consider the following states
\eqabegin
   V^+ I_0^- \ketp{j;m} \comma \qquad  V^- I_0^+ \ketp{j;m} \comma \qquad
   V^0 \ketp{j;m} \period \nn 
\eqaend
Here $ \ketp{j;m} $ are eigenstates with the Casimir 
$ \bfC = -j(j+1) $ and
$ I_0^0  = m $ (not necessarily  base states).
Moreover, assume they do not vanish. Then, by evaluating the matrix elements of
the Casimir operator, one finds that these states are decomposed into the
representations of $ \sltwo $ with the $j$-values $ j $ and $ j \pm 1 $.
\parmedskipn
3. \qquad
As a consequence of (2),
acting $ I^a_{-1 }$ $ N $ times on a base state $ \ket{j;m} $
yields $ 3^N $ independent states at grade $ N $ with $j$-values
ranging from $ j-N$ to $j+N$.
Let us call the states with $ j\pm N $
the ``extremal states" and denote them by $ \ket{E^\pm_N} $.
$ \ket{E^\pm_N} $ are physical if they satisfy the on-shell condition.
The reason is simple: Since the Casimir operator
commutes with $ L_n $, $ L_n \ket{E^\pm_N} $ have the same $j$-value as
$ \ket{E^\pm_N} $. However, $ L_n \ket{E^\pm_N} $ are at grade
$ N-n $, and thus
their $j$-values should range from $ j - (N-n) $ to $ j + (N-n) $.
Therefore,
one has $ L_n \ket{E^\pm_N} = 0 $ $\ (n >0)$;
together with the on-shell condition, they are physical.
\parmedskipn
4. \qquad Let $ \ket{\Psi} $ be a physical state.
Then the states obtained by acting
$ J^a_0 $ on $ \ket{\Psi} $ are also physical:
\eqabegin
  ( L_n -\delta_n ) J_0^a \cdots J_0^b \ket{\Psi} &=& \lbb ( L_n -\delta_n )  ,
J_0^a \cdots J_0^b \rbb \ket{\Psi}
  = 0  \quad \qquad (n \geq 0) \period \nn 
\eqaend
\parmedskipn
5. \qquad For the discrete series, one has a simple expression of the extremal
states, e.g.,
\eqabegin
  \ket{E^{d+}_N} & = & \lb I^+_{-1} \rb ^N \ket{j(N);j(N)} \comma \nn 
\eqaend
where $ \ket{j(N);j(N)} $ is a highest-weight state, namely
$ I^+_0 \ket{j(N);j(N)} = 0 $.
Then it is easy to obtain the norms of these states:
\eqabegin
    \bra{E^{d+}_N} \semiket{ E^{d+}_N} &=&  \bra{j(N);j(N)} \semiket{j(N);j(N)}
      (N !) \prod_{r=0}^{N-1} ( k + 2 j(N) + r )  \period
     \nn 
\eqaend
\parmedskip

From (1)-(5), one immediately finds physical states with negative
norms.
First, let us consider the case $ k < 2$.
{}From (1) and (3),  $ \ket{E^+_N} $ with $ j = j(N) $
at its base is a physical state.
At sufficiently large $ N $, $j(N)$
takes a value of the principal continuous series. On the other hand, the
$j$-value
of $ \ket{E^{+}_N} $ is $ j(N) + N $, but there is no unitary representations
with this $j$-value. Thus, the module
$ I_0^a \cdots I_0^b \ket{E^+_N} $ is physical, but forms a {\it non-unitary}
representation of $ \sltwo $.
\parsmallskip
Second, we consider the case $ k > 2 $. Again
$ \ket{E^{d+}_N} $
with $ j = j(N) $ at its base is a physical state.
In addition, one  finds that
\eqabegin
  && I_0^+ \ket{ E^{d+}_N }  = 0 \comma \qquad
   I^0_0 \ket{ E^{d+}_N } = \lb j(N) + N \rb \ket{ E^{d+}_N } \period \nn
\eqaend
Thus $ \ket{E^{d+}_N} $ is  a highest-weight state of
a highest-weight $ \sltwo $ representation like $ \ket{j(N) + N;j(N) +N} $.
However, the $ I^0 $-value becomes positive for large $ N $ .
Since there is no unitary representation of $ \sltwo $ with such a
highest weight state, the states in the module
$ I_0^a \cdots I_0^b \ket{E^{d+}_N} $
are physical but some have negative norms.
\parsmallskip
Although one can flip the sign of the norm of
$ \ket{ j(N);j(N) } $ so that $ \bra{ E^{d+}_N } \semiket{ E^{d+}_N } > 0 $
for arbitrary $ N $, it is impossible to
 remove physical states with negative norms.
This is because we have infinitely many physical states built on
$ \ket{ E^{d+}_N } $ as in (4),
and they form a non-unitary $ \sltwo $ representation.
\isubsubsection{Physical states up to grade 1}
Now we discuss the $ \cSLtwo / \Zphi $ orbifold case.
One difference from the previous discussion is the existence
of winding modes. Thus, for the twisted sectors,
(\ref{jN}) does not hold
and  the holomorphic and anti-holomorphic part are not independent.
Another important difference is that the Kac-Moody module
is restricted to the form (\ref{KMmodule}). We do not
have states of the type in (4) and (5) in the previous subsection.
Nevertheless, the discussion on the extremal
states is still valid,
so we will use them. 
\parsmallskip
To proceed, 
let us consider physical states up to grade one.
For the time being, we focus on the holomorphic part.
At grade one, we have three states for a fixed  $j$-, $ J^2_0 $- and
$\nw$- value;
\eqabegin
   &&
  \ket{ \pm } \equiv  K^\pm_{-1} \ket{j;\lambda,\nw } \comma \qquad
   \ket{ 2 } \equiv K^2_{-1} \ket{j;\lambda,\nw } \period  \nn 
\eqaend
From the argument in appendix B, 
these states are decomposed into the eigenstates
of the Casimir operator with $j$-values $ j $ and $ j \pm 1 $.
We denote them
by $ \ket{ \Phi^j (j;\lambda,\nw ) } $ and
$ \ket{ \Phi^{j\pm1}(j;\lambda,\nw ) } $. Note that 
$ \ket{ \Phi^{j\pm1} } $ are the extremal states.
Explicitly, they are
given by (up to normalization) \cite{NS}
\eqabegin
  \veciii{ \ket{ \Phi^{j+1} } }{ \ket{ \Phi^{j} } }{ \ket{ \Phi^{j-1} } }
  & \! \! = \! \! &
\matrixiii{j+1-i \lambda }{-(j+1 + i \lambda)}{2i \lb (j+1)^2 + \lambda^2 \rb}
{1}{1}{-2 \lambda}{-(j+i \lambda)}{j-i\lambda}{2i \lb j^2 + \lambda^2 \rb}
      \veciii{ \ket{+} }{ \ket{-} }{ \ket{2} }
   \period \nn 
\eqaend

At grade one, the conditions $ L_n = 0 \ \ ( n > 0 ) $  are reduced to
$ L_1 = 0 $.
This imposes one condition on a state given by a linear combination
of $ \ket{ \pm } $ and $ \ket{2} $.
Then the space of the solution has (complex)
two dimensions at a generic value of $ j $ and $ \lambda $.
Since we have the two
extremal states satisfying $ L_1 = 0 $, 
the solutions take the form
\eqabegin
   \alpha \ket{\Phi^{j+1}} + \beta \ket{\Phi^{j-1}}
   \period \label{phys}
\eqaend
At special values of $ \lambda $ and $ j $, we have extra solutions.
Similarly, we can get the states  satisfying 
$ \Lbar_1 = 0 $ at grade one. Hence, 
from the states of the type (\ref{phys})  and
base states, we obtain  the physical states up to grade one
by tensoring the holomorphic and the anti-holomorphic
sector so that they satisfy the on-shell condition
(\ref{L0n}) or (\ref{L00}).
\isubsubsection{Non-unitarity of a string on $ \cSLtwo / \Zphi $ orbifold}
Using the above physical states, we readily find
physical states with negative norms \cite{NS}.
First, let us discuss the case of real $ j $
(the complementary and the discrete series).
In this case, we have the following physical states, 
\eqabegin
 && \ket{ \Psi^d_1 } = \ket{ j;J_{L,1},1 }  \ket{j;J_{R,1},1}  \comma \quad
   \ket{ \Psi^{d}_2 } = \ket{ \Phi^{j+1} (j;J_{L,2},1) }  \ket{ j;J_{R,2},1 }
 \comma \nn
\eqaend
where $ m_{J,1} = 0 $, $ m_{J,2} = 1 $
and
\eqabegin
   \begin{array}{ll}
    {\displaystyle J_{L,1} =  - \frac{k}{4} \deltam  - \frac{1}{\deltam }
   \left( 1   + \frac{j(j+1)}{k-2} \right) \comma }  &
    {\displaystyle J_{R,1}  =    \frac{k}{4} \deltap  + \frac{1}{\deltap  }
   \left( 1   + \frac{j(j+1)}{k-2} \right) \comma } \\
  {\displaystyle J_{L,2} =   - \frac{k}{4} \deltam  - \frac{1}{\deltam }
    \frac{j(j+1)}{k-2}  \comma } &
  {\displaystyle J_{R,2} \ = \ J_{R,1} }  \period 
 \end{array} \label{JLR} 
\eqaend
Taking into account the Hermiticity (\ref{dagJ}) and the action of 
$ J_0^a (\Jbar _0^a)$, i.e., (\ref{Jpm}),
we get the norms of these states by explicit calculation:
\eqabegin
 \bra{ \Psi^d_1 } \semiket{ \Psi^d_1 } & = &
    \bra{ j;J_{L,1},1 } \semiket{ j;J_{L,1},1 }
    \bra{ j;J_{R,1},1 } \semiket{ j;J_{R,1},1 }  \comma \nn \\
  \bra{ \Psi^{d}_2 } \semiket{ \Psi^{d}_2 } & = &
   2  (j+1)(2j+1)(2j+k) \lb (j+1)^2 + J_{L,2}^2  \rb
   \nn \\
&& \qquad  \times
    \bra{ j;J_{L,2},1 } \semiket{ j;J_{L,2},1 }
    \bra{ j;J_{R,2},1 } \semiket{ j;J_{R,2},1 }
        \period \nn
\eqaend
 $  \bra{ j;J_{L,i},1 } \semiket{ j;J_{L,i},1 }
      \bra{ j;J_{R,i},1 } \semiket{ j;J_{R,i},1 }
$  $ (i = 1,2) $ take the same value 
if the bases of $ \ket{ \Psi^d_1 } $ and $ \ket{ \Psi^{d}_2 } $
belong to the same representation of $ \sltwo $. 
Thus, for a sufficiently large $ \abs{ j } $ (recall $ j \leq -1/2 $),
the latter norm behaves as $ 8 j^7/(k'\deltam)^2 $, and
the two norms have opposite signs. Although the $j$-value for the 
complementary series is restricted to $ -1 < j \leq -1/2 $, the discrete 
series appears by tensor products (see appendix C). In addition, 
bases with large $ \abs{ j } $ 
are generated from those with small values by tensor products
unless they decouple. Thus, if we include the
bases with real $ j $, our orbifold model cannot be unitary.
\parsmallskip
Next, we turn to the case of complex $ j $ (the principle
continuous series). Because $ j = -1/2 + i \nu \ (\nu > 0)$, the extremal
states at grade one have $ j = -1/2 \pm 1 + i \nu $.
These correspond to complex Casimir values and non-unitary $ \sltwo $
representations. This is not the end of the story however because
(i) infinite series of states build on these states
by the current zero-modes are not
allowed and (ii) the left and right sector are connected by the quantum
numbers $ \nw $ and $ \mj $.
Since the norm of $ \ket{ \Psi^{d}_2 } $ vanishes in this case, 
 we consider the following physical states instead:
\eqabegin
  && \ket{ \Psi^p_1 } = \ket{ j;J^1_L,1 }  \ket{ j;J^1_R,1 }  \comma
   \nn \\
 && \ket{\Psi^{p}_2} =
  \lb \ket{ \Phi^{j-1} (j;J_{L,2},1) }  - i \
   \ket{ \Phi^{j+ 1} (j;J_{L,2},1)}  \rb
   \ket{ j;J_{R,2},1 }
 \comma \nn
\eqaend
where $ J_{L(R),i} $ are given by (\ref{JLR}).
Again by explicit calculation, we get the norms of these states:
\eqabegin
  \bra{ \Psi^p_1 } \semiket{ \Psi^p_1 } &=&
    \bra{ j;J_{L,1},1 } \semiket{ j;J_{L,1},1 }
    \bra{ j;J_{R,1},1 } \semiket{ j;J_{R,1},1 }  \comma \nn \\
  \bra{ \Psi^{p}_2 } \semiket{ \Psi^{p}_2 } &=&
    - 4 \nu \lbb \lb J_{L,2}^2 - 1/4  - \nu^2  \rb
  \lb 4 \nu^2 - 3 k -1  \rb
     + 2 (1+k) J_{L,2}^2 - k  \rbb \nn \\
   &&  \qquad \qquad \qquad \times
    \bra{ j;J_{L,2},1 } \semiket{ j;J_{L,2},1 }
    \bra{ j;J_{R,2},1 } \semiket{ j;J_{R,2},1 }
    \period \nn
\eqaend
Then, for a sufficiently large $ \nu $, the latter norm
behaves as $ -16 \nu^7/(k'\deltam)^2 $. Thus,
the two norms have opposite signs if
the bases of $ \ket{ \Psi^p_1 } $ and $ \ket{ \Psi^{p}_2 } $
belong to the same representation of $ \sltwo $.
Since bases with large  $ \nu $
are generated from those with small values by tensor products,  
our orbifold model is again non-unitary if we include the bases with complex 
$ j $.
\parsmallskip
For the $ \SLtwo $ theory, a physical state at a sufficiently
high grade has large $ \abs{ j } $ at the base and it  caused the trouble.
In our case, some ghosts in the $ \SLtwo $ theory disappear,
but physical states with large $ \abs{ j } $ at the base exist
already at grade one owing to the winding modes. 
This is because the winding modes can produce negative Virasoro weight.
The existence of the ghost means that our model is not physical as it is.
However, 
we have still possibilities that the orbifold model becomes ghost-free,
for instance, by some truncation of the spectrum.
We will discuss this issue in section 4.5.
%
%
\csubsection{Tachyon and target-space geometry}
Before the consideration of the possibilities for a sensible theory, 
 we will discuss general properties of 
the tachyon propagation and the target-space geometry which are 
irrelevant to the details of the full spectrum \cite{NS}. 
We have worked in an abstract
framework based on representation theory so far, but we will find 
correspondences to the field theoretical approach in chapter 3.  
We see group theoretical meaning behind the black hole physics.
In addition, we find  properties 
similar to those in the $ \SLtwo /U(1) $ black hole theory because
both theories are based on the $ \SLtwo $ WZW model and 
closely related.
%
\isubsubsection{Tachyon in the untwisted sector}
First, we consider the tachyon in the untwisted sector.
It is expressed by the matrix elements of $ \cSLtwo $ in
unitary representations as (\ref{V0}). 
The matrix elements satisfy the differential equation \cite{VK}
\eqabegin
  \lbb  \Delta
   - j (j +1) \rbb  D^{j(\chi)}_{J_L,J_R} \lb g \rb
   &=& 0
  \comma \label{laplace}
\eqaend
where
$ \Delta $ is the Laplace operator on $ \SLtwo $.
Because the geometry of the black hole is
locally $ \SLtwo $, this equation is nothing
but the linearized tachyon equation in the black hole geometry
or the Klein-Gordon equation (\ref{TFE}) 
up to a factor. 
 For the untwisted
sector, the on-shell condition is 
$  -j(j+1)/ (k-2) = 1 $. Thus, at the critical value $ k = 52/23 $,
the $j$-value corresponds to the principal continuous series, and
the modes of the tachyon are given by (\ref{V0}) with 
 $ j = -1/2 + i/\sqrt{92} $.
\parsmallskip
Let us make an explicit correspondence between the tachyon and the scalar in 
chapter 3.
 Comparing (\ref{V0}) with (\ref{Tmodes2}), we find the correspondences
\eqabegin
   \begin{array}{lcllcl}
   f(x)  & \leftrightarrow  &
       \D{P} ^\chi_{J_L \pm,J_R \pm}(g) \comma \qquad  &
   T_{ \hatE \hatN }  & \leftrightarrow  &
    \D{P} ^\chi_{J_L \pm,J_R \pm}(g')\comma   \nn \\
       - \hatN & \leftrightarrow &  J_L + J_R\comma &
       \hatE & \leftrightarrow &   J_L - J_R  \comma \nn 
   \end{array}
\eqaend
where we have used (\ref{thetaLR}). Since $ \varphi $ has period
$ 2 \pi $, $ N =  \rmi \hatE + \rp \hatN \in \bfZ $.
This confirms the level matching condition (\ref{solLM}) with
$ \nw = 0 $.
\parsmallskip
As a further check, let us consider the matrix elements for
$ g' =
\matrixii{ \cosh \rho/2 }{ \sinh \rho/2 }{ \sinh \rho/2 }{ \cosh \rho/2 } $
$ \ ( \rho > 0 ) $; this corresponds to the region $ r > \rp $.
They are given by 
\eqabegin
  \D{P} ^\chi_{J_L +,J_R +} (g') & = & \frac{1}{2\pi} B(\mu_L,-\mu_L - 2j)
 \frac{\cosh^{2j + \mu_L + \mu_R} \rho/2 }{ \sinh^{\mu_L + \mu_R } \rho/2 }
   \ F \lb \mu_L, \mu_R; - 2 j ; - \sinh^{-2} \rho /2 \rb \comma
  \nn \\
 \D{P} ^\chi_{J_L -,J_R -} (g') & = &
  \frac{1}{2\pi} B\lb 1-\mu_R,\mu_R- 1 + 2(j+1) \rb
\frac{\cosh^{2j + \mu_L + \mu_R} \rho/2 }{ \sinh^{4 j + 2 +  \mu_L + \mu_R }
\rho/2 }
    \label{D--} \\
   && \qquad \quad \times  F \lb \mu_L+ 2j + 1 , \mu_R + 2j + 1 ;
 2 j + 2 ; - \sinh^{-2} \rho /2 \rb \comma \nn 
\eqaend
where $ \mu_{L,R} = i J_{L,R} - j $. $ F $ and $ B $ are the
hypergeometric function and the Euler beta function, respectively.
Then from $ - \sinh^2 \rho/2 = 1-\hatr^2  =  u $,
we find that these are nothing but
the mode functions in (\ref{modeUV}), i.e.,  $ U_{ \hatE \hatN }$ 
 and $ V_{ \hatE \hatN } $ up to a phase. The ``mass squared'' $ \mu $
and the $j$-value are related by $ \mu = 4j(j+1) $.
\parsmallskip
Generically, the untwisted tachyon behaves as
\eqabegin
  \begin{array}{llll}
    \D{P} ^\chi_{J_L +,J_R +} (g') & \sim & a_1 (r^2)^j
    & \mbox{as  } r \to \infty  \comma \\
    & \sim & b_1 \ e^{i(J_L-J_R) \ln \sqrt{r^2 - \rp ^2}}
             + b_2 \ e^{- i(J_L-J_R) \ln \sqrt{r^2 - \rp ^2}}
    & \mbox{as  } r \to \rp
     \comma \\
  \D{P} ^\chi_{J_L -,J_R -} (g') & \sim & a'_1 (r^2)^{-(j+1)}
    & \mbox{as  } r \to \infty
    \comma \\
    & \sim & b'_1 \ e^{i(J_L-J_R) \ln \sqrt{r^2 - \rp ^2}}
             + b'_2 \ e^{-i(J_L-J_R) \ln \sqrt{r^2 - \rp ^2}}
    & \mbox{as  } r \to \rp
   \comma \label{asymptD}
   \end{array}
\eqaend
where $ a^{(')}_{1} $ and $ b^{(')}_{1,2} $ are certain constants.
Since Re $ j = -1/2 $,
they behave like spherical waves asymptotically.
When $ J_L = J_R $, the hypergeometric function degenerates and 
the asymptotic behaviors as $ r \to \rp $ are different from
(\ref{asymptD}).
\isubsubsection{Tachyon in the twisted sectors}
Now we turn to the tachyon in the twisted sectors.
The twisted tachyon is given by the product of the matrix elements and
the twisting operator as (\ref{Vn}).
The twisting operator gives a phase to the tachyon.
In the twisted sectors, various $ j$-values are allowed from the on-shell
condition (\ref{L0n}) with $ \mj = N =\Nbar = 0 $ and $ \nw \neq 0$.
Thus the matrix elements of the complementary and the discrete   
series appear as well as those of
the principal continuous series. For the principal continuous and 
the complementary series,
the explicit forms and the asymptotic behaviors of the matrix elements
are given by the same expression as in (\ref{asymptD})
(although $j$-values are different).
\parsmallskip
For the discrete series,
only one linear combination of the solutions to (\ref{laplace}) appears.
As explained in appendix C, the matrix elements are obtained
from one of the matrix elements in the principal continuous series;
\eqabegin
  \D{L} ^j_{J_L,J_R } (g) & \propto  & \D{H} ^j_{J_L,J_R} (g) \ \propto \
  \D{P} ^\chi_{J_L +,J_R +} (g)  \period 
  \label{DLH} 
\eqaend
Thus we can read off the behaviors of $ \D{L,H} ^j_{J_L,J_R } (g') $
from $ \D{P} ^\chi_{J_L +,J_R +} (g') $. Note in particular that
$ \D{L,H} ^j_{J_L,J_R } (g') \to (r^2)^j$ as $ r \to \infty $ 
and $ j \leq -1/2 $. Therefore,
a tachyon state in the discrete series damps rapidly as one goes to infinity,
so this is a state
localized near the black hole. This is similar
to a winding state in the Euclidean $ \SLtwo /U(1)$ black hole
where one can regard it as a bound state in the dual geometry \cite{DVV}.
Consequently, we have three kinds of the tachyon:
One is from the principal continuous series and propagates like a wave,
and another is from the complementary series and asymptotically behaves
like $ r^{2j} $ or $ r^{-2(j+1)} $ $ (-1 < j \leq -1/2) $, and the other
is from the discrete series and is localized near the black hole.
\parsmallskip
For the untwisted tachyon, the tachyon scattering
and the Hawking radiation have been discussed in \cite{GL,NMS}. 
These arguments are also valid for the tachyon 
from the principal series in our case.
In addition, the tachyon modes from  
the discrete series are the same as 
$ U_{ \hatE \hatN } $ with $ \mu \geq -1 $.
Thus most of the discussion in chapter 3 is valid 
for the tachyon from the discrete series. 
Finally, the tachyon states
 satisfy the condition at infinity (\ref{BCAdS}) except 
for a part of the tachyon from the complementary series. 
\isubsubsection{Global properties}
So far we have not discussed the global properties of the tachyon, but
considered the tachyon propagation in one patch of the orbifold
(the region $ r > \rp $).
In order to discuss the tachyon propagation globally, we have to continue
it  from one region to another. Let us start with a tachyon
in region I ($ r > \rp $). 
Then the tachyon is given by a linear combination of
(\ref{D--}) or (\ref{DLH}) and is regular at infinity.
{}From the linear transformation formulas of the hypergeometric function,
we can obtain the expression around $ r = r_\pm $ as in (\ref{asymptD}).
We would like to continue it to the other regions.
\parsmallskip
Here we have two possible sources of obstacles. One is the complex
 power of  $ u $ or $ 1 - u $. 
The other is the logarithmic singularities like $ \ln  u $ or $ \ln  (1-u) $.
These cause troubles
as $ u \to 0 $ ($ r \to \rp $) or $ u \to 1 $ ($ r \to \rmi $).
The logarithmic singularity
at $ u = 0 $  arises when $ \mu_L - \mu_R \in \bfZ $, i.e.,
$ J_L - J_R = 0 $, and the one
at $ u = 1 $ arises when $ \mu_L + \mu_R + 2j \in \bfZ $, i.e.,
$ J_L + J_R = 0 $.
The latter corresponds to the case
of the $ \SLtwo/U(1) $ black hole in which the tachyon develops
a logarithmic singularity at the origin (singularity) \cite{DVV}.
This is natural because the inner horizon of the $ \cSLtwo/ \Zphi $ black hole
and the origin of the $ \SLtwo/U(1) $ black hole are the same point in
the $ \SLtwo $ group manifold.
\parsmallskip
Note that the matrix elements are continuous over the entire group
manifold. Thus if we consider a generalized function space including
distributions, we can continue the tachyon from one region to
another in any case.
Similarly, we can discuss the global properties of the scalar fields
in chapter 3 
by using representation theory of $ \cSLtwo $.
These considerations about tachyon fields may be useful 
for further investigations of the low energy theory. 
\isubsubsection{T-duality}
Finally, we will briefly discuss the properties under 
T-duality transformations.
The
$ \cSLtwo / \Zphi $ black hole has two Killing vectors $ \del_{\hatt} $
and
$ \del_{\hatphi} $. In the coordinate system $ (\hatt, \hatphi, \hatr) $,
the geometry is given by (\ref{GB}) and the dilaton
$ \phi = 0 $.
In order to deal with a general T-duality transformation,
let us define new coordinates $ x $ and $ y $ by
\eqabegin
  \vecii{\hatt}{\hatphi} &=& \matrixii{\alpha }{\beta}{\gamma}{\delta}
              \vecii{x}{y} \comma \quad \alpha \delta - \beta \gamma \neq 0
   \period \nn 
\eqaend
Then, the T-duality transformation with respect to $ \del_x $ covers
all the T-duality transformations.
\parsmallskip
First, let us consider the T-duality transformation
with respect to  $ \del_\varphi $. This has been discussed in \cite{HW}.
Setting $ x = \varphi $
and $ y = t $, the duals of the $ \cSLtwo / \Zphi $
black holes become in general black strings.
Thus this T-duality transformation is not self-dual.
\parsmallskip
Next, let us set $ x = \hatphi $ and $ y = \hatt - \hatphi $. In
these coordinates, the geometry is given by
\eqabegin
    ds^2_{string} &=& \alp k \lmb  dx^2
   + (1 - \hatr^2 ) dy^2
    + 2  ( 1 - \hatr ^2 ) d x d y
          +  ( \hatr ^2 - 1 )^{-1} d \hatr ^2 \rmb \comma \nn \\
    B  &=& \alp k \ \hatr ^2  d x \wedge d y
    \comma \qquad \phi \ = \ 0 \period \nn 
\eqaend
Then from the formula of T-duality transformations \cite{Buscher,GRV},
we get the following dual geometry \cite{NS}:
\eqabegin
 \tilde{ds}^2_{string}  &=&  \alp k \lmb dx^2
   + \hatr^2  dy^2
    + 2  \hatr ^2  d x d y
          +  ( \hatr ^2 - 1 )^{-1} d \hatr ^2 \rmb \comma \nn \\
   \tilde{B} &=& \alp k ( 1- \hatr ^2 ) d x \wedge d y
    \comma \qquad \tilde{\phi} \ = \ 0
   \period \nn 
\eqaend
This geometry is obtained from  the original one also via
$ \hatr ^2 \to 1 - \hatr ^2 $ or $ \hatt \leftrightarrow \hatphi $.
Thus, this T-duality transformation
is self-dual and interchanges  
the inside of the outer horizon $(\hatr ^2 < 1) $ and 
the outside of the inner horizon
(or the outside of the origin
for the non-rotating black hole) $ (\hatr ^2 > 0) $. In particular,
the outer and the inner horizon (or the origin) are interchanged.
Recall that translations of $ \hatt $ and $ \hatphi $ are the
vector and the axial symmetry. So,
the transformation $ \hatt \leftrightarrow \hatphi $ corresponds
to the T-duality transformation in the $ \SLtwo /U(1) $ black hole
which interchanges these symmetries and also the horizon and the singularity
\cite{DVV,Giveon}.
\parsmallskip
Since $ \varphi $
is periodic, we have to further
specify the periodicity of the dual coordinate.
In the above T-duality transformation, the period of $ x = \hatphi $ in the
dual geometry should be reciprocal of that in the original geometry
\cite{GRV}.
{}From (\ref{tphi}), we see that the periods of $ \hatt $ and $ \hatphi $
are not independent, so generically,
we cannot specify the period of $ \hatphi $ only.
However, for the non-rotating black hole ($ \rmi = 0 $), we have
$ \hatphi = \rp \varphi $ and the period of $ \hatphi $ in the original
geometry is equal to $ 2 \pi \rp $. Hence the period in the dual geometry
is $ 2 \pi/ (\rp k) $. This indicates that the black hole mass
is reversed under the T-duality transformation because $ \MBH = \rp ^2 $.
Since $ J_{L,R} $ take all real values, the spectrum of
$ L_0 $ and $ \Lbar _0 $ is
formally invariant under this T-duality transformation. But it is not bounded
from
below as in Minkowski spacetime, so we need some procedure
such as the Wick rotation for a rigorous argument.
%
%
\csubsection{Discussion}
In this chapter, we developed the string theory in the three dimensional 
black hole geometry in the framework of conformal field theory. 
This was the first attempt
to quantize a string in a black hole background with an infinite number of 
propagating modes.
The model was described 
by an orbifold of the $ \cSLtwo $ WZW model. 
 We  
constructed the orbifold. We discussed the spectrum by solving the level 
matching condition and obtained winding modes. We also analyzed the physical 
states and examined the ghost problem. We found explicit 
examples of negative-norm physical states. We then discussed the tachyon 
propagation and the target-space geometry. 
We found correspondence
between the group theoretical approach and the field theoretical one in 
the previous chapter. We also found a self-dual 
T-duality transformation reversing the black hole mass.
Although problems still remain, 
our results
may serve as a starting point for further investigations.
\parsmallskip
The existence of the negative-norm physical states indicates that 
our model is not physical  as it stands. Therefore, in the following, 
we will consider possibilities for obtaining a sensible theory
after a brief discussion on consistency conditions 
other than the ghost problem.  
%
\isubsubsection{Consistency conditions}
The basic physical consistency conditions for a string theory are not many.
In general, as a sensible physical theory, we must require Lorentz
invariance, a positive inner product for the observable Hilbert space and
the unitary transition amplitude. There are only a 
few in number, but these in turn
imply
various consistency conditions such as world-sheet diffeomorphism and Weyl
invariance, the absence of negative-norm states, closure of
OPE, level matching and modular invariance.  
Even though the absence of a tachyon might
also be added to the list, the presence of a tachyon in the bosonic string
does not indicate any fundamental inconsistency
in the theory.\footnote{See however Ref.~\cite{Joe},
which might imply that the bosonic
string does not exist nonperturbatively.}
In addition, for modular invariance, it is sufficient 
to check associativity of
OPE and modular invariance of the one-point amplitude at one-loop
\cite{Sonoda}.
\parsmallskip
It does not seem easy for a string theory to satisfy all these
requirements. However, there is a common belief that a world-sheet anomaly
(either local or global) always leads to a spacetime anomaly.\footnote{
Some works on this theme are as follows: the connection of the
modular invariance and spacetime anomalies are discussed in \cite{CaiJoe}
(for the type I) and \cite{SWarner} (for the type II and the heterotic
string); the connection between the modular invariance and unitarity are
discussed in \cite{CaiJoe,Big}.}
So, a string theory is likely to be automatically consistent once world-sheet
anomalies are removed. 
\parsmallskip
With these general remarks in mind, we comment on 
several consistency conditions in our case \cite{NS}.
\parbigskipn
{\sl Closure of OPE}
\parsmallskipn
Unitarity requires the closure of OPE,
and the fusion rules are determined by tensor products of the underlying
primaries and by non-trivial null states in the Kac-Moody and the 
Virasoro module.
We need detailed studies of these modules in order to find  
the condition from the non-trivial null states.
But it is easy to find that from the tensor products. 
The tensor products of the unitary representations are summarized
in appendix C. From them, we find that the tensor products
are closed if the content of the operators is given by
(i) only the highest (or the lowest) discrete series,
(ii)  the highest, lowest discrete series and
the principal continuous series, or (iii) 
all the unitary series, 
so that addition and subtraction
of the $j$-values are closed mod
$ \bfZ $. 
\parbigskipn
{\sl Partition function and modular invariance}
\parsmallskipn
Next, we turn to modular invariance.
{}From the spectrum in section 4.2, we get
\eqabegin
 L_0 - \Lbar_0  &=& - \nw \mj + N - \Nbar \comma \nn \\
 L_0 + \Lbar_0  &=&  \frac{-2 j(j+1)}{k-2} + N + \Nbar
      - \nw \lb \frac{k}{2} \deltap ^2 \nw - 2 \deltap J_R + \mj  \rb
     \period \nn 
\eqaend
Then the partition function diverges
since the Casimir $ -j(j+1)$, $ J_R $ and two integers $ \nw, \mj $ can
take arbitrarily large or small values.
In Minkowski spacetime, we can avoid the divergence
of the partition function by the Wick rotation, but
we have no analog in our case. Furthermore, our Kac-Moody
module is restricted to the states of the form (\ref{KMmodule}), so
we have to take this into account in the character calculation.
\parsmallskip
One resolution to this problem might be to find a subclass of the spectrum
and/or to develop an analog
of the Wick rotation so that we get a finite and
modular invariant partition function. This
might also solve the ghost problem.
For compact group manifolds \cite{GW}, the spectrum is
restricted to integrable representations
of the Kac-Moody algebra, so that one can get modular invariant partition
functions. Fields in non-integrable representations
decouple in correlators. However, the argument depends largely upon
compactness, so we have to take different strategies for non-compact
cases. So far, there is no general argument, but, for the $ \SLtwo $ theory,
there are a few attempts\cite{HHRS}-\cite{Huitu}. Besides group manifolds,
a partition function of a string theory in a curved spacetime is
discussed in \cite{RT}.
\isubsubsection{Toward a sensible theory}
Finally, let us discuss possibilities for obtaining a sensible 
string theory in the three dimensional black hole background \cite{NS}.
We can speculate various reasons why ghosts survive in our analysis:
\begin{enumerate}
\item  Further truncation might be necessary on the spectrum.
\item Modular invariance might fix the problem.
\item The theory based on $ \SLtwo $ might be sick. The $ \SLtwo $ WZW model
describes anti-de~Sitter space, so has unusual asymptotic properties.
\item One might have to use modified currents.
\item We might have to include non-unitary representations for base
representations of current algebras.
\end{enumerate}
All of the possibilities listed above appear in the discussion on 
the $ \SLtwo $ and the $ \SLtwo/U(1) $ theory
\cite{Petro,HHRS},\cite{Bars},\cite{DN}.
However, the possibility (5) does not work: even if we include
non-unitary representations, our argument in
section 4.3 does not change very much and we can easily find physical
states with negative norms.
From  general remarks in the previous subsection, the 
most plausible solution to our ghost problem is the possibility (2).
This might be related to (1).
However, the modular invariance for a string theory in
a curved spacetime is a hard problem as we saw in the above.
Here, 
we will discuss the possibility (1) which is different from 
previously discussed ones, and (4).
\parbigskipn
{\sl Discrete symmetries}
\parsmallskipn
One possibility to consistently truncate the spectrum
is further orbifolding besides that with respect to
$ \varphi \sim \varphi + 2 \pi $. As we will see,
only  a part of the $ \cSLtwo $ manifold is necessary for describing
the three dimensional black hole. Since we have started from
the $ \cSLtwo $ WZW model, the redundant part of the manifold
should be divided away by orbifolding. Let us discuss
the relevant discrete symmetries \cite{NS}.
\parsmallskip
In appendix C, we see that the $ \SLtwo $ manifold contains
sixteen domains denoted by $ \pm D^\pm_i $ $ \ (i = 1$-4$) $.
One correspondence between Region I-III and these domains is
\eqabegin
   && \mbox{Region I  } =  D_1^+ \comma \quad
      \mbox{Region II  } = D_2^- \cup \lb -D_3^+ \rb \comma \quad
      \mbox{Region III  } = -D_4^- \period \nn 
\eqaend
Here we have taken a parametrization in Region II and III slightly different
from the one in section 4.1, but the geometry is
the same. Thus we need only the universal covering space
of the region
$ \Omega_1 \equiv D_1^+ \cup D_2^- \cup \lb -D_3^+ \rb \cup \lb -D_4^- \rb $
to get the black hole geometry
as long as we do not consider its maximal extension.
Now let us define two transformations
by
\eqabegin
   \begin{array}{llllll}
  T_1 : &  g \to & g' = - g \comma
   &  &
    &   \\
  T_2 : &  g \to & g' = \calB g
   &  \mbox{ in } \pm D^\pm_{1,2}
  \comma &
   g' =  - \calB g
    &  \mbox{ in } \pm D^\pm_{3,4}
   \comma 
    \end{array} \nn 
\eqaend
where $ \calB $ is given by (\ref{Bauto})
and called Bargmann's automorphism of $ \SLtwo $.
$ T_{1,2} $ have the properties
\eqabegin
  \begin{array}{ll}
   T_1^2 = T_2^2 = 1 \comma &  \\
    T_1 : \ \Omega_{1 (2) } \to - \Omega_{1(2) } \comma \quad &
    T_2 : \ \Omega_{1 (2) } \to  \Omega_{2 (1) } \comma
  \end{array} \nn 
\eqaend
where $ \Omega_2  =  \lb  D_1^- \cup D_2^+ \cup  D_3^- \cup  D_4^+ \rb $.
Note that $ \pm \Omega_{1,2} $ cover all the sixteen domains of $ \SLtwo $
and have no overlap among them. Moreover we can obtain
the black hole geometry from each of the four sets as in section 4.1.
Thus we can divide $ \SLtwo $ by the $ \bfZ _2 $ symmetries,
$ T_1 $ and $ T_2 $, in order to drop redundant regions.
\parsmallskip
There is one more discrete symmetry.
This is related to
the problem of closed timelike curves.
Region I-III or
each of $ \pm \Omega_{1,2} $ includes
the region $ r^2 < 0 $ where closed timelike curves exist \cite{BTZ}.
This region corresponds to part of $ -D_4^- $ in $ \Omega_1 $
for the rotating case or the whole region for the non-rotating case.
Although we have no symmetry to remove this region only,
it is possible to drop it together with
the region $  (\rp^2 + \rmi^2 )/2 > r^2 > 0 $. The region
$  (\rp^2 + \rmi^2 )/2 > r^2  $ corresponds to
$ \lb -D_3^+ \rb \cup \lb -D_4^- \rb $ in $ \Omega_1 $, so we have only
to find a symmetry between $ D_1^+ \cup D_2^+ $
and $ \lb -D_3^+ \rb \cup \lb -D_4^- \rb $.
The symmetry is easy to find
in the coordinate system $ ( \hatt , \hatphi , \hatr ) $.
Let us define a $ \bfZ _2 $ transformation by
\eqabegin
  T_3 : && \lb \hatt \comma \hatphi \comma \hatr^2 - 1/2 \rb \quad
    \to  \lb \hatphi \comma \hatt \comma - (\hatr^2 - 1/2 ) \rb
  \period \nn 
\eqaend
Then 
the geometry given by (\ref{GB}) is  invariant under $ T_3 $.
This symmetry maps any point in $  D_1^+ \cup D_2^+ $ $(\hatr ^2 > 1/2 )$
to a point in $ \lb -D_3^+ \rb \cup \lb -D_4^- \rb $ $(\hatr ^2 < 1/2 )$ and
vice versa. Thus we can truncate both the spectrum and the region
with closed timelike curves by the orbifolding with respect to
$ T_3 $ at the expense of the additional dropped region.
Notice that a part of $ T_3 $, i.e., $ \hatr ^2 \to 1 -  \hatr ^2 $
or $ \hatt \leftrightarrow \hatphi $,  has
already appeared in the discussion of the T-duality in section 4.4.
\parbigskipn
{\sl The use of modified currents}
\parsmallskipn
Now we turn to another possibility.
In the flat theory, the no-ghost theorem has been proved \cite{noghost}.
Thus it seems useful to consider the flat limit of our model and 
observe how the ghosts disappear. However, we cannot take this limit: 
The three dimensional 
flat theory is described by three free bosons. Hence, e.g. for the left sector,  
there are three pairs of conjugate zero-modes, and the base states
are specified by three momenta as $ \ket{ p^0, p^1, p^2 } $. On the other 
hand, the base states of our model, e.g. in the left sector, 
have only two labels as $ \ket{ j;J } $ (although the total labels for both 
the left and the right sector are three).
Because of the deficiency of the zero-modes, we cannot get to the flat theory.
\parsmallskip
The deficiency of the zero-modes is observed from a different point of view.
Recall the Wakimoto realization of the $ \sltwo $ Kac-Moody algebra
\cite{Wakimoto}. 
It is realized by a free boson $ \phi $ and 
a $ \beta $-$ \gamma $ ghost system\footnote{ This representation 
 is slightly different 
from the one diagonalizing $ J^0 $. }:
\eqabegin
 i J^+(z) &=& \beta (z) \comma \nn \\
 i J^-(z) &=& \gamma^2 \beta (z) + \sqrt{2k'} \gamma \del \phi (z) 
      + k \del \gamma (z) \comma \nn \\
 i J^2 (z) &=& \gamma \beta (z) + \sqrt{k'/2} \del \phi (z) 
 \comma \nn
\eqaend
where $ k' \equiv k - 2 $ and 
\eqabegin
  \beta (z) \gamma (w) & = & - \gamma (z) \beta (w) \ \sim \ \frac{1}{z-w}
     \comma \nn \\
     \quad \phi (z) \phi (w) & \sim &  - \ln ( z - w )
   \period \nn 
\eqaend
The $ \beta $-$ \gamma $ ghosts can be bosonized by two free bosons
\cite{FMS}, but some of the zero-modes of these bosons 
are redundant.
The redundant zero-modes are related to
 the picture changing of the ghost system and absent from the original 
algebra.  
\parsmallskip
On the other hand, there is an argument based on effective action that 
a string in a nearly flat $ \AdS $ ($ \SLtwo $) with weak curvature
 must be unitary \cite{FT}.  
Therefore, it may be possible to construct 
a unitary $ \SLtwo $ and $ \cSLtwo/ \Zphi $ theory 
if we incorporate the deficient zero-modes so that the model has the flat limit. 
Indeed, we may re-interpret Bars' argument  for ghost-free 
spectrum of a $ \SLtwo $ theory \cite{Bars} along this line of thought.
He realizes the $ \beta $-$ \gamma $ ghosts by two free fields
\eqabegin
  && \beta = \del \phi^+ \comma \quad \gamma = \phi^- \comma  \nn 
\eqaend
where $ \phi^\pm = (1/\sqrt{2}) (\phi^0 \pm \phi^1)$ 
and $ \phi^i(z) \phi^j (w) \sim (-1)^i \delta^{ij} \ln (z-w) $ $(i = 0,1)$. 
Owing to the redundant
zero-modes, the currents are modified. However, by a careful treatment
of the zero-modes, one can show that the current algebra is maintained,  
the string on $ \SLtwo $ has no ghosts, and the flat theory is recovered
in the limit $ k \to \infty $. 
\parsmallskip
%
We cannot apply his realization to the string theories on $ \SLtwo/ \Zphi $
or \break 
$ \SLtwo/U(1) $: For the $ \SLtwo $ WZW model, the allowed states
in his realization  are
only certain combinations of the left and the right sector which 
diagonalize $ J^+_0 \ (\Jbar ^+_0 )$. 
For the  black hole cases, we need the states diagonalizing
$ J^2_0 \ (\Jbar _0^2)$.
However, 
it is interesting to generalize his argument and apply it to 
the black hole physics \cite{NS2}.
%
%
%
\parsmallskipn
\csection{CONCLUSION}
In this thesis, we discussed quantum aspects of the three dimensional 
black holes. In chapter 3, we considered scalar fields with a generic
 mass squared in the three dimensional black hole background, and 
discussed their thermodynamics in the framework of quantum 
field theory in curved spacetime. We took 
two approaches. One was based on mode expansion and summation over states.
 In the other approach, we used Hartle-Hawking
Green functions. We obtained exact expressions of mode functions, 
the Hartle-Hawking Green functions, Green functions on a cone geometry, 
and thermodynamic quantities. These constitute a reliable basis of the 
quantum field theory and the thermodynamics of scalar fields 
in the three dimensional black hole background.
Our results did not necessarily agree with those in the literature
and the thermodynamic quantities
depended largely upon their definitions, boundary conditions and
regularization schemes. These indicate the importance of curvature effects 
and precise discussions. We may need further investigations
of this issue in particular for the cases of finite black hole mass
(i.e., truly curved cases). Our model may be useful for 
this purpose.
\parsmallskip
In chapter 4, we considered the string theory in the three dimensional 
black hole geometry
 in the framework of conformal field theory. This was the first attempt
to quantize a string theory in a black hole background with 
an infinite number of propagating modes.
We constructed 
an orbifold of the $ \cSLtwo $ WZW model, which described the string in the 
three dimensional black hole geometry. 
 We discussed the spectrum by solving the level matching condition 
and obtained winding modes.
We also analyzed the physical states and found negative-norm physical states.
The tachyon and the target-space geometry were discussed. 
The existence of the negative-norm physical states implies that our model 
is not sensible as it stands.
Thus we discussed  possibilities to obtain a sensible string theory.
Our detailed analyses 
may serve as a basis for further investigations of this subject. 
\parsmallskip
We still have difficulties both in the thermodynamics and in the string theory.
Because of them, the analyses in chapter 3 and 4
are not fully connected yet. 
However, we believe that our results may provide
useful insights into quantum aspects of 
the three dimensional quantum black holes.
\parsmallskip
We are now  about to directly catch gravitational waves.
A large amount of data concerning cosmology is accumulating.
Moreover, we have seen an interesting  
result in super string theory that black holes work well as a probe 
into quantum gravity \cite{DEnt,DHrad}. 
I sincerely hope that, together with these developments, 
further investigations of quantum black holes
lead to a deeper understanding of quantum theory of gravity.
%
%
%
%
%
\parbigskipn
\parbigskipn
\csectionast{ACKNOWLEDGEMENTS}
I would like to thank my thesis advisor Y. Kazama for many valuable and 
instructive discussions, fruitful collaborations and a careful 
reading of the manuscript. I would also like to thank I. Ichinose 
for useful discussions, a fruitful collaboration and his constant 
encouragement during my course. It is my pleasure to acknowledge 
a fertile collaboration and useful discussions with S. Hirano, 
M. Natsuume and A. Tsuchiya.
I am grateful to I. Bars, K. Hori, H. Ishikawa, T. Izubuchi, D. Maison, 
Y. Matsuo, S. Mizoguchi, Y. Okawa, N. Sakai, K. Shiraishi, T. Tani, 
T. Yamamoto, especially to M. Kato, T. Oshima and T. Yoneya
for useful discussions. I would like to acknowledge 
useful comments from J. Polchinski, A.A. Tseytlin and J. Uglum.
Finally, I thank the Japan Society for Promotion of Science for 
financial support.
%
%
\newpage
%
%
\csectionast{APPENDIX}
\setcounter{section}{0}
\def\appsection#1{\addtocounter{section}{1} \setcounter{subsection}{0}
       \setcounter{equation}{0}
     {\large\bf \par \bigskip \parsmallskip \noindent \Alph{section} \quad  #1}
      \par \bigskip \noindent}
\def\appsubsection#1{\addtocounter{subsection}{1}
         \par \bigskip \noindent  {\normalsize\it
         \Alph{section}.\arabic{subsection} \quad #1} \par \medskip \noindent }
\renewcommand{\theequation}{\Alph{section}.\arabic{equation}} 
%
%
\appsection{The Feynman Green function in $\CAdS$}
In appendix A, we summarize the derivation of the Feynman Green function 
in the universal covering space of three dimensional anti-de Sitter space
 $ (\CAdS ) $. Quantization of a scalar field in $ \widetilde{AdS}_D  $
 has been discussed 
in \cite{AIS}-\cite{BL},
and the Feynman Green function has been obtained \cite{AIS,BL} 
in terms of the hypergeometric function. In 
the three dimensional case, the Feynman Green function is 
simplified and expressed in terms of elementary functions \cite{IS}.
\parsmallskip
$ \CAdS $ is defined by its embedding in a four dimensional 
flat space of signature $ (--++) $ through the equation
\eqabegin
&& -x_0^2 - x_1^2 + x_2^2 + x_3^2 = - l^{-2} \period \nn 
\eqaend
We parametrize this by 
\eqabegin
&& x_0 = l \sin \tau \sec \rho \comma \quad 
   x_1 = l \cos \tau \sec \rho \comma \nn \\
&& x_2 = l \sin \theta \tan \rho \comma \quad 
   x_3 = l \cos \theta \tan \rho \comma  \nn
\eqaend
where $ 0 \leq \rho < \pi/2 \comma  \ 0 \leq \theta <  2 \pi 
\comma \  -\infty  < \tau < \infty $.
Then the metric becomes
\eqabegin
  d s^2 &=&  l^2 \sec^2 \rho \lb  - \ d \tau^2 + d \rho^2 
                 + \sin^2 \rho \ d \theta^2 \rb
                 \period \nn 
\eqaend
The field equation for a scalar field is given by
\eqabegin
 \lb  \Box - \mu l^{-2} \rb  \psi (x) &=& 0 \comma \nn
\eqaend
where $ \Box = \frac{1}{\sqrt{-g}} \del_\mu \sqrt{-g} g^{\mu \nu} \del_\nu $.
Making the separation of variables 
\eqabegin
 && \psi(x) = \sum \psi_{m \omega} 
   = \sum \ e^{- i \omega \tau} \ e^{i m \theta} R_{m \omega} (\rho)
  \comma \qquad ( m \in {\bf Z} ) \comma \nn 
\eqaend
the equation for the radial function $ R_{m \omega} (\rho)$ is written as 
\eqabegin
 \lb \del_\rho^2 + \frac{1}{\sin \rho  \cos \rho} \del_\rho 
  + \omega^2 - \frac{m^2}{\sin^2 \rho} - \mu \sec^2 \rho \rb 
   R_{m \omega} (\rho) &=& 0
  \period \nn
\eqaend 
We make a further change of variables
$ v = \sin^2 \rho \comma $
and define a function $ f_{m \omega}(v) $ by 
\eqabegin
 R_{m \omega} (\rho) &=& v^{\abs{m} /2} ( 1-v )^{\lambda/2} 
   f_{m \omega}(v) \comma \nn
\eqaend
with
\eqabegin
 \lambda &=& \lambda_\pm \, \equiv \, 1 \pm \sqrt{1 + \mu} 
 \period \nn
\eqaend
Then the radial equation is reduced to the hypergeometric equation
\eqabegin
 \lbb v(1-v) \del_v^2 + \lmb  c - ( a + b + 1 )v \rmb  \del_v
  -ab \rbb \ f_{m \omega} (v) &=& 0 
  \comma \nn 
\eqaend
where
\eqabegin
 a &=& \half ( \lambda + \abs{m} - \omega ) \comma \nn \\
 b &=& \half ( \lambda + \abs{m} + \omega ) \comma \nn \\
 c &=& \abs{m} + 1 \period \nn 
\eqaend
(We will deal with the case of real $ \lambda $, i.e., $ \mu \geq -1 $.)
If we require the regularity at $ v = 0 $, the solution is expressed by 
the Gauss' hypergeometric function $ F $ as 
\eqabegin
 f_{m \omega} (v) &=& F(a, b; c; v)
 \period \nn
\eqaend

Since $ \CAdS $ is not globally hyperbolic, it is necessary to 
impose boundary conditions at spatial infinity. Following 
\cite{AIS,BFMT}, 
we impose the condition to conserve energy. 
This means that the surface integral
of the energy-momentum tensor at spatial infinity must vanish. This 
requirement leads to  
\eqabegin
 \abs{\omega} &=& \lambda + \abs{m} + 2 n \qquad ( n = 0, 1, 2, .., )
 \comma \nn 
\eqaend
where
\eqabegin
 \lambda &=& \lmb
    \begin{array}{ll}
       \lambda_\pm & {\rm for} \ 0 > \mu > -1 \comma  \\
       \lambda_+   & {\rm for} \ \mu \geq 0 \comma  \mu = -1
    \end{array}
    \right.
    \period \nn 
\eqaend
Then the value of $ a $ takes zero or a negative integer. 
By using a mathematical formula
\cite{AS}, one obtains
\eqabegin
 \psi (x) &=& \sum_{m, n} \lbb a_{mn} \psi_{m n} + 
 ( a_{mn} \psi_{m n} )^{\ast} \rbb
  \qquad ( m \in {\bf Z} , \quad n = 0, 1, 2, ... ) \comma \nn \\
 \psi_{m n} &=& C_{mn} \ e^{- i \omega \tau} \ e^{ i m \theta} 
      \lb  \sin \rho \rb ^{\abs{m} } \lb \cos \rho \rb ^{\lambda}
      P_n^{(\abs{m} , \lambda-1)} ( \cos 2 \rho )
      \comma \label{psimn} 
\eqaend
where $ P_n^{(\alpha,\beta)} $ are Jacobi Polynomials and $ C_{mn} $ 
are normalization constants.
\parsmallskip
For the positive frequency part $ \psi^{(+)} $ of the solution
one can define a positive definite scalar product by
\eqabegin
 \lb   \psi_1^{(+)} , \psi_2^{(+)} \rb 
 &\equiv& -i \int_{\Sigma} d^2 x \sqrt{-g} g^{0\nu}
       \psi_1^{(+)\ast} \mathop{\del_\nu}^{\leftrightarrow} \psi_2^{(+)}
         \comma \nn 
\eqaend
where $ \Sigma $ is a spacelike surface.
Then the normalization constant $ C_{mn} $ is determined by the condition
$ \lb  \psi_{mn}^{(+)} , \psi_{m'n'}^{(+)} \rb = 
\delta_{m m'} \delta_{n n'} $. 
By using the orthogonal relation with respect to the Jacobi Polynomials
\cite{AS},
\eqabegin
&& \int_0^{\pi/2} d \rho \tan \rho 
      \lb \sin \rho \rb ^{2 \abs{m} }  
      \lb \cos \rho \rb ^{2\lambda}
      P_n^{(\abs{m} , \lambda-1)}(\cos 2 \rho)
      \ P_{n'}^{(\abs{m} , \lambda-1) }(\cos 2 \rho) \nn \\
&=& \delta_{n n'} 
     \frac{1}{2(2n  +   \lambda  +   \abs{m} ) }
        \frac{\Gamma(n \! + \! \abs{m} \! + \! 1)\Gamma(n \!+ \! \lambda)}
             {n! \Gamma( n  +   \lambda  +   \abs{m} )} 
 \comma \label{orthoP}
\eqaend
one obtains 
\eqabegin
 C_{mn} &=& \lbb \frac{n! \ \Gamma(\abs{m}  +   \lambda  +   n )}
              {2 \pi l ( \abs{m}  +   n )! \ \Gamma(\lambda  +   n ) }
              \rbb ^{1/2} 
              \period \nn 
\eqaend

Now we quantize the scalar field by setting the commutation relation
\eqabegin
 \lbb a_{mn}, a_{m'n'}^\dagger \rbb = \delta_{mm'} \delta_{nn'} 
      \period \nn
\eqaend
Then one finds
\eqabegin
 \lbb \psi(x), \ \psi(x') \rbb _{\tau=\tau'} &=& 0 \comma \nn \\
 \lbb \psi(x), \ \del_{\tau'} \psi(x') \rbb _{\tau=\tau'} &=& 
 -i \frac{1}{g^{\tau\tau} \sqrt{-g}} \delta(\theta-\theta') 
    \delta(\rho - \rho')
  \period \label{comm}
\eqaend
Here we have used the orthogonal relation (\ref{orthoP}). The $ \delta $ 
function is defined for the space of functions of the form (\ref{psimn}).
\parsmallskip
We then define the Feynman Green function by
\eqabegin
 -i \Gf(x,x') &=& \bra{0} T \lmb \psi(x) \psi(x') \rmb  \ket{0} \nn \\
 &\equiv& \theta(\tau -\tau') \sum_{m \comma  n} 
  \psi_{mn}(x) \psi_{mn}^{\ast} (x')
         + \ ( x \leftrightarrow x' )
  \period \nn 
\eqaend
From (\ref{comm}), one can check 
\eqabegin
 \lb \Box - \mu l^{-2} \rb \Gf(x, x') &=&  \frac{1}{\sqrt{-g}} 
         \delta (x - x' ) \period \label{GfEq}
\eqaend

Furthermore, one can perform the summation with respect to $ m $ and $ n $.
First, we set $ x' = (\tau', \rho' , \theta') = (0,0,0) $ (i.e.,
$ (x_0', x_1', x_2', x_3') = (0,l,0,0) $) without loss of generality because
$ \CAdS $ is homogeneous. Then only the terms with $ m = 0 $ contribute 
to the summation; 
\eqabegin
 -i \Gf (x,0) &=& \frac{1}{2\pi l} \ e^{-i\lambda \abs{\tau} } 
  ( \cos \rho )^{\lambda}  
   \sum_{n= 0}^{\infty} \ e^{-2in\abs{\tau} } 
   \ P_n^{(0, \lambda-1)}( \cos 2 \rho)
   \period \nn 
\eqaend
By making use of the mathematical formula \cite{PBM}
\eqabegin
&& \sum_{k = 0}^{\infty} \frac{(\alpha  +   \beta  +   1)_k}
       {(\beta  +   1)_k} \ t^k \ P_k^{(\alpha, \beta)}(x) \nn \\
  && \qquad = ( 1 + t )^{-\alpha -\beta -1} 
   F \lb \frac{\alpha  +   \beta  +   1}{2}, 
     \frac{\alpha  +   \beta  +   2}{2}; \beta + 1 ; 
     \frac{2 t (x  +   1)}{(t  +   1)^2}
   \rb
   \comma \nn 
\eqaend
one obtains \cite{BL}
\eqabegin
 - i \Gf(x,0) \equiv - i \Gf(z) &=&
           \frac{l^{-1}}{2^{\lambda+1} \pi} z^{-\lambda} \ 
                  F \lb \half \lambda , \half (\lambda  +   1) ; 
                  \lambda ; z^{-2} \rb  
                     \period \nn 
\eqaend
Here $ z $ is defined by
\eqabegin
 z &=& 1 + l^{-2} \sigma (x, 0) + i \varepsilon  \period 
  \label{epGf} 
\eqaend
$ \sigma(x,x') $ is half of the distance between $ x $ and $ x'$ in the four 
dimensional flat space, 
\eqabegin
  \sigma (x, x') &=& \half \eta_{\alpha\beta} 
    ( x - x')^{\alpha}( x - x')^{\beta}
              \comma \nn 
\eqaend
where $ \eta_{\alpha\beta} $ and $ x^\alpha $  $(\alpha,\beta = 0$-$3)$
are the metric and the coordinates of the flat space, respectively. 
The infinitesimal imaginary part 
$ i \varepsilon \ ( \varepsilon > 0 ) $ 
in $ z $ was added so that the Green function looked locally like the Minkowski
one \cite{AIS}.  In the three dimensional case, by the formula,
\eqabegin
F\lb  a, \half  +   a ; 2 a ; z \rb &=& 
   2^{2a-1} (1-z)^{-1/2} \lbb  1 + ( 1-z)^{1/2} \rbb ^{1-2a}
   \comma \nn 
\eqaend
 the Feynman Green function is simplified to \cite{IS} 
\eqabegin
     - i \Gf(z) &=& \frac{l^{-1}}{4 \pi} ( z^2 -1)^{-1/2} 
                \lbb  z + (z^2-1)^{1/2} \rbb^{1-\lambda}
         \period \nn 
\eqaend
This result is obtained also by replacing $ \abs{\tau} $
with $ \abs{\tau} - i \varepsilon $ so that $ \abs{e^{-2in\abs{\tau} } } 
< 1 $ and by utilizing the generating function of  
the Jacobi Polynomials. 
\parsmallskip
For a generic $ x' $, we have only to replace $ \sigma(x, 0) $
with $\sigma(x, x') $.
%
%
%
\appsection{The Sommerfeld representation of Green functions}
In appendix B, we derive $ \Gfe (x,x'_n; \beta) $ in chapter 3 
and its derivative 
with respect to $ \beta $ \cite{IS}.
\appsubsection{Derivation of  $ \Gfe (x,x'_n; \beta) $ }
We begin with the definition
\eqabegin
  \tGf ( \zeta ; 2 \pi ) &\equiv& \Gf (z(\zeta, i\delphiEp_n; r,r'); \bH ) 
                    \when_{ \delphiEp_n, r, r' : \ {\rm fixed} } 
    \period \nn
\eqaend
By definition, $ \tGf (w_n; 2\pi) = \Gfe (x,x'_n;\bH) $
where $ w_n $ is given by (\ref{wn2}).
$ \tGf ( \zeta ; 2 \pi )  $ depends upon $ \zeta  $ through 
\eqabegin
  z(\zeta, i\delphiEp_n; r,r') - i \varepsilon &=& \frac{1}{\dH^2} 
         \lbb  \sqrt{r^2 -\rmi^2} \sqrt{r'^2 - \rmi^2} 
         \cosh \lb \frac{i \rp}{l} \delphiEp_n
               \rb  \right.
               \nn \\
      & & \left. \qquad \qquad -
          \sqrt{r^2 -\rp^2} \sqrt{r'^2 - \rp^2} 
         \cosh \lb i \zeta  \rb \rbb
               \period  
  \label{zzeta}
\eqaend
Thus $ \tGf ( \zeta ; 2 \pi )  $ is periodic under $ \zeta \to \zeta + 2 \pi $
($ \tau \to \tau + \bH $).
$ z = \pm 1 $ are the points of its singularities. On $ \zeta $-plane,  
there are four points corresponding to these singularities in the region
$ - \pi < $ Re $ \zeta $  $ \leq \pi $. 
They are indicated by crosses ($ \times $)
in Figure 2. 
 These points are symmetric with respect to the origin 
$ \zeta = 0 $, and infinitesimally close to the imaginary axis  
for $ \delphiEp _0 = 0 $, i.e.,
when we take the trace of Green functions.
The authors of \cite{SoCa} discussed how to construct 
a Green function with an arbitrary period for certain differential equations.
By following them, we get the Sommerfeld integral representation of the Green 
functions with an arbitrary period in our case \cite{IS}:
\eqabegin
  \tGf( w_n; \bbh ) 
  &=& \frac{\bH}{2 \pi \beta} \int_\Gamma d \zeta
  \tGf( \zeta ; 2 \pi ) 
  \frac{ e^{ i \bH \zeta/\beta}}
{ e^{ i \bH \zeta/\beta}- \ e^{ i \bH w_n /\beta}}
  \comma \label{intrep2}
\eqaend
where the contour $ \Gamma $  is given by the solid 
lines in Figure 2.
This contour consists of two parts and divides the four singularities into 
two pairs. Then by recovering other variables, we obtain
\eqabegin
   \Gfe (x,x'_n; \beta)  &=& \tGf(w_n;\bbh) \period \nn 
\eqaend
\parsmallskip
It is instructive to consider some special cases 
before we show the validity of the above expression.
First, we consider the case of $ \beta = \bH/q \comma \ (q = 1,2,...) $.
Notice that the contour $ \Gamma $ can be deformed into $ \Gamma'$ 
given by the dashed lines in Figure 2.
Since the integrand is of period $ 2 \pi $ in this case, 
the contributions from the path made
up of straight lines cancel with each other. Thus 
only the residues inside the circular path contribute to the integral. 
Then we get
\eqabegin
 \tGf (w_n ; 2 \pi/q) &=& \sum_{k} \tGf(w_n(k); 2 \pi)
   \label{2pq} 
 \comma
\eqaend
where $ w_n(k) $ and $ k ( \in {\bf Z}) $ are given by
$ w_n(k) =  w_n + 2\pi k/q $ and 
$ - \pi < w (k) \leq \pi $.
In this case, the method of images works and we can explicitly check the 
periodicity. Clearly, the right-hand side of 
(\ref{2pq}) reproduces $ \tGf(w_n ; 2 \pi) $ for $ q = 1 $.
\parsmallskip
Next, we consider the case $ \beta \to \infty $. 
In the limit $ \beta \to \infty $, the expression (\ref{intrep2})
is reduced  to
\eqabegin
   \tGf (w_n; \infty) &=& \frac{1}{2 \pi i } \int_\Gamma 
    \tGf(\zeta; 2 \pi)
    \frac{d \zeta}{\zeta - w_n } 
    \period \label{Ginfty}
\eqaend 
From the  formula 
$ \lim_{n \to \infty} \sum_{k = -n}^{n} 1/(x + k) = \pi \cot \pi x $, 
we obtain another expression of $ \tGf(w_n;\bbh) $ \cite{IS}:
\eqabegin
 \tGf(w_n;\bbh) &=& \sum_{k = - \infty }^{\infty} 
     \tGf(w_n + 2 \pi k \beta/\bH ;\infty) \nn \\
        &=&
        \frac{\bH}{ 4 \pi i \beta} \int_\Gamma  d \zeta \
       \tGf(\zeta; 2 \pi)
        \cot \lmb \frac{\bH}{2 \beta}( \zeta - w_n ) \rmb 
        \period \label{cotform2} 
\eqaend
The equivalence to the former expression is easily checked by noting
$ \tGf(w_n;\bH) = \tGf(w_n;- \bH) $.
\parsmallskip
Now we check the properties necessary for the Green function 
with a generic $ \beta $.
First, $ \tGf (w_n;\bbh) \ (\Gfe(x,x'_n;\beta))$
actually converges  because
$ \tGf(\zeta; 2\pi ) $ comes to vanish exponentially 
as $ \abs{ {\rm Im} \ \zeta} \to \infty  $. The periodicity of 
$ \tGf (w_n;\bbh ) $ is easily confirmed by (\ref{cotform2}).
Finally, let us check that $ \Gfe(x,x'; \beta) $ satisfies 
the inhomogeneous
equation.
Remember (the Euclidean version of) (\ref{GfEq}), 
then from (\ref{Ginfty}) we find  that
\eqabegin
( \Box^E - \mu  ) \Gfe(x,x'; \infty) &=& 
                \frac{a}{\sqrt{ \abs{g^E} }} \ \delta^E_{\infty} (x-x')
          \comma \nn
\eqaend
where $ a = -1 $ for $ \JBH = 0 $ and $ a = i $ for $ \JBH \neq 0 $.
Here we have explicitly denoted the period of the delta function
with respect to $ \tau $.
Then 
using $ \Gfe(x,x' ;\beta) = \sum_{k = -\infty}^{\infty}
\Gfe(x,x'; \infty)\Bigm{\vert}_{ \deltau \to \deltau + k \beta} $ ,
we get the desired result 
\eqabegin
( \Box ^E - \mu l^{-2} ) \Gfe(x,x'; \beta) &=& 
                \frac{a}{\sqrt{ \abs{g^E} }} \ \delta^E_{\beta} (x-x')
          \period \nn
\eqaend
\appsubsection{Derivation of 
$ \del_\beta \Gfe (x,x'_n; \beta) \vert_{\beta = \bH}$}
To calculate the entropy, 
we need $ \del_\beta \Gfe (x,x'_n; \beta) $.
From the integral representation (\ref{cotform2}), we have 
\eqabegin
 && \frac{\del}{\del \beta} \Gfe(x, x'_n;\beta) 
 = - \frac{1}{\beta} \Gfe(x,x'_n;\beta)  \label{delG} \\
  && \qquad \quad +
 \frac{\bH^2}{8 \pi i \beta^3} \int_\Gamma d \zeta \ \tGf(\zeta; 2 \pi)
          ( \zeta - w_n ) 
        {\rm cosec}^2 \lmb  \frac{\bH}{2 \beta} (\zeta - w_n ) \rmb  
   \period \nn 
\eqaend 
For $ \beta = \bH $, the above expression is fairly simplified.
First, we deform the contour $ \Gamma $ into $ \Gamma' $. 
Within the circular path, there is only  one singularity at $ \zeta = w_n $. 
The contribution from the
residue of this singularity cancels with
the first term in (\ref{delG}). 
Then by changing variables to $ i \zeta' = \zeta \pm \pi  $ according to 
the left and right straight path of $ \Gamma'$, we get
\eqabegin
  \frac{\del}{\del \beta} \Gfe(x,x'_n;\beta) 
&=& \frac{1}{4 \bH} \int_{- \infty}^{\infty} d \zeta' \ 
    \frac{\tGf( i \zeta' - \pi; 2 \pi )}
             {\cos^2 \lmb ( i \zeta' - w_n )/2 \rmb }
 \period \nn 
\eqaend
Note that $ \tGf( i \zeta' - \pi; \bH ) $ is a function of 
\eqabegin
   z ( \zeta' ) &\equiv & z(i\zeta' - \pi, i\delphiEp_n; r,r') 
   = \An + \Bn \cosh \zeta' 
  \comma \nn
\eqaend
where
\eqabegin
  \An &=& \frac{1}{\dH^2} \sqrt{r^2 - \rmi^2}\sqrt{r'^2 - \rmi^2}
    \cosh \lb \frac{i\rp}{l} \delphiEp \rb
  \comma \quad B \ = \ 
    \frac{1}{\dH^2} \sqrt{r^2 - \rp^2}\sqrt{r'^2 - \rp^2}
    \period \nn
\eqaend
We then make the further change of variables from $ \zeta' $ to $ z $, 
and use 
\eqabegin
 \frac{d z }{d \zeta'} &=& 
 \Bn \sinh \zeta'  
 \ = \ \pm \sqrt{( z - \An )^2 - \Bn^2}
 \comma \nn 
\eqaend
and 
\eqabegin
  2 \cos^2 \lmb \half (i \zeta' - w_n)) \rmb &=& 
    \cn \frac{z-\An}{\Bn} 
    \pm \sn \frac{\sqrt{( z - \An )^2 - \Bn^2}}{\Bn} \ + \ 1 
    \comma \nn
\eqaend
where
\eqabegin
 \cn &=& \cosh ( i w_n )
 \comma \quad 
 \sn = \sinh (i w_n)
 \period \nn
\eqaend
Consequently, we get the fairly simple expression \cite{IS}
\eqabegin
  && \frac{\del}{\del \beta} \Gfe(x,x'_n; \beta) \when_{\beta = \bH}  \nn \\
  &=& - \frac{\Bn}{\bH} \int_{\An + \Bn}^{\infty} d z 
   \ \Gfe (z; \bH) \frac{1}{\sqrt{( z  -  \An )^2  -  \Bn^2}}
   \frac{\cn (z  -  \An)  +   \Bn}{( z  -  \An  +   \cn \Bn )^2}
  \period \nn
\eqaend
%
%
\appsection{Representations of $ \SLtwo $}
In this appendix, we briefly summarize the
representation theory of $ \SLtwo $
(and of its universal covering group $ \cSLtwo $ ) and collect its useful
 properties for discussions in this thesis. For a review,
see \cite{VK} and \cite{Bargmann}-\cite{Wybourne}.
\appsubsection{$\SLtwo$}
%
{\sl Preliminary}
\parmedskipn
The group $\SLtwo$ is represented by real matrices
\eqabegin
  && g = \matrixii{a}{b}{c}{d} \comma \qquad ad -  bc = 1 \period \nn
\eqaend
It has one-parameter subgroups
\eqabegin
   \Omega_a &=& \lmb  g_a (t) = \ e^{ -i t \tau^a} \rmb \comma
  \qquad a = 0, 1, 2 \comma \nn 
\eqaend
where
\eqabegin
  \begin{array}{cccccl}
  \tau^0 &  =  &
 {\displaystyle -\frac{1}{2} \sigma_2 }
    & \to  &
  g_0 (t) & =  \matrixii{ \cos t/2 }{ \sin t/2 }{ - \sin t/2}{ \cos t/2}
    \comma \\
  \tau^1 &  = &
 {\displaystyle \frac{i}{2} \sigma_1 }
    & \to  &
   g_1 (t) & = \matrixii{\cosh t/2}{\sinh t/2 }{\sinh t/2}{\cosh t/2}
    \comma \\
 \tau^2 &   = & 
{\displaystyle \frac{i}{2} \sigma_3 }
    & \to  &
    g_2 (t) & =  \matrixii{ e^{t/2}}{0}{0}{ e^{-t/2}}
    \comma
  \end{array} \nn 
\eqaend
and $ \sigma_i \ (i = 1$-$3)$ are the Pauli matrices.
In $ \Omega_0 $, $ g_0 (0) $ and $ g_0(4 \pi) $ represent the same point
and $ g_0(t), \  t \in [0,4 \pi) $, 
traces an uncontractable loop in $ \SLtwo $.
If one decompactifies this loop and does not identify  
$ g_0 (0) $ and $ g_0 (4 \pi ) $,
one obtains the universal covering group $ \cSLtwo $.
 $ \tau^a (a = 0,1,2) $ have the properties
\eqabegin
  \left[ \tau^a , \tau^b \right] &=& i \epsilon^{ab}_{\ \ c} \tau^c
 \comma  \qquad
 \Tr \left( \tau^a \tau^b \right) = - \half \eta^{ab} \comma \nn
\eqaend
where $ \eta^{ab} = \mbox{ diag }(- 1, 1, 1) $.
These form a basis of $ \sltwo $.
\parsmallskip
$ \SLtwo $ is isomorphic to $ SU(1,1) $
(and so is $ \sltwo $ to $ su(1,1) $).
An isomorphism is given by
\eqabegin
  && \tilg = T^{-1} \ g \ T \comma \qquad
T = \frac{1}{\sqrt{2}} \matrixii{1}{i}{i}{1}
   \comma \nn
\eqaend
where $ \tilg \in SU(1,1) $ and $ g \in \SLtwo $.
Note that $ \tilg _0 $ is diagonal in $ SU(1,1) $, while so is $ g_2 $
in $ \SLtwo $.
\parbigskipn
{\sl Parametrization}
\parmedskipn
Any matrix $ g $ of $ \SLtwo $, with all its
elements being non-zero, can be represented as
\eqabegin
   g &=& d_1 \ (-e)^{\epsilon_1}\  s^{\epsilon_2} \ p \ d_2 \period \nn
\eqaend
Here, $\epsilon_{1,2} = 0 $ or $ 1 $; $ d_i = $ diag
$ ( e^{\psi_i/2}, e^{- \psi_i/2}) $ $ (i = 1,2 ) $;
\eqabegin
   -e &=&  \matrixii{-1}{0}{0}{-1} , \qquad
    s = \matrixii{0}{1}{-1}{0} ; \nn
\eqaend
and $p$ is one of the following matrices
\eqabegin
  p &=& g_1(\theta)
  \comma \qquad \ -\infty < \theta < + \infty \comma \nn \\
  p &=&  g_0(\theta)
  \comma \qquad -\pi /2 < \theta < + \pi /2
   \period \nn 
\eqaend
Thus, $ \SLtwo $ has eight domains given by
\eqabegin
 D_1 &=& \lmb A_1
  = \matrixii{ e^{\phi} \cosh \theta/2}{ e^{\psi} \sinh \theta/2}
{ e^{-\psi} \sinh \theta/2}{ e^{-\phi} \cosh \theta/2}
  \comma \  -\infty <  \theta < + \infty \rmb  \comma \nn \\
 D_2 &=& \lmb  A_2
  = \matrixii{ e^{\phi} \cos \theta/2}{ e^{\psi} \sin \theta/2}
{ - e^{-\psi} \sin \theta/2}{ e^{-\phi} \cos \theta/2}
   \comma
   \ - \frac{\pi}{2} < \theta < + \frac{\pi}{2}  \rmb  \comma \nn \\
D_3 &=& \lmb A_3
  = \matrixii{ - e^{\phi} \sin \theta/2}{ e^{\psi} \cos \theta/2}
{ - e^{-\psi} \cos \theta/2}{ - e^{-\phi} \sin \theta/2}
   \comma
   \ - \frac{\pi}{2} < \theta < + \frac{\pi}{2}  \rmb  \comma \nn \\
 D_4 &=& \lmb   A_4
  = \matrixii{ e^{\phi} \sinh \theta/2}{ e^{\psi} \cosh \theta/2}
{ - e^{-\psi} \cosh \theta/2}{ - e^{-\phi} \sinh \theta/2}
  \comma \ -\infty <  \theta < + \infty \rmb  \comma \nn \\
 - D_i &=& \lmb  - A_i \rmb \quad ( i = 1 \sim 4 ) \nn
\comma
\eqaend
where $  \ -\infty < \phi \comma \psi <   + \infty$.
One can further divide these domains according to the sign
of  $ \theta $. We denote the domains with positive $ \theta $ by
$ \pm  D_i^+ $ and those with negative $ \theta $ by $ \pm D_i^- $.
\parsmallskip
When a matrix element of $ g $ is zero, it is, for example, written by
$ \matrixii{a}{0}{b}{a^{-1}} $.
Taking appropriate limits of $ \pm A_i $ yields such a matrix.
\appsubsection{Unitary representations}
Let us denote the generators of
$ \sltwo $ by $ J^a $ and
consider the basis given by $ I^0 = J^0 $ and $ I^\pm = J^1 \pm i J^2 $.
In this basis, the non-trivial commutation relations read
\eqabegin
  \lbb I^0 , I^\pm \rbb &=& \pm I^\pm \comma \quad
 \lbb I^+ , I^- \rbb = - 2 I^0 \period  \nn 
\eqaend
This basis is natural from the $ su(1,1) $ point of view because
$ I^0 $ corresponds to diagonal elements and $ I^\pm $ are regarded
as ladder operators as in $su(2)$. Using this basis,
one can classify all unitary representations of $ \sltwo $ and hence
those of $ \SLtwo $ and $ \cSLtwo $ \cite{Bargmann},\cite{VK,DLP}.
There are five
classes of the unitary representations of $ \sltwo $ which are labeled by
the Casimir $ \bfC = \eta_{ab} J^a J^b $, $ I^0 $ and
a parameter $ m_0 \in [ \ 0 , 1 ) $:
\begin{enumerate}
 \item Principal continuous series $ T^P_\chi $ : \quad
    Representations realized in
     $ \left\{  \ket{j,m} \right\} $, $ m = m_0 + k $,
     $ 0 \leq m_0 < 1$, $k \in {\bf Z}$ and $ j = -1/2 + i \nu $,
     $ 0 < \nu $.
 \item Complementary (Supplementary) series $ T^C_\chi $ : $ \ $
   Representations  realized in
   $ \left\{  \ket{j,m} \right\}$, $ m = m_0 + k $,
     $ 0 \leq m_0 < 1$, $k \in {\bf Z}$,
     and $ \min \left\{ - m_0, m_0 -1 \right\} < j \leq -1/2 $.
 \item Highest weight discrete series  $ T^H_j $ : \quad
   Representations  realized in
     $\left\{  \ket{j,m} \right\} $, $ m = M_{max} - k$,
     $k \in {\bf Z}_{\geq 0} $ and
     $ j = M_{max} \leq -1/2  $ such that $ I^+ \ket{j,j} = 0 $.
 \item Lowest weight discrete series $ T^L_j $ : \quad
   Representations realized in
     $\left\{  \ket{j,m} \right\} $, $ m = M_{min} +  k$,
     $k \in {\bf Z}_{\geq 0} $ and
     $ j = - M_{min} \leq -1/2 $ such that $ I^- \ket{j,-j} = 0 $.
 \item Identity representation : \quad  The trivial representation
      $ \ket{-1,0} $.
\end{enumerate}
Here, $ \chi $ is the pair $(j,m_0)$;
$ {\bf Z}_{\geq 0} $ refers to  non-negative integers; and
we have denoted the value of
$ \bfC $ by $ -j(j+1) $. Note that $ j $ need not be real although
$ -j(j+1) $ should be and that one can restrict $ j $ to Im $ j > 0 $ for (1)
and $ j \leq -1/2 $ for  the others because $ j $ and $ -(j+1) $
represent the same Casimir.
\parsmallskip
Unitary representations of $ \cSLtwo $ are realized in the same space
$ \{ \ket{j,m} \}$.
For $ \SLtwo $,  the parameters are
further restricted to $ m_0 = 0 , 1/2 $ in  (1), $ m_0 = 0 $ in  (2)
and $ j = $ (half integers) in (3) and (4). We will use the same
notations for the groups as in $ \sltwo $.
\parsmallskip
{}From the harmonic analysis on $ \cSLtwo $, we find that 
a complete basis
for the square integrable functions on $ \cSLtwo $ is given by
the matrix elements of the principal continuous series, the highest
and lowest weight discrete series.
\newpage
\appsubsection{Tensor products}
Because one has various unitary representations, the decomposition of tensor
products is more complicated than $ SU(2) $. Basic strategy
to get the decomposition is to decompose the tensored representation spaces
into the eigenspaces of the Casimir operator \cite{Pukanszky,Neunhoffer}.
We are interested in tensor products among the unitary representations.
Then, for $ \SLtwo $, the decompositions are given as follows
\cite{VK,MR}:
\parmedskipn
1. \quad  For two discrete series of the same type,
\eqabegin
  T^{L,H}_{j_1} \ \otimes \ T^{L,H}_{j_2} &=&
   \sum_{n=0}^\infty \ \oplus \ T^{L,H}_{j_1 + j_2 - n }
 \period \nn 
\eqaend
\noindent
2. \quad  For two discrete series of different types,
\eqabegin
  T^L_{j_1} \ \otimes \ T^H_{j_2} &=&
  \int_0^\infty \ T^P_{(-1/2 + i \rho, m_0)} \ d \mu (\rho) \ \oplus \
   \sum_{ j = -m_0 -1}^{j_1-j_2} \ \lb  T^L_{j}  \oplus  T^H_{j} \rb
 \comma \nn 
\eqaend
where $ m_0 = j_1 - j_2 $ mod $ \bfZ $
and $ d \mu (\rho) $ is a continuous measure.
We have assumed
$ j_2 \geq j_1 $, but
the opposite case is obtained similarly.
We remark that $ j \leq -m_0 - 1 $ and the identity representation
does not appear in the right-hand side \cite{MR}.\footnote{
In \cite{HB}, it is claimed that the identity
representation does appear as an exceptional case. In our understanding,
they show just the existence of the solution to
the recursion equation for the Clebsch-Gordan coefficients.}

\parmedskipn
3. \quad  For a discrete and a principal continuous series,
\eqabegin
T^{L,H}_{j_1} \ \otimes \ T^P_{(-1/2 + i \rho', m'_0)} &=&
  \int_0^\infty \ T^P_{(-1/2 + i \rho, m_0)} \ d \mu (\rho) \ \oplus \
   \sum_{ j = -m_0 -1}^{-\infty} \  T^{L,H}_{j}
 \comma \nn 
\eqaend
where $ m_0 = m'_0 + j_1 $ mod $ \bfZ $.

\parmedskipn
4. \quad  For two principal continuous series,
\eqabegin
  && T^P_{(-1/2 + i \rho', m'_0)} \ \otimes
    \ T^P_{(-1/2 + i \rho'', m''_0)}  \label{CC} \\
  && \quad =
 \int_0^\infty \ T^P_{(-1/2 + i \rho, m_0)} \ d \mu_1 (\rho) \ \oplus \
  \int_0^\infty \ T^P_{(-1/2 + i \rho, m_0)} \ d \mu_2 (\rho) \ \oplus \
   \sum_{ j = -m_0 -1}^{-\infty}
   \ \lb  T^L_{j} \oplus  T^H_{j} \rb
  \comma \nn
\eqaend
where $ m_0 = m'_0 + m''_0 $ mod $ \bfZ $.
\parmedskipn
The tensor product of a principal and a complementary
series, or that of two complementary series is decomposed into principal
and discrete series like (\ref{CC}) \cite{Pukanszky,Neunhoffer}. 
In the latter,
one complementary series appears additionally in certain cases. 
The tensor product of
a complementary and a discrete series is similar to that of a principal
and a discrete series \cite{Neunhoffer}.
\parsmallskip
The decompositions are determined essentially by local properties of the group
as is clear from the consideration of tensor products of
$ \sltwo $. Thus the decompositions for $ \cSLtwo $ are obtained
by continuing the value of $ m_0 $ and $ j $.
\parsmallskip
The Clebsch-Gordan coefficients have been discussed in
\cite{HB,MR}, \cite{VK,Wybourne}, \cite{CL}.
\appsubsection{Representations in the hyperbolic basis}
In appendix C.2, we have discussed the representations in the basis
diagonalizing
$ J^0 = I^0 $ which is the compact direction of $ \SLtwo $.
One can also consider the basis diagonalizing
$ J^2 $ or $ J^- = J^0 - J^1 $ which are the non-compact
directions \cite{VK}, \cite{MR}, \cite{KMS}-\cite{BP}, \cite{DVV}.
The generators $ J^0 $, $ J^2 $ and $ J^- $
are called elliptic, hyperbolic and parabolic, respectively. An
outstanding feature of the non-compact generators is that they have
continuous spectrum. In what follows, we will
concentrate on representations in the hyperbolic basis.
\parsmallskip
In terms of $ J^\pm \equiv J^0 \pm J^1 $ and $ J^2 $, the commutation
relations are given by
\eqabegin
  \lbb J^+ , J^- \rbb &=& -2 i J^2 \comma \quad
  \lbb J^2 , J^\pm \rbb \ = \  \pm i J^\pm
   \period  
 \label{comrel} 
\eqaend
The latter equation indicates that
the ladder operators $ J^\pm $ change the eigenvalue of $ J^2 $ by
$ \pm i$. This seems to contradict the Hermiticity of $ J^2 $.
However, this is not the case \cite{KMS}:
In general, the eigenvalue of an Hermitian operator with {\it continuous }
spectrum need not be real \cite{AFIO}.
\parsmallskip
For our purpose, however,  it is
convenient to choose spectrum with real values.
Thus, we use the basis of a representation space given by
$ \{ \ket{ \lambda } \} $, where $ \lambda $ is the eigenvalue of
$ J^2 $ and runs through all the real number.
For the principal continuous and the complementary series,
the eigenvalue of $ J^2 $ has multiplicity two. So, the basis has an
index $ \pm $ to distinguish them and is given
by $ \{ \ket{ \lambda } _\pm \} $.
We will omit
this and the other indices to specify representations such as $ j , m_0 , L $
and $ H $ unless we need them.
In the above basis,
an element (a state) of the representation space is given by a
``wave packet"
\eqabegin
  \ket{ \phi } &=&
    \int_{-\infty}^\infty d \lambda \ \phi (\lambda) \ket{\lambda} \comma
    \quad
   \Vert \ \phi \ \Vert ^2 \ = \ \int_{-\infty}^\infty d \lambda  \
    \abs{\phi (\lambda)} ^2 \ < \ \infty \period \nn 
\eqaend
This is analogous to a state in  field theory
where one uses a plane wave basis in infinite space.
Then the generators act on the state as
\eqabegin
  J^2 \ket{\phi} &=&
     \int_{-\infty}^\infty d \lambda \ \lambda \phi (\lambda) \ket{\lambda}
 \comma \nn \\
   J^+ \ket{\phi} &=&  \int_{-\infty}^\infty d \lambda \
    f_+ ( \lambda ) \phi (\lambda - i) \ket{\lambda} \comma   
  \label{J+-}
   \\
  J^- \ket{\phi} &=&  \int_{-\infty}^\infty d \lambda \
    f_- ( \lambda + i ) \phi ( \lambda + i ) \ket{ \lambda }
   \period  \nn
\eqaend
$ f_\pm $ play the role of the matrix elements in this basis.
{}From the
above action,
the commutation rules are realized if
\eqabegin
    && f_+ (\lambda )  f_- (\lambda )
     -  f_- (\lambda + i )  f_+ (\lambda + i ) = - 2 i \lambda \period
  \label{fpfm}
\eqaend
An eigenstate $ \ket{\lambda '} $ is obtained in the limit
$ \phi (\lambda ) \to \delta ( \lambda - \lambda ') $.
\parsmallskip
It is possible to introduce $ \ket{\lambda \pm i} $ and write the action
of the generators as
\eqabegin
   J^+ \ket{\phi} &=&  \int_{-\infty}^\infty d \lambda \
    f_+ ( \lambda + i) \phi (\lambda) \ket{\lambda + i} \comma \nn \\
 J^- \ket{\phi} &=&  \int_{-\infty}^\infty d \lambda \
    f_- ( \lambda ) \phi (\lambda) \ket{\lambda - i} \comma \nn
   \\
   J^+ \ket{ \lambda } &=& f_+ ( \lambda + i) \ket{\lambda + i}
  \comma \quad
  J^- \ket{ \lambda } \ = \ f_-  ( \lambda ) \ket{\lambda - i}
 \period \nn
\eqaend
In this way, one can formally consider eigenstates $ \ket{ \lambda \pm i } $.
However, we should always understand them
in the sense of (\ref{J+-}). Note that $ \ket{\lambda \pm i} $ can be
``expanded" by the original basis $\{ \ket{ \lambda } \}$, where
$ \lambda \in \bfR $.
\parsmallskip
Now let us consider the matrix elements of $ J^\pm $. In the elliptic basis,
 the Casimir operator takes the form
\eqabegin
   \bfC & = &  \eta_{ab} J^a J^b \nn \\
    & = &  - I^0 ( I^0 + 1) + I^- I^+ \ = \ - I^0 ( I^0 - 1) + I^+ I^-
 \comma \nn 
\eqaend
and the actions of $ I^\pm I^\mp $ are given by
\eqabegin
   && I^- I^+ \ket{j;m} = \tilde{d}^2 (j,m) \ket{j;m} \comma \quad
      I^+ I^- \ket{j;m} = \tilde{d}^2 (j,m-1) \ket{j;m} \comma
  \nn 
\eqaend
where
$ \tilde{d}^2 (j,m) = -j(j+1) + m (m +1)$. Then one obtains the 
norms of $ I^\pm \ket{j;m}  $ and hence the matrix elements of 
$ I^\pm $.
In the hyperbolic basis, the final step 
dose not work because $ ( J^\pm ) ^{\dag} = J^\pm $.
In this case, the Casimir operator takes the form
\eqabegin
  \bfC  & = &  J^2 (J^2 + i) - J^- J^+ = J^2 (J^2 - i) - J^+ J^-
 \comma  \label{casimirh} 
\eqaend
and the actions of $ J^\pm J^\mp $ are given by
\eqabegin
  J^+ J^-  \ket{j;\lambda} &=& d^2 (j, \lambda - i) \ket{j;\lambda}
    \comma  \quad
  J^- J^+  \ket{j;\lambda}  \  = \ d^2 (j, \lambda ) \ket{j;\lambda}
 \comma \label{Jpm}
\eqaend
where 
\eqabegin
   d^2 (j, \lambda ) & \equiv & \lambda (\lambda - i) + j(j+1)
   \period \nn
\eqaend
Note that $ d^2 (j, \lambda - i) = \overline{ d^2 (j, \lambda ) }$ and 
these actions satisfy (\ref{fpfm}). 
One cannot determine the matrix elements of $ J^\pm $ (i.e., $ f_\pm $)
separately without additional conditions. 
We see that (\ref{comrel}), (\ref{casimirh}) and (\ref{Jpm})
are related to
the corresponding equations in the elliptic basis
by the ``analytic continuation"  $  J^\pm \to -i I^\pm $ and
$ J^2 \to i I^0  $ \cite{KMS}.
\newpage
\appsubsection{Matrix elements}
By explicit realization of the representations in spaces of functions, one
can calculate the matrix elements of $ \SLtwo $.
Here we consider the matrix elements in the hyperbolic basis
\cite{VK}, \cite{Mukunda,BP}, \cite{DVV}.
\parsmallskip
First, let us discuss the principal continuous series
$ T^P_\chi $ of $ \SLtwo $.
This representation is realized in a space of functions
on a real axis, $ \Ichi $.
The action of
the group element $ \matrixii{a}{b}{c}{d} \in \SLtwo $ and the inner
product are given by
\eqabegin
  \left(  T^P_\chi (g) f \right) (x) &=& \abs{b x + d} ^{2j}
     \mbox{ sign }^{2m_0} (bx + d) \ f \left(\frac{ax + c}{bx + d} \right)
    \comma \label{repIchi}  \\
 \lb f_1(x) , f_2(x) \rb &=&
   \int_{-\infty}^\infty d x \ \overline{f_1 (x)} f_2 (x)
  \period \label{inprod} 
\eqaend
Then one finds that
\eqabegin
 \psi^\chi_{\lambda \pm} (x) & \equiv &
      \frac{1}{\sqrt{2 \pi}} x^{- i \lambda + j } \theta (\pm x )
      \comma \quad \lambda \ \in {\bf R}  \comma  \label{psi2}
\eqaend
form an orthonormal basis diagonalizing the action of $ J^2 $, namely, 
\eqabegin
  \lb \psi^\chi_{\lambda \epsilon} (x) , \psi^\chi_{\mu \epsilon'} (x) \rb
   &=& \delta_{\epsilon \epsilon' } \delta (\lambda - \mu )  \comma \nn \\
\lbb T^P_\chi \lb g_2 (t) \rb \psi^\chi_{\lambda \pm} \rbb (x) &=&
   e^{-it \lambda } \psi^\chi_{\lambda \pm} (x) \comma
   \quad g_2 (t)\ \in \Omega_2 \comma \nn 
\eqaend
where $\epsilon \comma \epsilon' = \pm $.
$ \psi^\chi_{\lambda \pm} $ correspond to $ \ket{ \lambda } _\pm $ 
in the previous subsection and are not elements in $ \Ichi $.
\parsmallskip
One can calculate the
matrix elements in the basis (\ref{psi2}) using (\ref{repIchi}) and
(\ref{inprod}). For example, for $ t > 0 $
one has
\eqabegin
    \D{P} ^{\chi}_{\lambda +,\lambda' +} \lb g_1(t) \rb
  & = & \frac{1}{2 \pi } B \lb \mu, -\mu' -2j \rb
 \frac{\cosh^{2j + \mu + \mu'} t/2 }{ \sinh^{\mu + \mu' } t/2 }
    \label{g1++} \\
   &&  \qquad \qquad \times F \lb \mu, \mu'; - 2 j ; - \sinh^{-2} t /2 \rb
   \comma  \nn  \\
   \D{P} ^{\chi}_{\lambda -,\lambda' -} \lb g_1(t) \rb
 & = & \frac{1}{2 \pi } B \lb 1-\mu', \mu' -1 +2(j+1) \rb
 \frac{\cosh^{2j + \mu + \mu'} t/2 }{ \sinh^{4 j + 2 +  \mu + \mu' } t/2 }
    \label{g1--} \\
   & &  \times
  F \lb \mu+ 2j + 1 , \mu' + 2j + 1 ; 2 j + 2 ; - \sinh^{-2} t /2 \rb
  \comma  \nn \\
   \D{P} ^{\chi}_{\lambda \epsilon,\lambda' \epsilon'} \lb g_2(t) \rb & = &
   e^{-i t \lambda }\delta_{\epsilon \epsilon'}
   \delta ( \lambda - \lambda' )  \comma \label{g2}
\eqaend
where $ \mu^{(')} = i \lambda^{(')} - j $. $ F $ and $ B $ are
the hypergeometric and the Euler beta function, respectively.
For $ g_1(t)$,
$ \D{P} ^{\chi}_{\lambda -,\lambda' +} $ is given by a linear
combination of (\ref{g1++}) and (\ref{g1--}), and
$ \D{P} ^{\chi}_{\lambda +,\lambda' -} $ vanishes.
\parsmallskip
The matrix elements for the complementary series are obtained
by analytically continuing the value of $ j $ \cite{Mukunda}.
\parsmallskip
Let us turn to the discrete series $ T^L_j $. This is realized in a
space of analytic functions on
$ {\bf C}_+ $ (the upper half-plane).
(This can also be embedded in the principal continuous series.)
The action of $ g = \matrixii{a}{b}{c}{d} \in \SLtwo $ and the
inner product are given by
\footnote{$ j = -1/2 $ case needs special treatment, but the
matrix elements take the same forms as in  $ j < -1/2 $ cases
\cite{Bargmann,Mukunda}. }
\eqabegin
 \lb T_j^L (g) f \rb (w) &=& \lb b w + d \rb^{2j}
  f \lb \frac{a w + c}{b w + d} \rb \comma \nn \\
 \lb f_1(w) , f_2 (w) \rb &=&
   \frac{i}{2 \Gamma (-2j-1)} \int_{\bfC _+} d w d \wbar \ y^{-2j-2}
   \overline{f_1 (w)} f_2 (w) \comma \nn 
\eqaend
where $ w = x + i y $ and $dw d \wbar= -2i dx dy $.
One then finds that
\eqabegin
\varphi^j_{\lambda}(w) &=& \frac{1}{2^{(j+1)} \pi }
     e^{-\lambda \pi/2}
   \Gamma (-i\lambda -j) \  w^{-i \lambda + j}
  \comma \quad \lambda \ \in \bfR \comma \nn 
\eqaend
form an orthonormal basis diagonalizing $ J^2 $. Thus similarly
to the previous case (or using the fact that $ f(w) $ is determined by
its values on the semi-axis $ w = i y \ (y > 0)$), one obtains the matrix
elements.
$  \D{L} ^{j}_{\lambda,\lambda'} \lb g_1(t) \rb $ is  the same up to
a numerical factor as (\ref{g1++}) and
$ \D{L} ^{j}_{\lambda,\lambda'} \lb g_2(t) \rb $ is given by
  (\ref{g2})  without $ \delta_{\epsilon \epsilon'}$.
\parsmallskip
For the highest weight series $ T^H_j $, one can get the matrix elements from
those of the lowest weight series. By utilizing an automorphism of $ \SLtwo $
called Bargmann's
automorphism of $ \SLtwo $
\eqabegin
 {\cal B} : && \matrixii{a}{b}{c}{d} \to \matrixii{a}{-b}{-c}{d}
 \comma \label{Bauto}
\eqaend
the matrix elements of the highest weight series
 are given
by \cite{VK,Mukunda}
\eqabegin
  \D{H} ^{j}_{\lambda,\lambda'} \lb g \rb &=&
 \D{L} ^{j}_{\lambda,\lambda'} \lb {\cal B} g \rb
 \period \nn 
\eqaend
\parsmallskip
All the matrix elements satisfy the
differential equation
\eqabegin
  \lbb  \Delta - j (j +1) \rbb  D^{j(\chi)}_{\lambda,\lambda'} \lb g \rb
   &=& 0
  \comma \nn 
\eqaend
where
$ \Delta $ is the Laplace operator on $ \SLtwo $ and they are characterized
essentially by local properties of $ \SLtwo $.
Hence, the matrix elements of $ \cSLtwo $ are obtained by continuing the values
of
$ j $ and $ m_0 $.
%
%
\appsection{Decomposition of the Kac-Moody module}
The Clebsch-Gordan decomposition
similar to $ su(2) $ holds for $ \sltwo $ $ (su(1,1)) $ in the
elliptic basis \cite{BOFW}. This argument is valid for the hyperbolic basis as
well with
a slight modification. In appendix D, we will show this \cite{NS}.
\parsmallskip
Let $ V^a $ be a vector operator, i.e.,
\eqabegin
  \lbb J_0^a , V^b \rbb &=& i \epsilon^{ab}_{\ \ c} V^c \comma \nn
\eqaend
and $ \ketp{j;\lambda} $ be an eigenstate of $ \bfC $ and $ J^2 $.
An example is $ V^a = J^a_{-n}$.
$  \ketp{j;\lambda} $ need not be a base state of the Kac-Moody module.
Then let us consider states
\eqabegin
   V^+ J_0^- \ketp{j;\lambda} \comma \qquad
   V^- J_0^+ \ketp{j;\lambda} \comma \qquad
   V^2 \ketp{j;\lambda} \period \label{VJ}   
\eqaend
In the hyperbolic basis,
these states do not vanish in any unitary representation.
{}From (\ref{casimirh}),
the matrix elements of the Casimir operator with respect to
these states are
\eqabegin
  \bfC &= & \matrixiii{c + 2 i \lambda }{0}{i}{0}{c - 2 i \lambda}{-i}
{-2i d^2(j,\lambda -i )}{ 2 i d^2 (j,\lambda)}{ c-2 }
 \comma  \quad \mbox{where} \quad c = -j(j+1) \period \nn 
\eqaend
The trace and determinant in this subspace are given by
\eqabegin
  && \Tr  \bfC = 3 c - 2 \comma \qquad \det \bfC = c^2 (c + 2)
 \period \nn 
\eqaend
In addition, 
it is easy to see that the state $ ( 1,1, -2 \lambda ) $ is an eigenvector
with the Casimir $ \bfC = -j(j+1) $. Thus, the other eigenvalues are
$ -j(j-1) $ and $ -(j+1)(j+2) $ and 
the states in (\ref{VJ}) are decomposed into the $ \sltwo $
representations with $j$-values $ j $ and $ j \pm 1 $.
The corresponding eigenvectors $ \psi_j $ and $ \psi_{j\pm1}$
are given by
\eqabegin
  \psi_j &=& ( 1,1, -2 \lambda )  \comma \nn \\
  \psi_{j-1} &=& \lb -(j + i \lambda ) \comma
    j - i \lambda \comma 2 i (j^2 + \lambda^2) \rb
   \comma \nn \\
    \psi_{j+1} &=& \lb j + 1 - i \lambda \comma  -(j + 1 + i \lambda )
   \comma  2 i ( (j+1)^2 + \lambda^2) \rb   \period \nn
\eqaend
Note that $ \psi_{j+1} $ is obtained
from $ \psi_{j-1} $
by the replacement $ j \to -j-1 $.
\parsmallskip
Here, it may be useful to remark
on the norm of states \cite{BOFW}. Consider
representations where the Casimir operator is Hermitian. The representations
need not be unitary. Furthermore, let 
$ \ket{ \Psi_1} $ and $ \ket{ \Psi_2 } $ be
eigenstates with the Casimir values $ c_1 $ and $ c_2 $, respectively.
Then by evaluating the matrix element
$ \lb \Psi_1 \comma \bfC \Psi_2 \rb
  = \lb \bfC \Psi_1 \comma  \Psi_2 \rb $, one obtains
\eqabegin
  && \lb  \bar{c}_1 - c_2 \rb \bra{\Psi_1} \semiket{\Psi_2} = 0
  \period \nn 
\eqaend
Therefore, for complex $ c_1 $ and $ c_2 $,
the norm vanishes when $ c_1 = c_2 $.
It can be non-zero only when $ c_1 $ and $ c_2 $ are complex conjugate.
Since the extremal states built on a principal continuous series
have complex Casimir values (see section 4.3),
they become physical states with zero norm.
On the other hand, $ \bra{E^+_N} \semiket{E^-_N} $
can be non-zero because their Casimir
values are complex conjugate. (Thus these extremal states are not null.)
%
%
\newpage
\def\thebibliography#1{\list
 {[\arabic{enumi}]}{\settowidth\labelwidth{[#1]}\leftmargin\labelwidth
  \advance\leftmargin\labelsep
  \usecounter{enumi}}
  \def\newblock{\hskip .11em plus .33em minus .07em}
  \sloppy\clubpenalty4000\widowpenalty4000
  \sfcode`\.=1000\relax}
 \let\endthebibliography=\endlist
%
\csectionast{REFERENCES}
%

%
%
\end{document}